\documentclass[%
 reprint,
 amsmath,amssymb,
pra,
]{revtex4-2}

\usepackage{graphicx}
\usepackage{dcolumn}
\usepackage{bm}

\usepackage{multirow}
\usepackage{graphics}
\usepackage{xcolor}
\usepackage{amssymb}
\usepackage{braket}
\usepackage{longtable}
\usepackage{rotating}
\usepackage{diagbox}

\usepackage{scalerel}

\newcommand{\phase}[1]{ \mbox{$ \scaleto{(-1)}{9pt}^{\scaleto{#1\mathstrut}{6pt}} $}}

\newcommand{\MEred}[3]{ \mbox{$\langle #1\,||\,#2\,||\,#3\rangle $} }
\newcommand{\CGC}[6]{ \mbox{$ ( #1 #2 , #3 #4\,|\, #5 #6 ) $} }
\newcommand{\SechsJ}[6]{ \mbox{$ %
            \arraycolsep0.25ex %
            \left\{ \begin{array}{ccc} %
                       #1 & #2 & #3 \vspace{0.5ex}\\%
                       #4 & #5 & #6 %
                   \end{array} \right\} $} }
\newcommand{\NeunJ}[9]{ \mbox{$ %
            \arraycolsep0.25ex %
            \left\{ \begin{array}{ccc} %
                       #1 & #2 & #3 \vspace{0.5ex}\\%
                       #4 & #5 & #6 \vspace{0.5ex}\\%
                       #7 & #8 & #9 %
                   \end{array} \right\} $} }
\begin{document}

\preprint{APS/123-QED}

\title{Evolution of the ionic polarization in multiple sequential ionization: general equations and an illustrative example}

\author{Elena V. Gryzlova}
\affiliation{ 
Skobeltsyn Institute of Nuclear Physics, Lomonosov Moscow State University, 119991 Moscow, Russia
}
\email{gryzlova@gmail.com}
\author{Maksim D. Kiselev}
 \affiliation{ 
Skobeltsyn Institute of Nuclear Physics, Lomonosov Moscow State University, 119991 Moscow, Russia
}
 \affiliation{ 
 Faculty of Physics, Lomonosov Moscow State University, Moscow 119991, Russia
}
 \affiliation{ 
Laboratory for Modeling of Quantum Processes, Pacific National University, 680035 Khabarovsk, Russia
}
\author{Maria M. Popova}
 \affiliation{ 
Skobeltsyn Institute of Nuclear Physics, Lomonosov Moscow State University, 119991 Moscow, Russia
}
 \affiliation{ 
 Faculty of Physics, Lomonosov Moscow State University, Moscow 119991, Russia
}
\author{Alexei N. Grum-Grzhimailo}
\affiliation{ 
Skobeltsyn Institute of Nuclear Physics, Lomonosov Moscow State University, 119991 Moscow, Russia
}

\date{\today}

\begin{abstract}
The modern Free-Electron-Lasers generate a highly intense polarized radiation which initiate a sequence of ionization and decay events. Their probability  depends on the polarization of each state as function of time. Its complete accounting is limited by the fact that a state can be formed in various ways. Here we present the equivalent of rate equations for population that completely accounts polarization of radiation and formulated in terms of the statistical tensors. To illustrate our approach we theoretically consider sequential photo\-ionization of krypton by an intense extreme ultraviolet femto\-second pulse for the photon energies below the $3d$-shell excitation threshold. The calculations of
the ion yields, photo\-electron spectra  and ionic polarization for various photon fluence are presented and role of polarization is discussed.
\end{abstract}


\maketitle


\section{\label{sec:i} Introduction}

When an atom is irradiated by intense electromagnetic field generated by free-electron laser (FEL) operating in the extreme ultraviolet (XUV), the first photo\-ionization act initiates the variety of competitive processes, such as sequential ionization, Auger decay, radiation decay and others. The sample evolution depends on the radiation parameters: intensity, pulse duration and polarization. The last is often left behind the scenes, in particular, because accounting for the polarization increases number of degrees of freedom enormously.
The knowledge about charge and state evolution of an irradiated sample is crucially important for a number of applications such as modeling the radioactive damage of biological samples for coherent diffraction imaging  and as a fundamental test of the photo\-ionization description basis
\cite{Nass2015,Galli2015}.  

Multiple ionization of atoms by FELs  has been 
subject of numerous and extensive investigations since the first observation at the Free-electron LASer in Hamburg 
(FLASH)~\cite{Sorokin2007}. 
Roughly multiple ionization may proceed in the two regimes: (a) strong field regime involving multiphoton direct single or multiple ionization \cite{Kubel2016,Moshammer2007,Kanter2011,Fushitani2020}; and (b) multi\-photon sequential regime proceeding with creation and subsequent ionization of different intermediate ion(s) with their possible excitation to discrete or auto\-ionizing states \cite{Young2010,Gerken2014,Richter2010,Klumpp2017,Berrah2014,Fukuzawa2013,Southworth2019,Kurka2009,Braune2016,Carpeggiani2019,Mazza2020}.  Current research belongs to the second group. In general, both regimes may co-exist and question to attribute a process to the first regime or to the second depends on a region of a considered  photo\-electron spectrum rather than on the pulse parameters.

One of the advantage of FELs is that generated radiation is highly polarized either linearly or circularly. There are a variety of researches devoted to an appearance of the polarization effects in the differential observable characteristics of sequential ionization: from photo-electron angular distribution \cite{Kheifets2007,Grum2016,Mondal_2013,Braune2016,Ilchen2018} and angular correlation \cite{Kurka2009,Augustin2018}  up to recent realization of {\it complete/perfect} experiment \cite{Carpeggiani2019} or ion-ion correlation \cite{Fushitani2020}. We are not aware about angle-resolved experiments proceeding with  forming more than triple-charged ions \cite{Rouzee2011,Grum2016}.  

On the other side, there are many researches of polarization effects in integral cross-section mainly in a integral linear or circular dichroism within pump-probe scheme \cite{Kabachnik2007,OKeeffe2013,Wernet2001,Meyer2011,Wedowski1997,Ilchen2017}. Polarization of an ionized state may appear as variation of ionization probability up to complete suppression.   The dynamically  quasiforbidden transitions in photo\-ionization of pumped open-shell atoms were found and interpreted \cite{Cubaynes2004}.  

In spite of numerous investigations of dichroism have shown that polarization may affect integral characteristics such as ionic yield and photo\-electron spectrum, to the best of our knowledge there are no studies investigating multiple ionization with complete accounting of intermediate states polarization. These type of researches is quite resource-consuming because they suppose to solve a system of the rate equations for all affected magnetic sub-levels instead of having one equation for one state that increase number of equations significantly \cite{Nortershauser2021}. Therefore, most of researches use relevant ionization/excitation cross sections by polarized radiation neglecting influence of sub-sequined steps to former one which may cause depletion of target magnetic sub-levels.  

In the manuscript we developed method suitable for an atom (ion)  in linearly or circularly polarized radiation under assumption that the levels are separated and populated incoherently. The method is based on solution the system of equations for statistical tensors similar to the system of rate equations for populations which is widely used in the description atom-field interactions \cite{Karamatskos2013,Nakajima2002,Makris2009,Son2011,Son2012,Lorenz2012,Lunin2015,Serkez2018,Buth2018} but presented in terms of statistical tensors.

As an illustrative example we consider sequential ionization of krypton by electromagnetic pulse with photon energy 60--80 eV that is below the lowest excitation energy of $3d$ shell ($3d^{-1}5p[5/2]_1$ 91.2~eV~\cite{King1977}). For both single or double $4s$-vacancy states the Auger decay is energetically forbidden, therefore the dominant processes govern over temporal dynamics  are subsequent photo\-ionization from the $4s$ and the $4p$ orbitals with emission of a photo\-electron $e_{ph}$:

\begin{equation}\label{eq:pr1}
\gamma + {\rm Kr^{n+}} \, 4s^k4p^m \longrightarrow  \bigg\{  { {\rm Kr^{(n+1)+}}\, 4s^{k-1}4p^m + e_{ph} \atop {\rm Kr^{(n+1)+}} \, 4s^k4p^{m-1} + e_{ph} } 
\end{equation}
Relaxation of $4s$-vacancies proceeds via radiative transition from the $4p$ to the $4s$ level with fluorescence of a  photon $\gamma_{fl}$
\begin{eqnarray} \label{eq:pr2}
 {\rm Kr^{n+}} \, 4s^k4p^m \longrightarrow {\rm Kr^{n+}} \, 4s^{k+1}4p^{m-1} + \gamma_{fl}
\end{eqnarray}
We consider relatively short femto\-second pulses and the relaxation transitions can be neglected.

\begin{figure}
\centering
\includegraphics[width=0.49\textwidth]{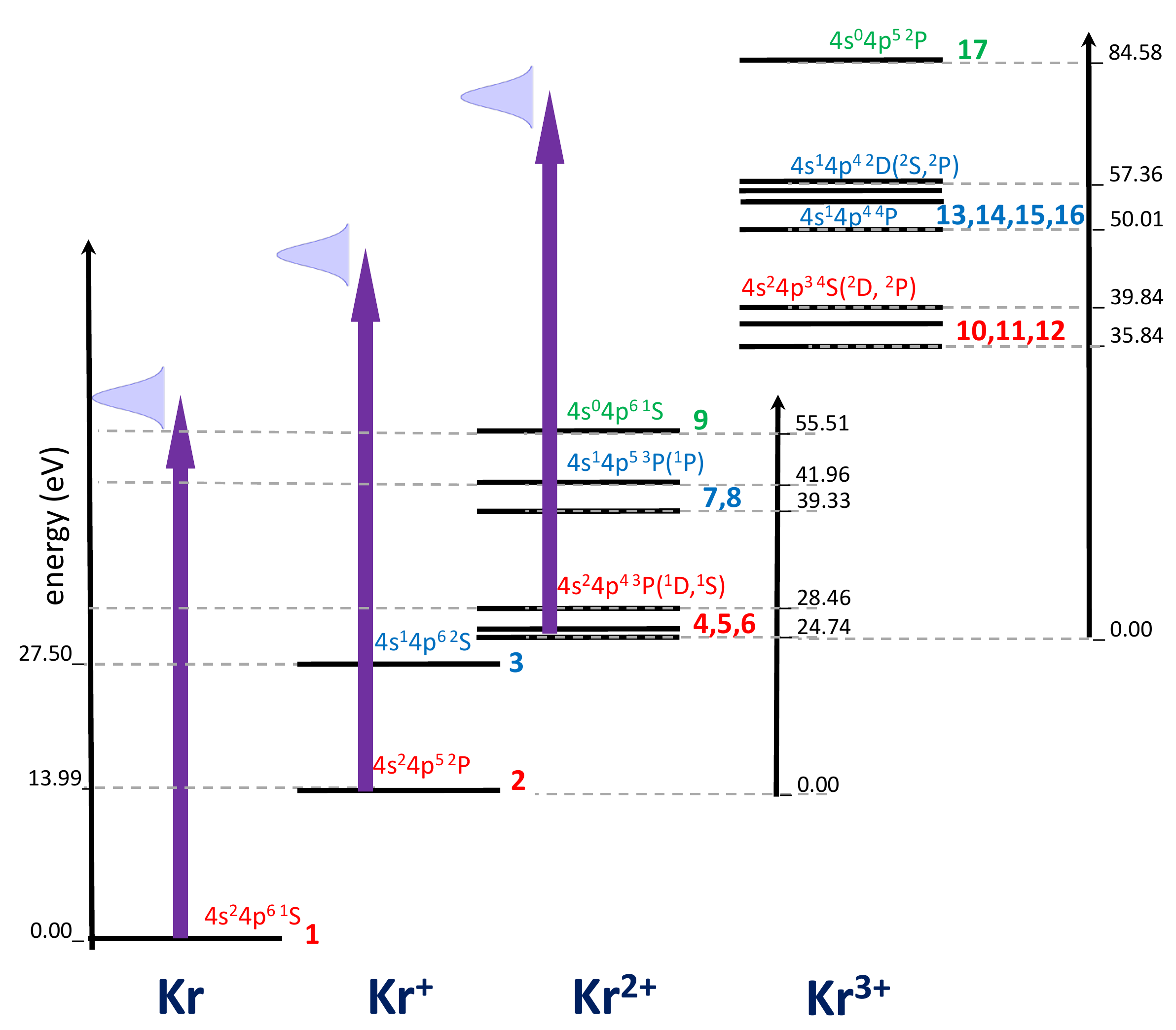} 
\caption{The scheme of transitions in sequential multi\-photon ionization of Kr. The experimental ionization thresholds were taken from~\cite{NIST} and averaged over a multiplet.}
\label{fig:1}
\end{figure}

Sequential ionization from neutral Kr to the triply charged ion Kr$^{3+}$ involving $4s$ and $4p$ shells of Kr and its ions is shown 
in Figure~\ref{fig:1}, where the configurations, terms and relative energies are indicated. The additional details needed for interpretation of the photo-electron spectrum are presented in Table~\ref{tab1}.
   We neglect fine-structure splitting, do not include shake-up, direct two-photon and one-photon
double ionization channels~\cite{Ilchen2016} and consider the sequential ionization up to triple three-photon ionization exactly in the same setup as we did in \cite{Gryzlova2020} for {\it unpolarized} radiation.

In the next section we outline a theoretical approach 
for modeling the atom-field interaction based on the solution of a system of the rate equations for statistical tensors. In Section~\ref{sec:4}, we present throughout analysis of the ionic polarization.  In Section~\ref{sec:5}, the results on the time evolution  of the Kr target under the FEL pulse (ionic yields) and the resulting photo\-electron spectra are presented and discussed.
We use atomic units until otherwise~indicated.

\section{\label{sec:3} System of equations for evolution of the statistical tensors}

The goal of this section is to formulate equivalent of rate equations accounting polarization of the states in terms of the statistical tensors for linearly and circularly polarized radiation in dipole approximation. Under these conditions statistical tensor of photons $\rho_{{k_{\gamma}}q_{\gamma}}\neq 0$ for $q_{\gamma}=0$ and $k_{\gamma} \leq 2$, and leaving treating of an electron-ionic correlation out, only zero rank statistical tensors of an electron $\rho_{k_e=0,q_e=0}$ exist. Therefore all non-zero statistical tensors of ionic states have projection $q=0$. 

It is well established \cite{Balashov2000,Blum96} that if a state $i$ of a $n$-charged ion $\rm A^{n+}$ with total angular momentum $J_i$ absorbs a (dipole) photon and ionizes into a state $f$ of  $\rm A^{(n+1)+}$ with total momentum $J_f$, the statistical tensors of initial $\rho_{k_i0}(J_i)$ and final $\rho_{k_f0}(J_f)$ states are connected by the equation:
\begin{eqnarray}\label{eq:rho1}
 \rho_{k_f0}(J_f)&=&\sum_{k_{\gamma}} \rho_{k_i0}(J_i) S[k_i,k_{\gamma},k_f]\,,
\end{eqnarray}
where transition parameter $S[k_i,k_{\gamma},k_f]$  is presented in terms of reduced ionization amplitudes as

\begin{eqnarray}\label{eq:S}
S[k_i,k_{\gamma},k_f]&=&4\pi^2\alpha\omega\rho_{k_{\gamma}0}B[k_i,k_{\gamma},k_f]\,;\\
  B[k_i,k_{\gamma},k_f]&=&\hat{k}_i\hat{k}_{\gamma}\CGC{k_i}{0}{k_{\gamma}}{0}{k_f}{0}\frac{\hat{J}_f}{\hat{J}_i}\phase{J_f+k_f}\nonumber\\
 && \sum_{jJJ'}\hat{J}\hat{J}'\phase{J'+j}\SechsJ{J_f}{J_f}{k_f}{J}{J'}{j}\NeunJ{J_i}{1}{J}{J_i}{1}{J}{k_i}{k_{\gamma}}{k_f}\nonumber\\
 &&\MEred{(J_fj)J}{\hat{D}}{J_i}\MEred{(J_fj)J'}{\hat{D}}{J_i}^{\ast}\,.   \label{eq:B}
\end{eqnarray}
Here $\hat{a}=\sqrt{2a+1}$, $\hat{D}$ is the operator of the atomic electric dipole momentum, and usual notation for Clebsch-Gordan coefficients, 6j- and 9j-symbols are used. Throughout the manuscript we use non-conventional normalization of the statistical tensors: if state $J_a$ is completely populated, $\rho_{00}(J_a)=1$ (instead $\rho_{00}(J_a)=1/\hat{J_a}$) that allows to consider zero rank statistical tensors as population in percents (summed over all states population $\sum_a\rho_{00}(J_a)=1$). 
Parameter $S[0,0,0]$ is the ionization cross-section of an unpolarized state with $J_i$ to an ion with $J_f$.

The transparent way to get time-dependent form of (\ref{eq:rho1})  is to start with conventional rate equations for the level populations:  
\begin{eqnarray} \label{eq:rate}
\frac{dN_{aM_a}(t)}{dt} &=& j(t) \hspace{-0.2cm} \sum_{b \neq a,M_b}^L \hspace{-0.2cm} 
\left[ \sigma_{bM_b \rightarrow aM_a} \, N_{bM_b}(t) - \right.\nonumber\\
&&\hspace{25pt}-\left.\sigma_{aM_a \rightarrow bM_b}  \, N_{aM_a}(t) \right] \,,
\end{eqnarray}
where $N_{aM_a}(t)$ is the population of sub\-level $a$ with magnetic quantum number $M_a$ and $j(t)$ is the time-dependent intensity of the incident radiation (envelope), $L$ is the number of the accounted states,  $\sigma_{aM_a \rightarrow bM_b}$ is the photo\-ionization cross section from sub\-level $aM_a$ of $\rm A^{n+}$ to sub\-level $bM_b$ of $\rm A^{(n+1)+}$ which is connected with transition parameter (\ref{eq:S}):
\begin{eqnarray}\label{eq:sech}
\sigma_{iM_i \rightarrow fM_f}\hspace{-0.1cm}&=& \hspace{-0.05cm}\frac{\hat{J}_i}{\hat{J}_f}\hspace{-0.1cm}\sum_{k_ik_{\gamma}{k_f}} \hspace{-0.15cm}\phase{J_i-M_i+J_f-M_f}\CGC{J_i}{M_i}{J_i}{-\hspace{-0.1cm}M_i}{k_i}{0}\nonumber\\
&&\hspace{20pt}\CGC{J_f}{M_f}{J_f}{-\hspace{-0.1cm}M_f}{k_f}{0} S[k_i,k_{\gamma},k_f]\,.
\end{eqnarray}
The total ionization cross section of an unpolarized state by unpolarized radiation is an averaged sum of cross sections from magnetic sub\-levels: 
\begin{eqnarray}
\sigma&=\dfrac{\sum_{M_iM_f}\sigma_{iM_i \rightarrow fM_f}}{2J_i+1}&\equiv S[0,0,0]\,.
\end{eqnarray}

The statistical tensor corresponding to a state $a$ is constructed from its population $N_{aM_a}(t)$  straightforward by the definition:
\begin{eqnarray}\label{eq:rho2}
 \hspace{-0.5cm}\rho_{k_a0}(J_a)\hspace{-0.05cm}=\hspace{-0.05cm}\hat{J}_a\hspace{-0.1cm}\sum_{M_a}\hspace{-0.1cm}\phase{J_a-M_a}\CGC{J_a}{M_a}{J_a}{-\hspace{-0.1cm}M_a}{k_a}{0}N_{aM_a}\!;\end{eqnarray}
\begin{eqnarray}
 \hspace{-0.5cm}N_{aM_a}\hspace{-0.1cm}=\hspace{-0.1cm}\frac{1}{\hat{J}_a}\hspace{-0.15cm}\sum_{k_a}\hspace{-0.1cm}\phase{J_a-M_a}\CGC{J_a}{M_a}{J_a}{-\hspace{-0.1cm}M_a}{k_a}{0}\rho_{k_a0}(J_a)\!.
\end{eqnarray}
Collecting the all above, we are now able to write down the analogue of the rate equations (\ref{eq:rate}) in terms of statistical tensor:
\begin{eqnarray} \label{eq:drho0}
\frac{d\rho_{k_a0}(J_a)}{dt}=\frac{d\rho_{k_a0}(J_a)}{dt}\Bigg{|}_{\rm in}-\frac{d\rho_{k_a0}(J_a)}{dt}\Bigg{|}_{\rm out}\,,
\end{eqnarray}
where the term describing pumping of population ($b=i$ is initial, $a=f$ is final) is quite simple:
\begin{widetext}
\begin{eqnarray} \label{eq:drho1}
&&\hspace{-0.4cm}\frac{d\rho_{k_a0}(J_a)}{dt}\Bigg{|}_{\rm in}=j(t)\hat{J}_a\hspace{-0.2cm}\sum_{M_abM_b}\hspace{-0.2cm}\phase{J_a-M_a}\CGC{J_a}{M_a}{J_a}{-\hspace{-0.1cm}M_a}{k_a}{0}\sigma_{bM_b \rightarrow aM_a}N_{bMb}\nonumber\\
&=&j(t)\hspace{-0.3cm}\sum_{M_abM_b\atop k'_ak_{\gamma}k_bk'_b}\hspace{-0.3cm}\CGC{J_a}{M_a}{J_a}{-\hspace{-0.1cm}M_a}{k_a}{0}\CGC{J_a}{M_a}{J_a}{-\hspace{-0.1cm}M_a}{k'_a}{0}\CGC{J_b}{M_b}{J_b}{-\hspace{-0.1cm}M_b}{k_b}{0}\CGC{J_b}{M_b}{J_b}{-\hspace{-0.1cm}M_b}{k'_b}{0}S[k_b,k_{\gamma,k_a'}]\rho_{k'_b0}(J_b)\nonumber\\
&=&j(t)\hspace{-0.1cm} \sum_{k_{\gamma}bk_b}\hspace{-0.1cm} S[k_b,k_{\gamma},k_a]\rho_{k_b0}(J_b)\,.
\end{eqnarray}
\end{widetext}

The term describing the leaking of population ($a=i$ is initial, $b=f$ is final) is much more trickier:
\begin{widetext}
\begin{eqnarray}\label{eq:drho2}
&&\frac{d\rho_{k_a0}(J_a)}{dt}\Bigg{|}_{\rm out}=j(t)\hat{J}_a\hspace{-0.2cm}\sum_{M_abM_b}\hspace{-0.2cm}\phase{J_a-M_a}\CGC{J_a}{M_a}{J_a}{-\hspace{-0.1cm}M_a}{k_a}{0}\sigma_{aM_a \rightarrow bM_b}N_{aMa}\nonumber\\
&&\hspace{2.2cm}=j(t)\frac{\hat{J}_a}{\hat{J}_b}\hspace{-0.3cm}\sum_{M_abM_b\atop k'_ak''_ak_{\gamma}k_b}\hspace{-0.3cm}\phase{J_a-M_a}\CGC{J_a}{M_a}{J_a}{-\hspace{-0.1cm}M_a}{k_a}{0}\CGC{J_a}{M_a}{J_a}{-\hspace{-0.1cm}M_a}{k'_a}{0}
\CGC{J_a}{M_a}{J_a}{-\hspace{-0.1cm}M_a}{k''_a}{0}\nonumber\\
&&\hspace{2.5cm}\phase{J_b-M_b}\CGC{J_b}{M_b}{J_b}{-\hspace{-0.1cm}M_b}{k_b}{0} S[k'_a,k_{\gamma},k_b]\rho_{k''_a0}(J_a)\nonumber\\
&&\hspace{2.2cm}=j(t) \hat{J}_a \hspace{-0.2cm}\sum_{k'_ak''_ak_{\gamma}}\hspace{-0.2cm} \phase{k_a''}\hat{k}'_a\CGC{k_a}{0}{k'_a}{0}{k''_a}{0} \SechsJ{k_a}{k'_a}{k''_a}{J_a}{J_a}{J_a}S[k_a',k_{\gamma},0]\rho_{k''_a0}(J_a)\,.
\end{eqnarray}
\end{widetext}
The feature of this equation is that, while in the rate equations for population decreasing of a population is always proportional to it, in the rate equations for statistical tensors decreasing of a tensor depends on other tensors of the state.

Having the equations (\ref{eq:drho0})--(\ref{eq:drho2}), one can solve the system for statistical tensors which is completely similar to the conventional rate equations apart from the specific form of the coefficients. The advantage of this approach is that with more states are involved, the  size of system (\ref{eq:drho0}) increases slower than the size of system  (\ref{eq:rate}) where magnetic quantum numbers are directly accounted. In our case (see Table~\ref{tab1}) $L=17$, the number of magnetic sub\-levels is $\sum_a(2J_a+1)=45$, the number of statistical tensors is 31.

Time-dependent alignment and orientation of an ion are the ratio of the corresponding statistical tensors:
\begin{eqnarray}\label{eq:A2}
A_{2}(J_f)=\frac{\rho_{20}(J_f)}{\rho_{00}(J_f)}\,,~A_{1}(J_f)=\frac{\rho_{10}(J_f)}{\rho_{00}(J_f)}\,.
\end{eqnarray}

For forthcoming discussion let's introduce conventional {\it stationary} alignment of a state $f$:
\begin{eqnarray}\label{eq:A2st}
\mathcal A_f=\frac{-\sqrt{2}B[0,2,2]+\mathcal A_i(B[2,0,2]-\sqrt{2}B[2,2,2])}{B[0,0,0]-\sqrt{2}B[2,2,0]}\,.
\end{eqnarray}
 By {\it stationary} alignment we named the alignment obtained in some branch $i\rightarrow f$ neglecting  all other pathways as well as depletion/saturation of the considered state. 
 
 There are maximal and minimal values of an alignment directly following from the statistical tensor definition:
 \begin{eqnarray}
 A_2(J)&=&\frac{\sum_M\phase{J_M}\CGC{J}{M}{J}{-\hspace{-0.1cm}M}{2}{0}N_{JM}}{\sum_M\phase{J_M}\CGC{J}{M}{J}{-\hspace{-0.1cm}M}{0}{0}N_{JM}}\,;
 \end{eqnarray}
that gives:
\begin{eqnarray}
 \hspace{-0.4cm}A_2(P)&=&\hspace{-0.1cm}\frac{1}{\sqrt{2}}\frac{N_{11}+N_{1-1}-2N_{10}}{N_{11}+N_{1-1}+N_{10}}\,;\label{eq:limP}\\
 \hspace{-0.4cm}A_2(D)&=&\hspace{-0.1cm}\sqrt{\frac{10}{7}}\frac{N_{22}+N_{2-2}-(N_{21}+N_{2-1})/2-N_{20}}{N_{22}+N_{2-2}+N_{21}+N_{2-1}+N_{10}}\!.\label{eq:limD}
 \end{eqnarray}
It is obvious that for $P$-term states maximal and minimal alignment is $1/\sqrt{2}$ and $-\sqrt{2}$, correspondingly (\ref{eq:limP}); for $D$-term they are $\sqrt{10/7}$ and $-\sqrt{10/7}$ (\ref{eq:limD}).
 
The Gaussian distribution of the incident {\it photon flux
density} is assumed:
\begin{equation} \label{eq:pulse}
j(t) =  j_0 \, \exp (- t^2/t_p^2) \,,
\end{equation}
thus the pulse full width at half maximum ${\rm FWHM = 2\sqrt{\ln 2}} \; t_p$. 
The integral number of photons per 1 \AA$^2$ in the entire pulse i.e. {\it fluence} $F$
 is related to the intensity as
\begin{equation} \label{eq:fluence}
j_0=
\frac{2 \sqrt{\ln 2} \; F}{\sqrt{\pi} \; {\rm FWHM}}
= 0.0063634 \, \frac{F \, [{\rm ph/\rm \mathring{A}^2}]}{\rm FWHM \, [fs]} \,.    
\end{equation}

The typical pulse duration obtained at the seeded FEL FERMI~\cite{Allaria2012,Finetti2017} is around 50--100~fs; we set the pulse duration to $t_p = 60$ fs (FWHM = 100 fs).

\section{\label{sec:4} Analysis of the ions alignment}

Photo\-ionization cross sections of $4s^k4p^{m \; 2S+1}L$ multiplets for Kr ions in different charge states were calculated by means of the B-spline R-matrix code~\cite{Zatsarinny2006}. For each of the ionization steps the basis wave functions of the initial nine and the final seventeen states listed in Table~\ref{tab1} were obtained in the way described in \cite{Gryzlova2020}, where one can also find the photo\-ionization cross sections curves for each step and comparison with another theories and experiments for neutral krypton ionization.

\begin{figure}
\centering
\includegraphics[width=0.49\textwidth]{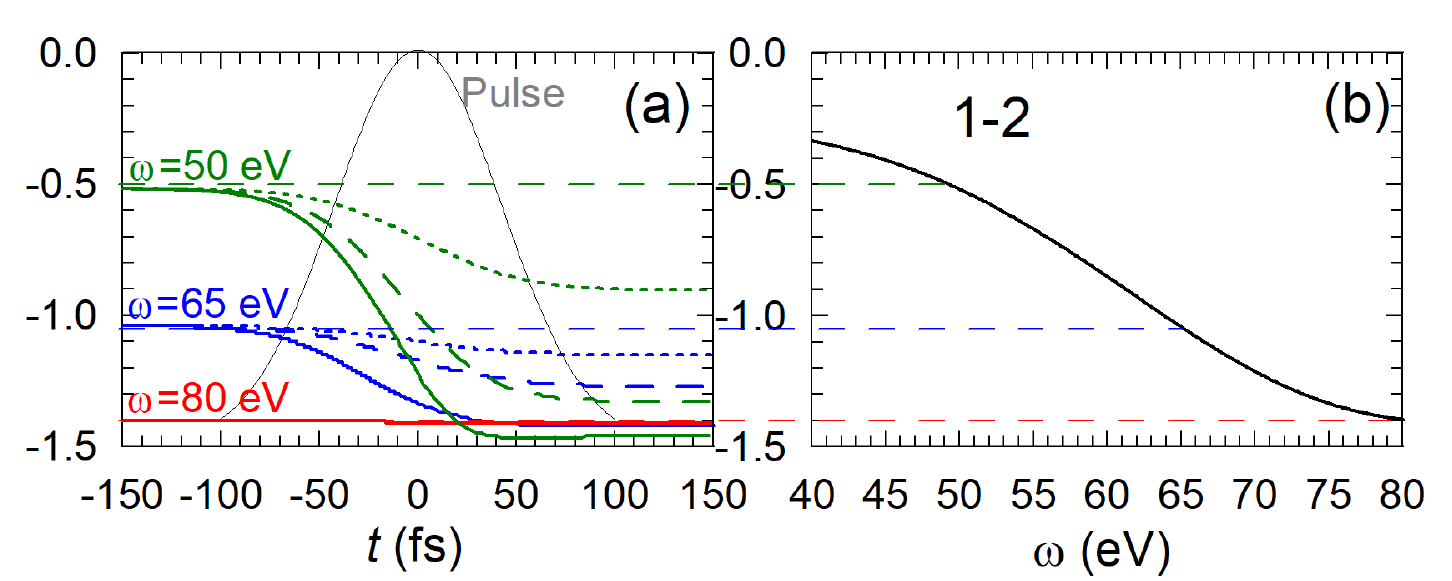}
\caption{(a) The alignment of Kr$^+4s^24p^{5\,2}P$ ion calculated as function of time at $\omega=50$, 65 and 80~eV. The solid, dashed and dotted lines correspond to fluence $F=3000$, 1000 and 400 ph/\AA$^2$. (b) Conventional alignment of the same ion as function of energy (accounting difference in definitions this data are in accordance with  \cite{KLEIMAN200329}). Here '1-2' and after 'a-b'  indicates the transition between the states labeled according to Table~\ref{tab1}.}
\label{fig:2}
\end{figure}

Figure~\ref{fig:2} illustrates the main idea of the investigation: to account dynamical change of polarization and its impact to the observable values. It shows alignment of the 'first' ion Kr$^+$\,$4s^24p^{5\;2}P$ as function of time  (eqn.~\ref{eq:A2}) and its stationary alignment (eqn.~\ref{eq:A2st}). The conventional alignment (figure~\ref{fig:2}(b)) as function of energy has a broad maximum of absolute value caused by Cooper minimum of $4p\rightarrow \varepsilon d$ ionization amplitude. At the Cooper minimum alignment approaches to $-\sqrt{2}$ that is minimal possible value allowed by eqn.~(\ref{eq:limP}). 
One can see (figure~\ref{fig:2}(a)) that at the pulse beginning the alignment is equal to the conventional value at corresponding photon energy: $-0.5$ ($\omega=50$ eV), $-1$ ($\omega=65$ eV) and $-1.41$ ($\omega=80$ eV). Then as linearly polarized pulse ionizes the ion preferably from $|m|=1$, the alignment tends to get stronger, and it is clearly seen for $\omega=50$ and $\omega=65$~eV. Moreover, originally smaller alignment at $\omega=50$~eV may became stronger than at $\omega=65$~eV due to dynamical effects. At $\omega=80$~eV the tendency is suppressed by the complete depletion of $|m|=1$ sub\-levels.

\begin{table*}
\begin{center}
\caption{List of the pathways in sequential three-photon ionization of Kr  within the LS-coupling scheme. The columns and lines correspond to the initial and final states of the pathway, respectively. We denote different transitions contributed to a one photo\-electron line  by a capital letter. Numbers are experimental ionization thresholds~\cite{NIST}  for the corresponding transitions in~eV, averaged over a multiplet. Unmarked transitions are weak and their contribution to the photo\-electron spectra is negligible.}

\label{tab1}
\scalebox{0.78}[0.78]{

\begin{tabular}{rlccccccccc}
\noalign{\hrule height 1pt} 
\textbf{N}&  \diagbox[width=2.8cm, height=0.9cm]{\textbf{Final State}}{\textbf{Initial State}}  &
\begin{turn}{90}{ \boldmath{$4s^24p^{6\; 1}\!S$}} \end{turn}& \begin{turn}{90}{\boldmath{$4s^24p^{5\;2}\!P^o$} }\end{turn}& 
\begin{turn}{90}{\boldmath{$4s^14p^{6\; 2}\!S$}}\end{turn} &
\begin{turn}{90}{\boldmath{$4s^24p^{4\;3}\!P$} }\end{turn}&
 \begin{turn}{90}{\boldmath{$4s^24p^{4\;1}\!D$}}\end{turn}& 
\begin{turn}{90}{\boldmath{$4s^24p^{4\;1}\!S$} }\end{turn}& \begin{turn}{90}{\boldmath{$4s^14p^{5\;3}\!P^o$}} \end{turn}&
\begin{turn}{90}{ \boldmath{$4s^14p^{5\;1}\!P^o$}}\end{turn} &
\begin{turn}{90}{ \boldmath{$4s^04p^{6\;1}\!S$}}\end{turn} \\
\noalign{\hrule height .5pt}
  1&$4s^24p^6 \; ^1\!S$&  -    & -         & -        & -     & -     & -     &   -  &  -   & - \\
  2&$4s^24p^{5\;2}\!P^o$ &A, 14.0& -         & -        & -     & -     & -     & -    & -    & - \\
  3&$4s^14p^{6\;2}\!S$ &E, 27.5& -         & -        & -     & -     & -     & - & - & - \\
  4&$4s^24p^{4\;3}\!P$ & -     & B, 24.4   & -        & -     & -     & -     & - & - & - \\
  5&$4s^24p^{4\;1}\!D$ & -     & D, 26.2   & -        & -     & -     & -     & - & - & - \\
  6&$4s^24p^{4\;1}\!S$ & -     & F, 28.5   & -        & -     & -     & -     & - & - & - \\
  7&$4s^14p^{5\;3}\!P^o$ & -     & I, 38.7   &C, 25.2   & -     & -     & -     & -     & -    & - \\
  8&$4s^14p^{5\;1}\!P^o$ & -     & K, 42.0   &F, 28.4   & -     & -     & -     & -     & -    & - \\
  9&$4s^04p^{6\;1}\!S$ & -     & -         &K, 42.1   & -     & -     & -     & -     & -    & - \\
  10&$4s^24p^{3\;4}\!S^o$ & -    & -         & -        &G, 35.8& -     & -     & -     & -    & - \\
  11&$4s^24p^{3\;2}\!D^o$ & -    & -         & -        &J, 38.0&H, 36.1&   33.8&-      &-          &-\\
  12&$4s^24p^{3\;2}\!P^o$ & -    & -         & -        &K, 39.7&I, 37.9&G, 35.6& -     & -         & - \\
  13&$4s^14p^{4\;4}\!P$ & -    & -         & -        &M, 50.6& -     & -     &G, 35.5& - & - \\
  14&$4s^14p^{4\;2}\!D$ & -    & -         & -        &O, 53.9&N, 52.1&   49.8&J, 38.8&H, 36.3&- \\
  15&$4s^14p^{4\;2}\!S$ & -    & -         & -        &S, 57.4&  Q, 55.6&   53.3&L, 42.3&K, 39.8& - \\
  16&$4s^14p^{4\;2}\!P$ & -    & -         & -        &R, 56.4&P, 54.6&N, 52.3&L, 41.4&J, 38.8&-\\
  17&$4s^04p^{5\;2}\!P^o$ & -    & -         & -        & -     & -     & -     &P, 54.9&N, 52.3&J, 38.7\\
\end{tabular}}
\end{center}
\end{table*}

\begin{figure*}
\begin{center}
\includegraphics[width=0.9\textwidth]{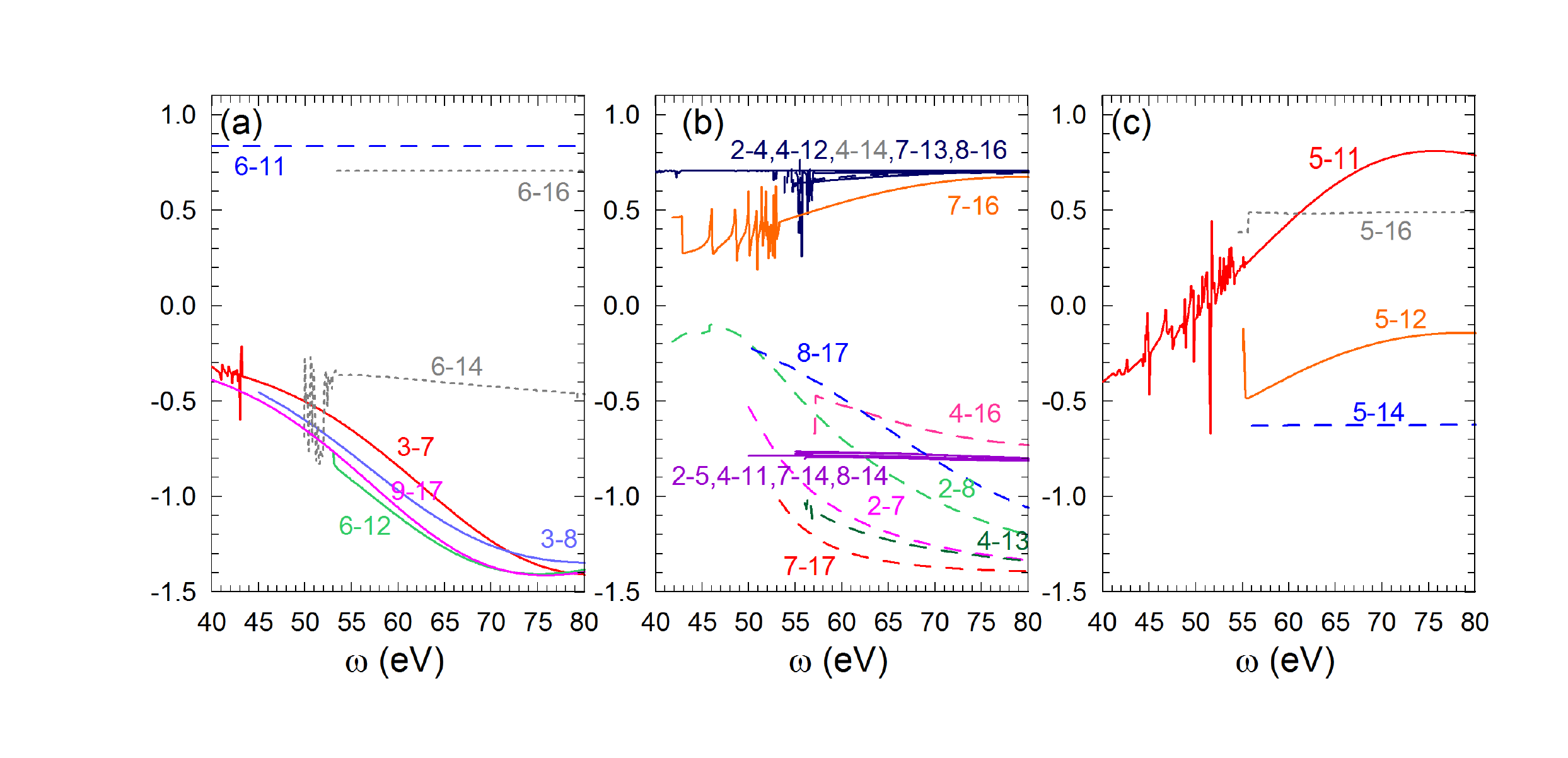}
\caption{Stationary alignment as function of energy for different ionization pathways: \textbf{(a)} from ions with $4s^n4p^{m\;2S_i+1}S$ term, \textbf{(b)} from $4s^n4p^{m\;2S_i+1}P$ term, \textbf{(c)} from $4s^n4p^{m\;2S_i+1}D$ term.}
\label{fig:3}
\end{center}
\end{figure*}

Figure~\ref{fig:3} shows stationary alignment (eqn.~\ref{eq:A2st}) of the other ions assuming that they are created from an {\it unpolarized} state $\mathcal A_i=0$. There are three types of features one can see: (a) the series of sharp structures at the lower-energy edge; (b) deep minima or high maxima placed around $\omega=80$ eV; (c) no energy dependency for some ionization pathways.

(a) The sharp structures emerge due to Rydberg auto\-ionization series and observed in all spectra except ones corresponding to highest allowed threshold, e.g. $1-3$, $2-8$, $3-9$, $\{4,5,6\}-16$, $\{7,8,9\}-17$. The resonance structure is hardly can be resolved in the modern experiments and therefore it is not a subject of current investigation. Here we cut off most of the resonances leaving a few just for illustration.

(b) The minima and maxima are connected with the Cooper minimum in $4p\rightarrow \varepsilon d$ ionization amplitudes. Note that most of them are quite close to the allowed limits: $-\sqrt{2}$ and $1/\sqrt{2}$ for $P$-terms (eqn.~\ref{eq:limP}) and $\pm\sqrt{7/10}$ for $D$-terms (eqn.~\ref{eq:limD}).

(c) Weak dependency on energy is explained by domination of a particular channel. In order to explore the issue further it is constructive to present eqn.~(\ref{eq:B}) in simpler form via single-electron transition amplitude $d_{l_il}$ from $l_i$-shell to $\varepsilon l$-continuum i.e. neglecting term dependency of the amplitudes:
\begin{eqnarray}\label{eq:BSim}
\bar{B}[k_i,k_{\gamma},k_f]&=&\hat{k}_i\hat{k}_{\gamma}\CGC{k_i}{0}{k_{\gamma}}{0}{k_f}{0} \hat{L}_i\hat{L}_f\phase{k_{\gamma}}\nonumber\\
&&\sum_{l}\!\SechsJ{1}{1}{k_{\gamma}}{l_i}{l_i}{l}\NeunJ{l_i}{L_i}{L_f}{l_i}{L_i}{L_f}{k_{\gamma}}{k_i}{k_f}|d_{l_il}|^2\!.
\end{eqnarray}
We will address to this simplification as {\it no correlation} NC-model. 

{\bf Ionized shell is $\bm{l_i=4s}$.} Within NC-model  $L_i=L_f$ and polarization of an initial state simply transfers to a final ion: $\mathcal A_f=\mathcal A_i$. That means that non-zero alignment of these branches (marked by dashed lines in figure \ref{fig:3}) is result of the correlation effects originating from strong term mixing. Moreover, branches $4-14$, $4-15$, $5-15$, $5-16$, $6-14$, $6-16$ are not allowed in the simple NC-model at all, because without correlation ionization of $s$-shell cannot change term of an ion. Their cross-sections are much lower than others \cite{Gryzlova2020}. Branch $6-16$ is governed by eqn.~(\ref{eq:A2st}) and (\ref{eq:B}) and only allowed due to the correlations channel  $\MEred{^2P\varepsilon p \,^1P}{\hat{D}}{^1S}$ leads to $\mathcal A_f=1/\sqrt{2}$.

{\bf Ionized shell is $\bm{l_i=4p}$.} Lets introduce $\kappa=|d_{p d}|^2/|d_{p s}|^2$ and re-arrange eqn.~(\ref{eq:A2st}) using (\ref{eq:BSim}) as:

\begin{eqnarray}
&&\hspace{-0.25cm}\mathcal A_f=-\sqrt{2}\frac{1+\frac{\kappa}{10}}{1+\kappa}\,,\hspace{3.2cm}\scaleto{\substack{L_i=S\\ L_f=P}}{17pt}\,,\label{ruleSP}\\
&&\hspace{-0.25cm}\mathcal A_f=\frac{\frac{1}{\sqrt{2}}(1+\frac{\kappa}{10})+\mathcal A_i(\frac{1}{2}-\frac{2}{5}\kappa)}{1+\kappa+\mathcal A_i/\sqrt{2}(1+\frac{\kappa}{10})}\,,\hspace{1.1cm}\scaleto{\substack{L_i=P\\ L_f=P}}{17pt}\,,\label{rulePP}\\
&&\hspace{-0.25cm}\mathcal A_f\hspace{-0.05cm}=\hspace{-0.05cm}-\sqrt{\frac{7}{10}}\frac{1+\frac{\kappa}{10}-\mathcal A_i\sqrt{2}(\frac{5}{14}+\frac{17}{35}\kappa)}{1+\kappa\hspace{-0.05cm}-\hspace{-0.05cm}\mathcal A_i\frac{\sqrt{2}}{10}       (1+\frac{\kappa}{10})}\!,\hspace{0.2cm}\scaleto{\substack{L_i=P\\ L_f=D}}{17pt}\,.\label{rulePD}
\end{eqnarray}

Equation (\ref{ruleSP}) governs $1-2$, $3-7$, $3-8$, $6-12$, $9-17$ branches. They generally demonstrate essential energy-dependence because of interplay between $d_s$ and $d_d$ amplitudes (see figure \ref{fig:3}(a)). Branch $6-11$ is not allowed in NC-model, governed by eqn.~(\ref{eq:A2st}) and (\ref{eq:B}), hence the alignment determined by the only allowed channel $\MEred{^2D\varepsilon d\,^1P}{\hat{D}}{^1S}$ is $\mathcal A_f=\sqrt{7/10}$ .

Equation (\ref{rulePP}) governs $2-4$, $4-12$, $7-13$, $7-16$, $8-16$ branches. All of them practically independent on energy and very close to $\mathcal A_f=1/\sqrt{2}$ indicating domination of $s$-wave  ($\kappa \approx 0$). The same is correct for $2-5$ and $4-11$ governed by eqn.~(\ref{rulePD}): for them alignment values are close to $\mathcal A_f=-\sqrt{7/10}$.

In figure~\ref{fig:3} all of initial states are supposed to be unpolarized. {\it Ab initio} it is correct only for $S$-terms (panel a). Below we discuss how panels b and c affected by polarization.

Equation~(\ref{rulePP}) shows that while $\kappa$ is small $P-P$ ionization causes the maximal possible positive alignment $\mathcal A_f=1/\sqrt{2}$ independently on alignment of initial state. But the closer polarization of initial state $\mathcal A_i$ to $-\sqrt{2}$ the lower $s$-wave contribution, finally at $\mathcal A_i=-\sqrt{2}$ $s$-wave contributions are completely compensated both in numerator and denominator and the channel is closed. At this situation (as well as if $d$-wave dominates) alignment of final state tends to edge negative value $\mathcal A_f=-\sqrt{2}$.

Equation~(\ref{rulePD}) shows that where $s$-wave dominates the alignment of final state varies from $\mathcal A_f=-\sqrt{10/7}$ (minimal possible) at initial state $\mathcal A_i=-\sqrt{2}$ to twice smaller $\mathcal A_f=-\sqrt{5/14}$ at $\mathcal A_i=1/\sqrt{2}$. At negative $\mathcal A_i$ $d$-wave weakly modifies the alignment because works "to the same side" (sign before $\kappa$ is always positive). For example at $\mathcal A_i=-\sqrt{2}$ ratio $\mathcal A_f$ at $\kappa=0$ ($s$-wave dominates) and $\mathcal A_f$ at $\kappa\longrightarrow \infty$ ($d$-wave dominates) is 25/22 $\approx 1.14$.

From the above one can see that while the NC-model describes ionization from valence $4p$-shell pretty well, it completely fails for $4s$-shell indicating much stronger correlations in ionization of the sub-valence shell.

\section{\label{sec:5} Results and Discussion}

\begin{figure}
\centering
\includegraphics[width=0.49\textwidth]{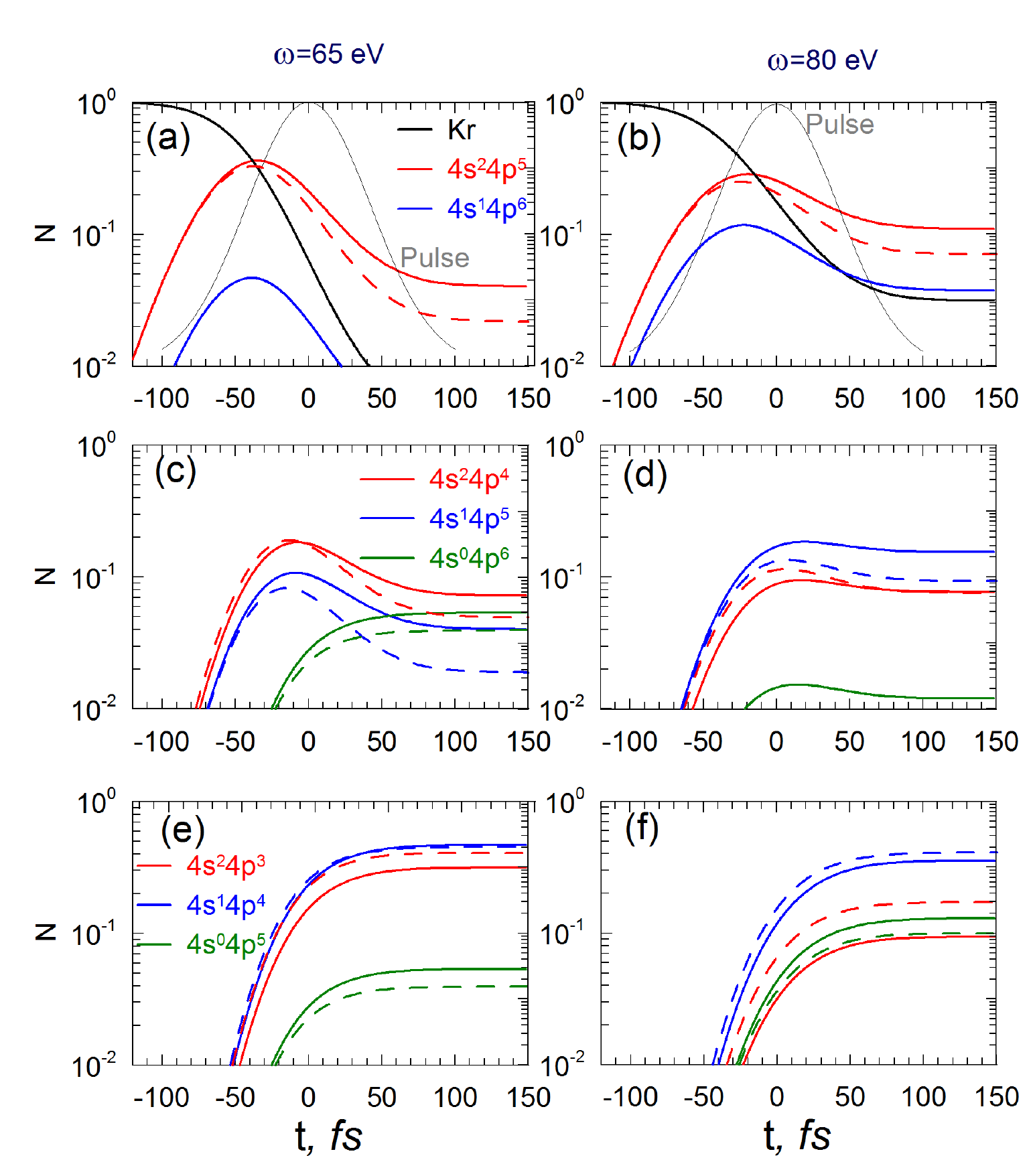}
\caption{Population of the different ion charge states and 
configurations for fluences $F=1000$ ph/\AA$^2$
  at the photon energies
65~eV (\textbf{a,c,e}) and 80~eV (\textbf{b,d,f}) calculated with  (solid lines) and without (dashed lines) accounting polarization of the ionic states. The population of the neutral Kr and Kr$^+$,  Kr$^{2+}$ and Kr$^{3+}$ are in the 
first, second and third rows.
Black lines --- yield of neutral Kr;
red lines --- yield of ions with $4s^2 4p^{6-n}$ configuration, where $n$ is the ion 
charge; blue lines --- ions with $4s^{1}4p^{6-n+1}$ configuration, green lines --- ions with $4s^{0}4p^{6-n+2}$ configuration. The pulse envelope (grey line) is indicated 
in the upper panels.}
\label{fig:4}
\end{figure}

\begin{figure}
\begin{center}
\includegraphics[width=0.49\textwidth]{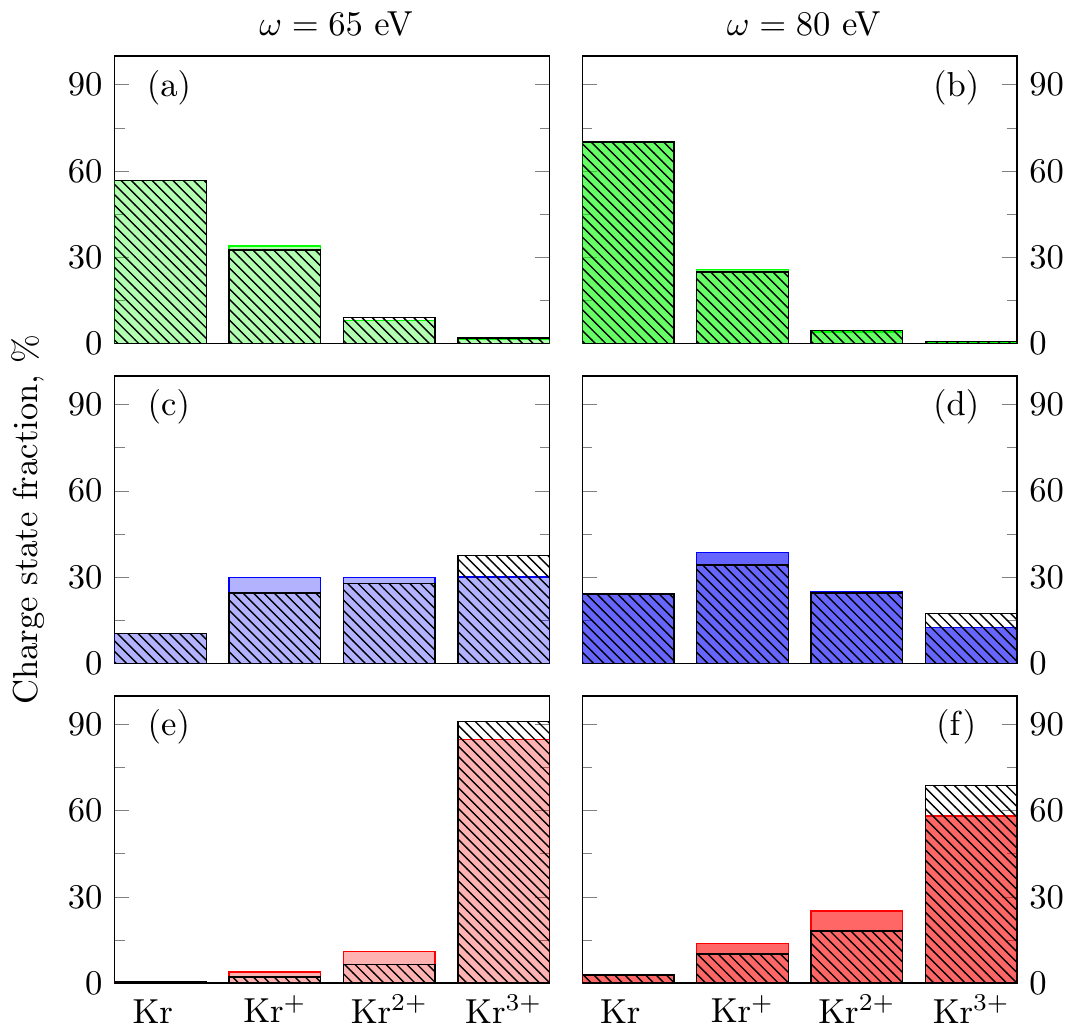}
\caption{The charge-state yields for three fluences: $F=100$ ph/\AA$^2$ (\textbf{a,b}), 
$F=400$ ph/\AA$^2$ (\textbf{c,d}), and $F=1000$ ph/\AA$^2$ (\textbf{e,f})
for the photon energies 65~eV (\textbf{a,c,e}) and 80~eV (\textbf{b,d,f}). The shaded areas show the results for unpolarized radiation.}
\label{fig:5}
\end{center}
\end{figure}

In this section we present and discuss results for observable values: population of different ionic species (fig.~\ref{fig:4}), ionic yields (fig.~\ref{fig:5}) and photo\-electron spectra (fig.~\ref{fig:6}). As was discussed in \cite{Gryzlova2020}, the curves in figures~\ref{fig:4}--\ref{fig:6} remain the same for
fixed values of fluence $F$ if we change pulse duration and do appropriate scaling of the timescale. The reason of the scaling is absence of any natural timetick (like the Auger decay) in the system under consideration.

\subsection{The ionic yields}
In figure~\ref{fig:4} we present results  for linearly polarized and unpolarized radiation at two photon energies (65~eV and 80~eV) and fluence $F = 1000$ ph/\AA$^2$ corresponding to intensity $2.2\cdot 10^{15}$ W/cm$^2$ (at $t_p = 60$ fs). We present population  as a function of time  summed up within one configuration  over possible terms.

The~concentration of neutral atoms monotonically decreases with time (figure~\ref{fig:3}(a,b)). The number of singly and doubly charged ions first increases with time, but then it may drop down (figures~\ref{fig:3}(c,d)), because of their further ionization  to Kr$^{3+}$. The last monotonically increases with time (figure~\ref{fig:3}(e,f)). 
At any time the sum of all populations presented in figure~\ref{fig:4} corresponding to the same fluence, photon energy and polarization equals unity.  

At first step (figure~\ref{fig:4}(a,b)) the alignment of Kr$^+$\,$4s^24p^{5\, 2}P$ ion (red curves) prevents its further ionization triply increasing its final population. In low intensity regime this tendency would keep: polarization of an ion should simply suppress some of ionization channels, leaving the other channels unaffected. If intensity is high enough to involve the saturation and depletion effects the tendency violates. For example, alignment increases number of double $4s$-hole Kr$^{2+}4s^{0}4p^{6\, 1}S$ ions (green lines in figure \ref{fig:4}(c)) in spite of the ions are created in the pathway affecting only $S$-terms.
The tendency is also relevant for the next step (figure~\ref{fig:4}(e)) and at $\omega=80$ eV corresponding higher polarization may rearrange ionic yields: for unpolarized radiation output of $4s^{2}4p^{3}$ is higher than $4s^{0}4p^{5}$, while for polarized radiation output of $4s^{0}4p^{5}$ is higher (red and green lines in figure~\ref{fig:4}(f)).

Such a high yields of $4s$ single- and double-hole ions at $\omega=65$~eV are really surprising accounting that there are the Cooper minima in $4s$-shell ionization of both Kr and Kr$^+$ \cite{Gryzlova2020} situated in the region of 45-50~eV and, moreover, including the fact that $4s$-ionization cross section at $\omega=65$~eV is still one order of magnitude lower than $4p$-ionization cross section.

There is the Cooper minimum in the $4p$-shell ionization amplitude of neutral Kr near 80~eV, aside cross sections from $4p$-ionization of other ions at $\omega=80$~eV are smaller than at $\omega=65$~eV. As result, yield of single and double $4s$-hole states are higher than for 65~eV (compare blue lines in figure~\ref{fig:4}(c) and \ref{fig:4}(d), green lines in \ref{fig:4}(e) and \ref{fig:4}(f)). Results for the first Kr$^+4s^24p^5$ ion may look contradicted: in spite of cross section drops down by 3 times, the yield increases by 5 times between 65 and 80 eV (red lines in fig. \ref{fig:4}ab). That is because very high (minimal) alignment prohibited the next ionization step, and the ions accumulated in this state.

In practice the different ionic configurations  are not distinguished. In figure~\ref{fig:5} the overall ionic yields at the
fluences $F=100$ ph/\AA$^2$, $F=400$ ph/\AA$^2$ and $F=1000$ ph/\AA$^2$ 
and the photon energies $\omega=65$~eV and $\omega=80$~eV are presented.
The relative populations of the different ionic states  change with intensity, switching from perturbative (panels a,b) to saturation (e,f) patterns. As expected, summing over the configurations decreases difference between polarized and unpolarized cases. At lower intensity (figure~\ref{fig:5}(a,b)) there are no polarization effects, but at higher they become more essential for final Kr$^{3+}$ ions and may decrease the ionic yield up to  10\% of total ionic number (figure~\ref{fig:5}(c,e,f)).

The next step, i.e. ionization of Kr$^{3+}$
 for 80~eV photons is energetically possible and
 the last column in figure~\ref{fig:5}(b,d,f) actually presents the sum yield of ions with charges three and higher.  Nevertheless that cannot affect presented below photo\-electron spectra.

\subsection{Photoelectron Spectra}

Photo\-electron spectrum can be cast as function of the probability $P_{ab}(F)$ of an ion (atom) in a state $a$ is ionized into the ion in a state $b$ over the entire pulse and energy of this transition:
\begin{equation} \label{eq:spec}
f_F(\varepsilon) = \sum_{ab} P_{ab} (F)
\exp[-(\varepsilon + I_{ab}-\omega)^2/\Gamma^2]\,,
\end{equation}
where $\varepsilon$ is the photo\-electron kinetic energy, $I_{ab}$ is the binding energy of ionization of the state $a$ to the state $b$  and $\Gamma$ is the resolution  of the electron detector. 
As in \cite{Gryzlova2020}, we set value $\Gamma =0.42$~eV to leave the fine structure of levels unresolved.

\begin{figure*}
\begin{center}
\includegraphics[width=0.8\textwidth]{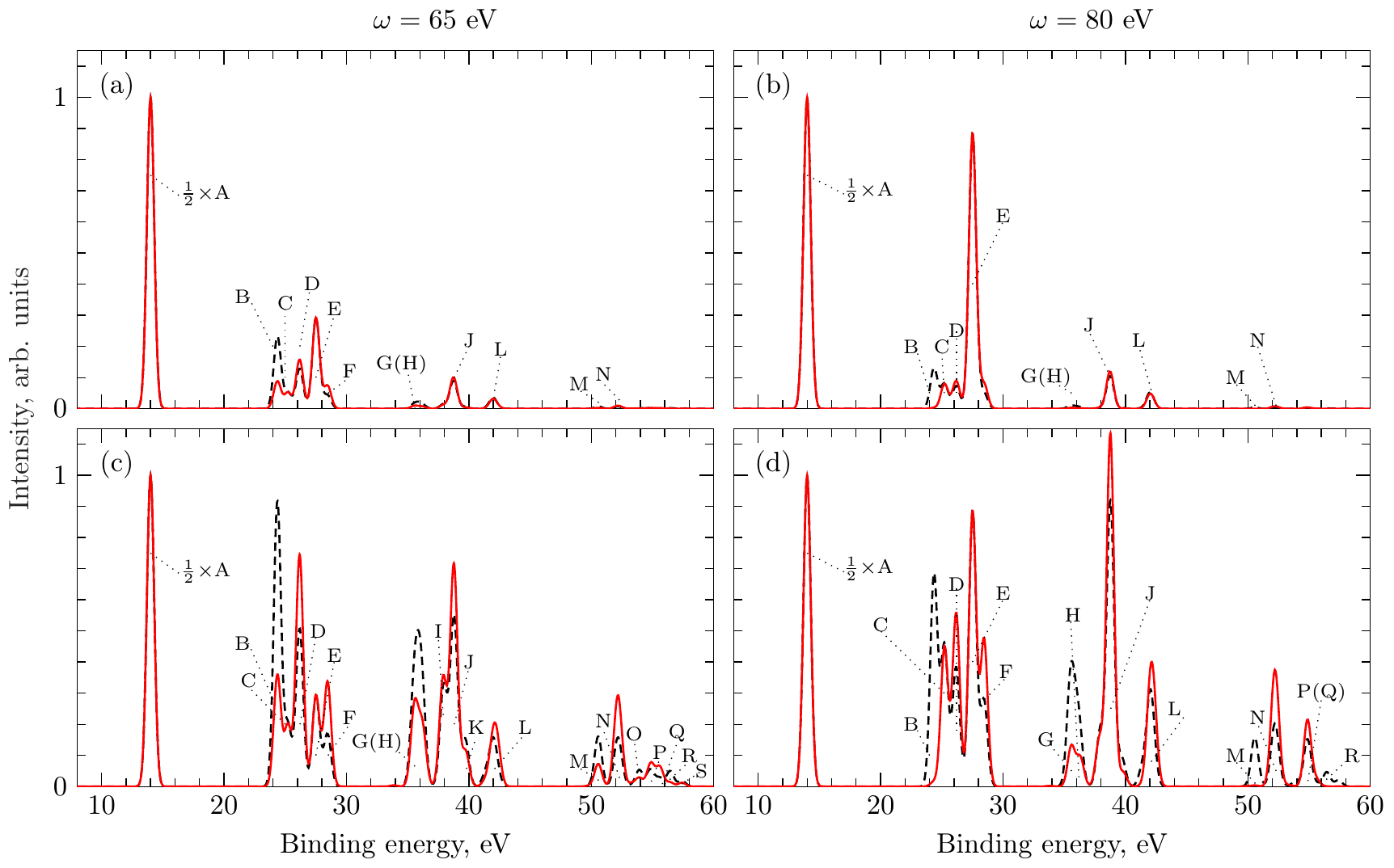}
\caption{Photo\-electron spectrum for different photon energy:
$\omega=65$~eV (\textbf{(a)} for $F=100$ ph/\AA$^2$, \textbf{(c)} for $F=1000$ ph/\AA$^2$,), and $\omega=80$~eV (\textbf{(b)} for $F=100$ ph/\AA$^2$, \textbf{(d)} for $F=1000$ ph/\AA$^2$).
The solid lines correspond to calculations for linearly polarized radiation, the dashed lines -- to un\-polarized radiation.
The spectra are normalized in such a way that $1/2$ of the main line A equals unity. 
The spectral features are indicated by capital letters in accordance with Table~\ref{tab1}. }
\label{fig:6}
\end{center}
\end{figure*}

The photo\-electron spectrum provides more detailed information on the pathways of the sequential ionization than the ion yield because it keeps memory on the relative population of the intermediate states of the process (see figure~\ref{fig:1} and Table~\ref{tab1}).

The generated photo\-electron spectra for two photon energies, 65~eV and 80~eV, are displayed in figures~\ref{fig:6}(a,c) and \ref{fig:6}(b,d), correspondingly. We consider two values of fluence: low $F=100$ ph/\AA$^2$ (figure~\ref{fig:6}(a,b)) and high $F=1000$ ph/\AA$^2$ (figure~\ref{fig:6}(c,d)). The dashed lines show the results obtained for unpolarized radiation.

In figures~\ref{fig:6}(a--d) the lines are concentrated in  four groups: the main photo\-line A from the $4p$-shell ionization of neutral Kr; lines from ionization of ${\rm Kr^+} \, 4s^24p^5$ mostly (B--D,F); from ionization of 
${\rm Kr^{2+}} \, 4s^24p^4$ mostly (G--L); ionization from $4s$-shell of 
${\rm Kr^{2+}} \, 4s^24p^4$ and ${\rm Kr^{2+}} \, 4s^14p^5$ (M--S). 

For low-intensity regime (figure~\ref{fig:6}(a)) polarization suppresses line B and completely demolish it at $\omega=80$~eV (figure~\ref{fig:6}(b)). That is because Kr$^+$\,$4s^24p^{5\, 2}P$ is completely polarized ($A_2(P)=-\sqrt{2}$) in the region of Cooper minimum and the polarization does not allow ionization to $s$-wave (rule~\ref{rulePP}). Lines G, H and M demonstrate similar behaviour, but at this intensity they are difficult to see.

The case of higher fluence (figure~\ref{fig:6}(c,d)) is more interesting. Besides overall increasing of the multiple ionization contributions which appears in enlarging peaks for binding energy above 30~eV and decreasing above discussed lines B, G, H and M, there are some lines which increase or even appear in comparison with unpolarized case. That are line F which can be distinguished only for polarized radiation, and lines J and N. The re-distribution is caused by enhanced contributions from ionization of $S$-terms, when ionization of $P$-terms suppresed by polarization. 

\begin{figure}
\begin{center}
\includegraphics[width=0.49\textwidth]{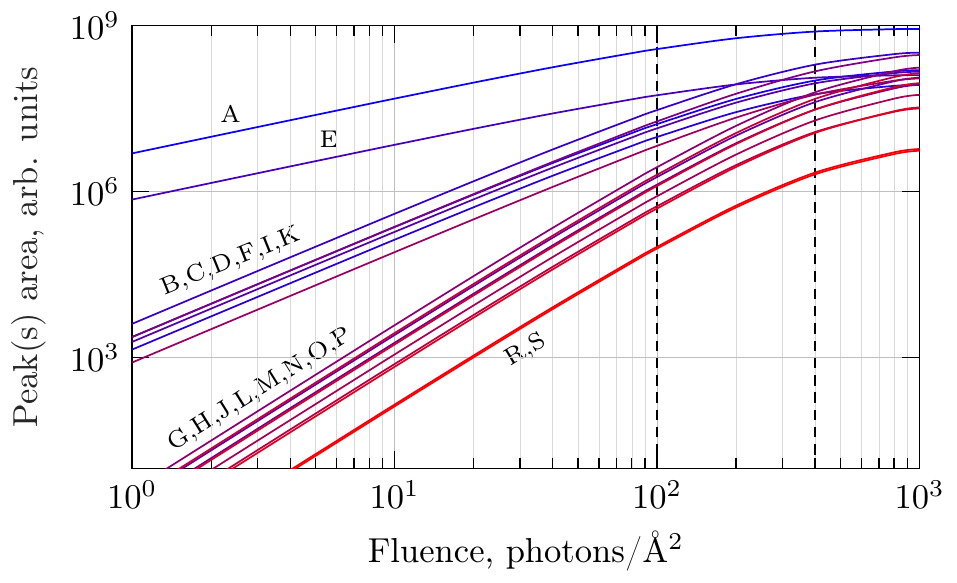}
\caption{Intensity dependence of different photo\-electron lines on the fluence $F$ at the photon energy $\omega=65$~eV. The vertical dashed lines indicate fluences related to figures~\ref{fig:3}--\ref{fig:6}:
100~ph/\AA$^2$ and~400~ph/\AA$^2$.}
\label{fig:7}
\end{center}
\end{figure}

Figure~\ref{fig:7} shows the fluence dependence of the intensities
of the spectral lines. The curves clearly indicate the one-, two- and three-photon
origin of the spectral features A and E;  B--D, F, I, K; and G, H, J, L--S, 
respectively. The saturation appears at fluence above 100 ph/\AA$^2$, and the more pronounced the more number of photons involved. The figure shows that accounting of polarization does not change general multi-photon behaviour with intensity.

\section{Conclusions}

The role of the polarization both of the radiation and ionic states in sequential multiple ionization is studied theoretically. The system of equations similar to conventional rate equations for states population is formulated in terms of the statistical tensors.  It is applicable to linearly and circularly polarized pulses for systems with well-separated (excited incoherently) levels within dipole approximation.

The method was applied to multiple ionization of Kr in the 65--80 eV region. Ionic states evolution, yields and photo\-electron spectrum were calculated and compared for the linearly polarized and unpolarized radiation. Redistribution of ionic yields up to 10\% of total amount and noticeable re-arrangement of photo\-electron spectrum caused by polarization are predicted. In low-intensity regime polarization suppresses or even demolishes some photo\-lines; in high-intensity regime polarization may emphasize structure of spectra and some lines become distinguishable only for polarized radiation.

It is shown that doubly hollow ionic states  which can decay only via spontaneous emission are created more efficiently if a sample is irradiated by polarized radiation than by un\-polarized. These hollow ion states  are of particular interest because they can serve as a target to create autoionizing states of very exotic configuration.

The present study is a necessary point before accounting for the complex evolution of the system due to different opening Auger decay channels which become possible when $3d$-shell ionization comes into play.

The research was funded by the Russian Foundation for Basic Research (RFBR) under project No. 20-52-12023 and Ministry of Science and Higher Education of the Russian Federation grant No. 075-15-2021-1353. The work of M.D.K. is supported by the Ministry of Science and Higher Education of the Russian Federation (project No. 0818-2020-0005) using resources of the Shared Services “Data Center of the Far-Eastern Branch of the Russian Academy of Sciences”.



\nocite{*}
\bibliography{references}

\begin{thebibliography}{74}%
\makeatletter
\providecommand \@ifxundefined [1]{%
 \@ifx{#1\undefined}
}%
\providecommand \@ifnum [1]{%
 \ifnum #1\expandafter \@firstoftwo
 \else \expandafter \@secondoftwo
 \fi
}%
\providecommand \@ifx [1]{%
 \ifx #1\expandafter \@firstoftwo
 \else \expandafter \@secondoftwo
 \fi
}%
\providecommand \natexlab [1]{#1}%
\providecommand \enquote  [1]{``#1''}%
\providecommand \bibnamefont  [1]{#1}%
\providecommand \bibfnamefont [1]{#1}%
\providecommand \citenamefont [1]{#1}%
\providecommand \href@noop [0]{\@secondoftwo}%
\providecommand \href [0]{\begingroup \@sanitize@url \@href}%
\providecommand \@href[1]{\@@startlink{#1}\@@href}%
\providecommand \@@href[1]{\endgroup#1\@@endlink}%
\providecommand \@sanitize@url [0]{\catcode `\\12\catcode `\$12\catcode
  `\&12\catcode `\#12\catcode `\^12\catcode `\_12\catcode `\%12\relax}%
\providecommand \@@startlink[1]{}%
\providecommand \@@endlink[0]{}%
\providecommand \url  [0]{\begingroup\@sanitize@url \@url }%
\providecommand \@url [1]{\endgroup\@href {#1}{\urlprefix }}%
\providecommand \urlprefix  [0]{URL }%
\providecommand \Eprint [0]{\href }%
\providecommand \doibase [0]{https://doi.org/}%
\providecommand \selectlanguage [0]{\@gobble}%
\providecommand \bibinfo  [0]{\@secondoftwo}%
\providecommand \bibfield  [0]{\@secondoftwo}%
\providecommand \translation [1]{[#1]}%
\providecommand \BibitemOpen [0]{}%
\providecommand \bibitemStop [0]{}%
\providecommand \bibitemNoStop [0]{.\EOS\space}%
\providecommand \EOS [0]{\spacefactor3000\relax}%
\providecommand \BibitemShut  [1]{\csname bibitem#1\endcsname}%
\let\auto@bib@innerbib\@empty
\bibitem [{\citenamefont {Nass}\ \emph {et~al.}(2015)\citenamefont {Nass},
  \citenamefont {Foucar}, \citenamefont {Barends}, \citenamefont {Hartmann},
  \citenamefont {Botha}, \citenamefont {Shoeman}, \citenamefont {Doak},
  \citenamefont {Alonso-Mori}, \citenamefont {Aquila}, \citenamefont {Bajt},
  \citenamefont {Barty}, \citenamefont {Bean}, \citenamefont {Beyerlein},
  \citenamefont {Bublitz}, \citenamefont {Drachmann}, \citenamefont
  {Gregersen}, \citenamefont {Jonsson}, \citenamefont {Kabsch}, \citenamefont
  {Kassemeyer}, \citenamefont {Koglin}, \citenamefont {Krumrey}, \citenamefont
  {Mattle}, \citenamefont {Messerschmidt}, \citenamefont {Nissen},
  \citenamefont {Reinhard}, \citenamefont {Sitsel}, \citenamefont {Sokaras},
  \citenamefont {Williams}, \citenamefont {Hau-Riege}, \citenamefont
  {Timneanu}, \citenamefont {Caleman}, \citenamefont {Chapman}, \citenamefont
  {Boutet},\ and\ \citenamefont {Schlichting}}]{Nass2015}%
  \BibitemOpen
  \bibfield  {author} {\bibinfo {author} {\bibfnamefont {K.}~\bibnamefont
  {Nass}}, \bibinfo {author} {\bibfnamefont {L.}~\bibnamefont {Foucar}},
  \bibinfo {author} {\bibfnamefont {T.~R.~M.}\ \bibnamefont {Barends}},
  \bibinfo {author} {\bibfnamefont {E.}~\bibnamefont {Hartmann}}, \bibinfo
  {author} {\bibfnamefont {S.}~\bibnamefont {Botha}}, \bibinfo {author}
  {\bibfnamefont {R.~L.}\ \bibnamefont {Shoeman}}, \bibinfo {author}
  {\bibfnamefont {R.~B.}\ \bibnamefont {Doak}}, \bibinfo {author}
  {\bibfnamefont {R.}~\bibnamefont {Alonso-Mori}}, \bibinfo {author}
  {\bibfnamefont {A.}~\bibnamefont {Aquila}}, \bibinfo {author} {\bibfnamefont
  {S.}~\bibnamefont {Bajt}}, \bibinfo {author} {\bibfnamefont {A.}~\bibnamefont
  {Barty}}, \bibinfo {author} {\bibfnamefont {R.}~\bibnamefont {Bean}},
  \bibinfo {author} {\bibfnamefont {K.~R.}\ \bibnamefont {Beyerlein}}, \bibinfo
  {author} {\bibfnamefont {M.}~\bibnamefont {Bublitz}}, \bibinfo {author}
  {\bibfnamefont {N.}~\bibnamefont {Drachmann}}, \bibinfo {author}
  {\bibfnamefont {J.}~\bibnamefont {Gregersen}}, \bibinfo {author}
  {\bibfnamefont {H.~O.}\ \bibnamefont {Jonsson}}, \bibinfo {author}
  {\bibfnamefont {W.}~\bibnamefont {Kabsch}}, \bibinfo {author} {\bibfnamefont
  {S.}~\bibnamefont {Kassemeyer}}, \bibinfo {author} {\bibfnamefont {J.~E.}\
  \bibnamefont {Koglin}}, \bibinfo {author} {\bibfnamefont {M.}~\bibnamefont
  {Krumrey}}, \bibinfo {author} {\bibfnamefont {D.}~\bibnamefont {Mattle}},
  \bibinfo {author} {\bibfnamefont {M.}~\bibnamefont {Messerschmidt}}, \bibinfo
  {author} {\bibfnamefont {P.}~\bibnamefont {Nissen}}, \bibinfo {author}
  {\bibfnamefont {L.}~\bibnamefont {Reinhard}}, \bibinfo {author}
  {\bibfnamefont {O.}~\bibnamefont {Sitsel}}, \bibinfo {author} {\bibfnamefont
  {D.}~\bibnamefont {Sokaras}}, \bibinfo {author} {\bibfnamefont {G.~J.}\
  \bibnamefont {Williams}}, \bibinfo {author} {\bibfnamefont {S.}~\bibnamefont
  {Hau-Riege}}, \bibinfo {author} {\bibfnamefont {N.}~\bibnamefont {Timneanu}},
  \bibinfo {author} {\bibfnamefont {C.}~\bibnamefont {Caleman}}, \bibinfo
  {author} {\bibfnamefont {H.~N.}\ \bibnamefont {Chapman}}, \bibinfo {author}
  {\bibfnamefont {S.}~\bibnamefont {Boutet}},\ and\ \bibinfo {author}
  {\bibfnamefont {I.}~\bibnamefont {Schlichting}},\ }\bibfield  {title}
  {\bibinfo {title} {{Indications of radiation damage in ferredoxin
  microcrystals using high-intensity X-FEL beams}},\ }\href
  {https://doi.org/10.1107/S1600577515002349} {\bibfield  {journal} {\bibinfo
  {journal} {Journal of Synchrotron Radiation}\ }\textbf {\bibinfo {volume}
  {22}},\ \bibinfo {pages} {225} (\bibinfo {year} {2015})}\BibitemShut
  {NoStop}%
\bibitem [{\citenamefont {Galli}\ \emph {et~al.}(2015)\citenamefont {Galli},
  \citenamefont {Son}, \citenamefont {Klinge}, \citenamefont {Bajt},
  \citenamefont {Barty}, \citenamefont {Bean}, \citenamefont {Betzel},
  \citenamefont {Beyerlein}, \citenamefont {Caleman}, \citenamefont {Doak},
  \citenamefont {Duszenko}, \citenamefont {Fleckenstein}, \citenamefont {Gati},
  \citenamefont {Hunt}, \citenamefont {Kirian}, \citenamefont {Liang},
  \citenamefont {Nanao}, \citenamefont {Nass}, \citenamefont {Oberthur},
  \citenamefont {Redecke}, \citenamefont {Shoeman}, \citenamefont {Stellato},
  \citenamefont {Yoon}, \citenamefont {White}, \citenamefont {Yefanov},
  \citenamefont {Spence},\ and\ \citenamefont {Chapman}}]{Galli2015}%
  \BibitemOpen
  \bibfield  {author} {\bibinfo {author} {\bibfnamefont {L.}~\bibnamefont
  {Galli}}, \bibinfo {author} {\bibfnamefont {S.-K.}\ \bibnamefont {Son}},
  \bibinfo {author} {\bibfnamefont {M.}~\bibnamefont {Klinge}}, \bibinfo
  {author} {\bibfnamefont {S.}~\bibnamefont {Bajt}}, \bibinfo {author}
  {\bibfnamefont {A.}~\bibnamefont {Barty}}, \bibinfo {author} {\bibfnamefont
  {R.}~\bibnamefont {Bean}}, \bibinfo {author} {\bibfnamefont {C.}~\bibnamefont
  {Betzel}}, \bibinfo {author} {\bibfnamefont {K.~R.}\ \bibnamefont
  {Beyerlein}}, \bibinfo {author} {\bibfnamefont {C.}~\bibnamefont {Caleman}},
  \bibinfo {author} {\bibfnamefont {R.~B.}\ \bibnamefont {Doak}}, \bibinfo
  {author} {\bibfnamefont {M.}~\bibnamefont {Duszenko}}, \bibinfo {author}
  {\bibfnamefont {H.}~\bibnamefont {Fleckenstein}}, \bibinfo {author}
  {\bibfnamefont {C.}~\bibnamefont {Gati}}, \bibinfo {author} {\bibfnamefont
  {B.}~\bibnamefont {Hunt}}, \bibinfo {author} {\bibfnamefont {R.~A.}\
  \bibnamefont {Kirian}}, \bibinfo {author} {\bibfnamefont {M.}~\bibnamefont
  {Liang}}, \bibinfo {author} {\bibfnamefont {M.~H.}\ \bibnamefont {Nanao}},
  \bibinfo {author} {\bibfnamefont {K.}~\bibnamefont {Nass}}, \bibinfo {author}
  {\bibfnamefont {D.}~\bibnamefont {Oberthur}}, \bibinfo {author}
  {\bibfnamefont {L.}~\bibnamefont {Redecke}}, \bibinfo {author} {\bibfnamefont
  {R.}~\bibnamefont {Shoeman}}, \bibinfo {author} {\bibfnamefont
  {F.}~\bibnamefont {Stellato}}, \bibinfo {author} {\bibfnamefont {C.~H.}\
  \bibnamefont {Yoon}}, \bibinfo {author} {\bibfnamefont {T.~A.}\ \bibnamefont
  {White}}, \bibinfo {author} {\bibfnamefont {O.}~\bibnamefont {Yefanov}},
  \bibinfo {author} {\bibfnamefont {J.}~\bibnamefont {Spence}},\ and\ \bibinfo
  {author} {\bibfnamefont {H.~N.}\ \bibnamefont {Chapman}},\ }\bibfield
  {title} {\bibinfo {title} {Electronic damage in s atoms in a native protein
  crystal induced by an intense x-ray free-electron laser pulse},\ }\href
  {https://doi.org/10.1063/1.4919398} {\bibfield  {journal} {\bibinfo
  {journal} {Structural Dynamics}\ }\textbf {\bibinfo {volume} {2}},\ \bibinfo
  {pages} {041703} (\bibinfo {year} {2015})},\ \Eprint
  {https://arxiv.org/abs/https://doi.org/10.1063/1.4919398}
  {https://doi.org/10.1063/1.4919398} \BibitemShut {NoStop}%
\bibitem [{\citenamefont {Sorokin}\ \emph {et~al.}(2007)\citenamefont
  {Sorokin}, \citenamefont {Bobashev}, \citenamefont {Feigl}, \citenamefont
  {Tiedtke}, \citenamefont {Wabnitz},\ and\ \citenamefont
  {Richter}}]{Sorokin2007}%
  \BibitemOpen
  \bibfield  {author} {\bibinfo {author} {\bibfnamefont {A.~A.}\ \bibnamefont
  {Sorokin}}, \bibinfo {author} {\bibfnamefont {S.~V.}\ \bibnamefont
  {Bobashev}}, \bibinfo {author} {\bibfnamefont {T.}~\bibnamefont {Feigl}},
  \bibinfo {author} {\bibfnamefont {K.}~\bibnamefont {Tiedtke}}, \bibinfo
  {author} {\bibfnamefont {H.}~\bibnamefont {Wabnitz}},\ and\ \bibinfo {author}
  {\bibfnamefont {M.}~\bibnamefont {Richter}},\ }\bibfield  {title} {\bibinfo
  {title} {Photoelectric effect at ultrahigh intensities},\ }\href
  {https://doi.org/10.1103/PhysRevLett.99.213002} {\bibfield  {journal}
  {\bibinfo  {journal} {Phys. Rev. Lett.}\ }\textbf {\bibinfo {volume} {99}},\
  \bibinfo {pages} {213002} (\bibinfo {year} {2007})}\BibitemShut {NoStop}%
\bibitem [{\citenamefont {K\"ubel}\ \emph {et~al.}(2016)\citenamefont
  {K\"ubel}, \citenamefont {Burger}, \citenamefont {Kling}, \citenamefont
  {Pischke}, \citenamefont {Beaufore}, \citenamefont {Ben-Itzhak},
  \citenamefont {Paulus}, \citenamefont {Ullrich}, \citenamefont {Pfeifer},
  \citenamefont {Moshammer}, \citenamefont {Kling},\ and\ \citenamefont
  {Bergues}}]{Kubel2016}%
  \BibitemOpen
  \bibfield  {author} {\bibinfo {author} {\bibfnamefont {M.}~\bibnamefont
  {K\"ubel}}, \bibinfo {author} {\bibfnamefont {C.}~\bibnamefont {Burger}},
  \bibinfo {author} {\bibfnamefont {N.~G.}\ \bibnamefont {Kling}}, \bibinfo
  {author} {\bibfnamefont {T.}~\bibnamefont {Pischke}}, \bibinfo {author}
  {\bibfnamefont {L.}~\bibnamefont {Beaufore}}, \bibinfo {author}
  {\bibfnamefont {I.}~\bibnamefont {Ben-Itzhak}}, \bibinfo {author}
  {\bibfnamefont {G.~G.}\ \bibnamefont {Paulus}}, \bibinfo {author}
  {\bibfnamefont {J.}~\bibnamefont {Ullrich}}, \bibinfo {author} {\bibfnamefont
  {T.}~\bibnamefont {Pfeifer}}, \bibinfo {author} {\bibfnamefont
  {R.}~\bibnamefont {Moshammer}}, \bibinfo {author} {\bibfnamefont {M.~F.}\
  \bibnamefont {Kling}},\ and\ \bibinfo {author} {\bibfnamefont
  {B.}~\bibnamefont {Bergues}},\ }\bibfield  {title} {\bibinfo {title}
  {Complete characterization of single-cycle double ionization of argon from
  the nonsequential to the sequential ionization regime},\ }\href
  {https://doi.org/10.1103/PhysRevA.93.053422} {\bibfield  {journal} {\bibinfo
  {journal} {Phys. Rev. A}\ }\textbf {\bibinfo {volume} {93}},\ \bibinfo
  {pages} {053422} (\bibinfo {year} {2016})}\BibitemShut {NoStop}%
\bibitem [{\citenamefont {Moshammer}\ \emph {et~al.}(2007)\citenamefont
  {Moshammer}, \citenamefont {Jiang}, \citenamefont {Foucar}, \citenamefont
  {Rudenko}, \citenamefont {Ergler}, \citenamefont {Schr\"oter}, \citenamefont
  {L\"udemann}, \citenamefont {Zrost}, \citenamefont {Fischer}, \citenamefont
  {Titze}, \citenamefont {Jahnke}, \citenamefont {Sch\"offler}, \citenamefont
  {Weber}, \citenamefont {D\"orner}, \citenamefont {Zouros}, \citenamefont
  {Dorn}, \citenamefont {Ferger}, \citenamefont {K\"uhnel}, \citenamefont
  {D\"usterer}, \citenamefont {Treusch}, \citenamefont {Radcliffe},
  \citenamefont {Pl\"onjes},\ and\ \citenamefont {Ullrich}}]{Moshammer2007}%
  \BibitemOpen
  \bibfield  {author} {\bibinfo {author} {\bibfnamefont {R.}~\bibnamefont
  {Moshammer}}, \bibinfo {author} {\bibfnamefont {Y.~H.}\ \bibnamefont
  {Jiang}}, \bibinfo {author} {\bibfnamefont {L.}~\bibnamefont {Foucar}},
  \bibinfo {author} {\bibfnamefont {A.}~\bibnamefont {Rudenko}}, \bibinfo
  {author} {\bibfnamefont {T.}~\bibnamefont {Ergler}}, \bibinfo {author}
  {\bibfnamefont {C.~D.}\ \bibnamefont {Schr\"oter}}, \bibinfo {author}
  {\bibfnamefont {S.}~\bibnamefont {L\"udemann}}, \bibinfo {author}
  {\bibfnamefont {K.}~\bibnamefont {Zrost}}, \bibinfo {author} {\bibfnamefont
  {D.}~\bibnamefont {Fischer}}, \bibinfo {author} {\bibfnamefont
  {J.}~\bibnamefont {Titze}}, \bibinfo {author} {\bibfnamefont
  {T.}~\bibnamefont {Jahnke}}, \bibinfo {author} {\bibfnamefont
  {M.}~\bibnamefont {Sch\"offler}}, \bibinfo {author} {\bibfnamefont
  {T.}~\bibnamefont {Weber}}, \bibinfo {author} {\bibfnamefont
  {R.}~\bibnamefont {D\"orner}}, \bibinfo {author} {\bibfnamefont {T.~J.~M.}\
  \bibnamefont {Zouros}}, \bibinfo {author} {\bibfnamefont {A.}~\bibnamefont
  {Dorn}}, \bibinfo {author} {\bibfnamefont {T.}~\bibnamefont {Ferger}},
  \bibinfo {author} {\bibfnamefont {K.~U.}\ \bibnamefont {K\"uhnel}}, \bibinfo
  {author} {\bibfnamefont {S.}~\bibnamefont {D\"usterer}}, \bibinfo {author}
  {\bibfnamefont {R.}~\bibnamefont {Treusch}}, \bibinfo {author} {\bibfnamefont
  {P.}~\bibnamefont {Radcliffe}}, \bibinfo {author} {\bibfnamefont
  {E.}~\bibnamefont {Pl\"onjes}},\ and\ \bibinfo {author} {\bibfnamefont
  {J.}~\bibnamefont {Ullrich}},\ }\bibfield  {title} {\bibinfo {title}
  {Few-photon multiple ionization of ne and ar by strong free-electron-laser
  pulses},\ }\href {https://doi.org/10.1103/PhysRevLett.98.203001} {\bibfield
  {journal} {\bibinfo  {journal} {Phys. Rev. Lett.}\ }\textbf {\bibinfo
  {volume} {98}},\ \bibinfo {pages} {203001} (\bibinfo {year}
  {2007})}\BibitemShut {NoStop}%
\bibitem [{\citenamefont {Kanter}\ \emph {et~al.}(2011)\citenamefont {Kanter},
  \citenamefont {Kr\"assig}, \citenamefont {Li}, \citenamefont {March},
  \citenamefont {Ho}, \citenamefont {Rohringer}, \citenamefont {Santra},
  \citenamefont {Southworth}, \citenamefont {DiMauro}, \citenamefont {Doumy},
  \citenamefont {Roedig}, \citenamefont {Berrah}, \citenamefont {Fang},
  \citenamefont {Hoener}, \citenamefont {Bucksbaum}, \citenamefont {Ghimire},
  \citenamefont {Reis}, \citenamefont {Bozek}, \citenamefont {Bostedt},
  \citenamefont {Messerschmidt},\ and\ \citenamefont {Young}}]{Kanter2011}%
  \BibitemOpen
  \bibfield  {author} {\bibinfo {author} {\bibfnamefont {E.~P.}\ \bibnamefont
  {Kanter}}, \bibinfo {author} {\bibfnamefont {B.}~\bibnamefont {Kr\"assig}},
  \bibinfo {author} {\bibfnamefont {Y.}~\bibnamefont {Li}}, \bibinfo {author}
  {\bibfnamefont {A.~M.}\ \bibnamefont {March}}, \bibinfo {author}
  {\bibfnamefont {P.}~\bibnamefont {Ho}}, \bibinfo {author} {\bibfnamefont
  {N.}~\bibnamefont {Rohringer}}, \bibinfo {author} {\bibfnamefont
  {R.}~\bibnamefont {Santra}}, \bibinfo {author} {\bibfnamefont {S.~H.}\
  \bibnamefont {Southworth}}, \bibinfo {author} {\bibfnamefont {L.~F.}\
  \bibnamefont {DiMauro}}, \bibinfo {author} {\bibfnamefont {G.}~\bibnamefont
  {Doumy}}, \bibinfo {author} {\bibfnamefont {C.~A.}\ \bibnamefont {Roedig}},
  \bibinfo {author} {\bibfnamefont {N.}~\bibnamefont {Berrah}}, \bibinfo
  {author} {\bibfnamefont {L.}~\bibnamefont {Fang}}, \bibinfo {author}
  {\bibfnamefont {M.}~\bibnamefont {Hoener}}, \bibinfo {author} {\bibfnamefont
  {P.~H.}\ \bibnamefont {Bucksbaum}}, \bibinfo {author} {\bibfnamefont
  {S.}~\bibnamefont {Ghimire}}, \bibinfo {author} {\bibfnamefont {D.~A.}\
  \bibnamefont {Reis}}, \bibinfo {author} {\bibfnamefont {J.~D.}\ \bibnamefont
  {Bozek}}, \bibinfo {author} {\bibfnamefont {C.}~\bibnamefont {Bostedt}},
  \bibinfo {author} {\bibfnamefont {M.}~\bibnamefont {Messerschmidt}},\ and\
  \bibinfo {author} {\bibfnamefont {L.}~\bibnamefont {Young}},\ }\bibfield
  {title} {\bibinfo {title} {Unveiling and driving hidden resonances with
  high-fluence, high-intensity x-ray pulses},\ }\href
  {https://doi.org/10.1103/PhysRevLett.107.233001} {\bibfield  {journal}
  {\bibinfo  {journal} {Phys. Rev. Lett.}\ }\textbf {\bibinfo {volume} {107}},\
  \bibinfo {pages} {233001} (\bibinfo {year} {2011})}\BibitemShut {NoStop}%
\bibitem [{\citenamefont {Fushitani}\ \emph {et~al.}(2020)\citenamefont
  {Fushitani}, \citenamefont {Sasaki}, \citenamefont {Matsuda}, \citenamefont
  {Fujise}, \citenamefont {Kawabe}, \citenamefont {Hashigaya}, \citenamefont
  {Owada}, \citenamefont {Togashi}, \citenamefont {Nakajima}, \citenamefont
  {Yabashi}, \citenamefont {Hikosaka},\ and\ \citenamefont
  {Hishikawa}}]{Fushitani2020}%
  \BibitemOpen
  \bibfield  {author} {\bibinfo {author} {\bibfnamefont {M.}~\bibnamefont
  {Fushitani}}, \bibinfo {author} {\bibfnamefont {Y.}~\bibnamefont {Sasaki}},
  \bibinfo {author} {\bibfnamefont {A.}~\bibnamefont {Matsuda}}, \bibinfo
  {author} {\bibfnamefont {H.}~\bibnamefont {Fujise}}, \bibinfo {author}
  {\bibfnamefont {Y.}~\bibnamefont {Kawabe}}, \bibinfo {author} {\bibfnamefont
  {K.}~\bibnamefont {Hashigaya}}, \bibinfo {author} {\bibfnamefont
  {S.}~\bibnamefont {Owada}}, \bibinfo {author} {\bibfnamefont
  {T.}~\bibnamefont {Togashi}}, \bibinfo {author} {\bibfnamefont
  {K.}~\bibnamefont {Nakajima}}, \bibinfo {author} {\bibfnamefont
  {M.}~\bibnamefont {Yabashi}}, \bibinfo {author} {\bibfnamefont
  {Y.}~\bibnamefont {Hikosaka}},\ and\ \bibinfo {author} {\bibfnamefont
  {A.}~\bibnamefont {Hishikawa}},\ }\bibfield  {title} {\bibinfo {title}
  {Multielectron-ion coincidence spectroscopy of xe in extreme ultraviolet
  laser fields: Nonlinear multiple ionization via double core-hole states},\
  }\href {https://doi.org/10.1103/PhysRevLett.124.193201} {\bibfield  {journal}
  {\bibinfo  {journal} {Phys. Rev. Lett.}\ }\textbf {\bibinfo {volume} {124}},\
  \bibinfo {pages} {193201} (\bibinfo {year} {2020})}\BibitemShut {NoStop}%
\bibitem [{\citenamefont {Young}\ \emph {et~al.}(2010)\citenamefont {Young},
  \citenamefont {Kanter}, \citenamefont {Kr{\"a}ssig}, \citenamefont {Li},
  \citenamefont {March}, \citenamefont {Pratt}, \citenamefont {Santra},
  \citenamefont {Southworth}, \citenamefont {Rohringer}, \citenamefont
  {DiMauro}, \citenamefont {Doumy}, \citenamefont {Roedig}, \citenamefont
  {Berrah}, \citenamefont {Fang}, \citenamefont {Hoener}, \citenamefont
  {Bucksbaum}, \citenamefont {Cryan}, \citenamefont {Ghimire}, \citenamefont
  {Glownia}, \citenamefont {Reis}, \citenamefont {Bozek}, \citenamefont
  {Bostedt},\ and\ \citenamefont {Messerschmidt}}]{Young2010}%
  \BibitemOpen
  \bibfield  {author} {\bibinfo {author} {\bibfnamefont {L.}~\bibnamefont
  {Young}}, \bibinfo {author} {\bibfnamefont {E.~P.}\ \bibnamefont {Kanter}},
  \bibinfo {author} {\bibfnamefont {B.}~\bibnamefont {Kr{\"a}ssig}}, \bibinfo
  {author} {\bibfnamefont {Y.}~\bibnamefont {Li}}, \bibinfo {author}
  {\bibfnamefont {A.~M.}\ \bibnamefont {March}}, \bibinfo {author}
  {\bibfnamefont {S.~T.}\ \bibnamefont {Pratt}}, \bibinfo {author}
  {\bibfnamefont {R.}~\bibnamefont {Santra}}, \bibinfo {author} {\bibfnamefont
  {S.~H.}\ \bibnamefont {Southworth}}, \bibinfo {author} {\bibfnamefont
  {N.}~\bibnamefont {Rohringer}}, \bibinfo {author} {\bibfnamefont {L.~F.}\
  \bibnamefont {DiMauro}}, \bibinfo {author} {\bibfnamefont {G.}~\bibnamefont
  {Doumy}}, \bibinfo {author} {\bibfnamefont {C.~A.}\ \bibnamefont {Roedig}},
  \bibinfo {author} {\bibfnamefont {N.}~\bibnamefont {Berrah}}, \bibinfo
  {author} {\bibfnamefont {L.}~\bibnamefont {Fang}}, \bibinfo {author}
  {\bibfnamefont {M.}~\bibnamefont {Hoener}}, \bibinfo {author} {\bibfnamefont
  {P.~H.}\ \bibnamefont {Bucksbaum}}, \bibinfo {author} {\bibfnamefont {J.~P.}\
  \bibnamefont {Cryan}}, \bibinfo {author} {\bibfnamefont {S.}~\bibnamefont
  {Ghimire}}, \bibinfo {author} {\bibfnamefont {J.~M.}\ \bibnamefont
  {Glownia}}, \bibinfo {author} {\bibfnamefont {D.~A.}\ \bibnamefont {Reis}},
  \bibinfo {author} {\bibfnamefont {J.~D.}\ \bibnamefont {Bozek}}, \bibinfo
  {author} {\bibfnamefont {C.}~\bibnamefont {Bostedt}},\ and\ \bibinfo {author}
  {\bibfnamefont {M.}~\bibnamefont {Messerschmidt}},\ }\bibfield  {title}
  {\bibinfo {title} {Femtosecond electronic response of atoms to ultra-intense
  {X}-rays},\ }\href {https://doi.org/10.1038/nature09177} {\bibfield
  {journal} {\bibinfo  {journal} {Nat.}\ }\textbf {\bibinfo {volume} {466}},\
  \bibinfo {pages} {56} (\bibinfo {year} {2010})}\BibitemShut {NoStop}%
\bibitem [{\citenamefont {Gerken}\ \emph {et~al.}(2014)\citenamefont {Gerken},
  \citenamefont {Klumpp}, \citenamefont {Sorokin}, \citenamefont {Tiedtke},
  \citenamefont {Richter}, \citenamefont {B\"urk}, \citenamefont {Mertens},
  \citenamefont {Jurani\ifmmode~\acute{c}\else \'{c}\fi{}},\ and\ \citenamefont
  {Martins}}]{Gerken2014}%
  \BibitemOpen
  \bibfield  {author} {\bibinfo {author} {\bibfnamefont {N.}~\bibnamefont
  {Gerken}}, \bibinfo {author} {\bibfnamefont {S.}~\bibnamefont {Klumpp}},
  \bibinfo {author} {\bibfnamefont {A.~A.}\ \bibnamefont {Sorokin}}, \bibinfo
  {author} {\bibfnamefont {K.}~\bibnamefont {Tiedtke}}, \bibinfo {author}
  {\bibfnamefont {M.}~\bibnamefont {Richter}}, \bibinfo {author} {\bibfnamefont
  {V.}~\bibnamefont {B\"urk}}, \bibinfo {author} {\bibfnamefont
  {K.}~\bibnamefont {Mertens}}, \bibinfo {author} {\bibfnamefont
  {P.}~\bibnamefont {Jurani\ifmmode~\acute{c}\else \'{c}\fi{}}},\ and\ \bibinfo
  {author} {\bibfnamefont {M.}~\bibnamefont {Martins}},\ }\bibfield  {title}
  {\bibinfo {title} {Time-dependent multiphoton ionization of xenon in the
  soft-x-ray regime},\ }\href {https://doi.org/10.1103/PhysRevLett.112.213002}
  {\bibfield  {journal} {\bibinfo  {journal} {Phys. Rev. Lett.}\ }\textbf
  {\bibinfo {volume} {112}},\ \bibinfo {pages} {213002} (\bibinfo {year}
  {2014})}\BibitemShut {NoStop}%
\bibitem [{\citenamefont {Richter}\ \emph {et~al.}(2010)\citenamefont
  {Richter}, \citenamefont {Bobashev}, \citenamefont {Sorokin},\ and\
  \citenamefont {Tiedtke}}]{Richter2010}%
  \BibitemOpen
  \bibfield  {author} {\bibinfo {author} {\bibfnamefont {M.}~\bibnamefont
  {Richter}}, \bibinfo {author} {\bibfnamefont {S.~V.}\ \bibnamefont
  {Bobashev}}, \bibinfo {author} {\bibfnamefont {A.~A.}\ \bibnamefont
  {Sorokin}},\ and\ \bibinfo {author} {\bibfnamefont {K.}~\bibnamefont
  {Tiedtke}},\ }\bibfield  {title} {\bibinfo {title} {Multiphoton ionization of
  atoms with soft x-ray pulses},\ }\href
  {https://doi.org/10.1088/0953-4075/43/19/194005} {\bibfield  {journal}
  {\bibinfo  {journal} {Journal of Physics B: Atomic, Molecular and Optical
  Physics}\ }\textbf {\bibinfo {volume} {43}},\ \bibinfo {pages} {194005}
  (\bibinfo {year} {2010})}\BibitemShut {NoStop}%
\bibitem [{\citenamefont {Klumpp}\ \emph {et~al.}(2017)\citenamefont {Klumpp},
  \citenamefont {Gerken}, \citenamefont {Mertens}, \citenamefont {Richter},
  \citenamefont {Sonntag}, \citenamefont {Sorokin}, \citenamefont {Braune},
  \citenamefont {Tiedtke}, \citenamefont {Zimmermann},\ and\ \citenamefont
  {Martins}}]{Klumpp2017}%
  \BibitemOpen
  \bibfield  {author} {\bibinfo {author} {\bibfnamefont {S.}~\bibnamefont
  {Klumpp}}, \bibinfo {author} {\bibfnamefont {N.}~\bibnamefont {Gerken}},
  \bibinfo {author} {\bibfnamefont {K.}~\bibnamefont {Mertens}}, \bibinfo
  {author} {\bibfnamefont {M.}~\bibnamefont {Richter}}, \bibinfo {author}
  {\bibfnamefont {B.}~\bibnamefont {Sonntag}}, \bibinfo {author} {\bibfnamefont
  {A.~A.}\ \bibnamefont {Sorokin}}, \bibinfo {author} {\bibfnamefont
  {M.}~\bibnamefont {Braune}}, \bibinfo {author} {\bibfnamefont
  {K.}~\bibnamefont {Tiedtke}}, \bibinfo {author} {\bibfnamefont
  {P.}~\bibnamefont {Zimmermann}},\ and\ \bibinfo {author} {\bibfnamefont
  {M.}~\bibnamefont {Martins}},\ }\bibfield  {title} {\bibinfo {title}
  {Multiple auger cycle photoionisation of manganese atoms by short soft x-ray
  pulses},\ }\href {https://doi.org/10.1088/1367-2630/aa660a} {\bibfield
  {journal} {\bibinfo  {journal} {New Journal of Physics}\ }\textbf {\bibinfo
  {volume} {19}},\ \bibinfo {pages} {043002} (\bibinfo {year}
  {2017})}\BibitemShut {NoStop}%
\bibitem [{\citenamefont {Berrah}\ \emph {et~al.}(2014)\citenamefont {Berrah},
  \citenamefont {Fang}, \citenamefont {Osipov}, \citenamefont {Murphy},
  \citenamefont {Bostedtc},\ and\ \citenamefont {Bozek}}]{Berrah2014}%
  \BibitemOpen
  \bibfield  {author} {\bibinfo {author} {\bibfnamefont {N.}~\bibnamefont
  {Berrah}}, \bibinfo {author} {\bibfnamefont {L.}~\bibnamefont {Fang}},
  \bibinfo {author} {\bibfnamefont {T.}~\bibnamefont {Osipov}}, \bibinfo
  {author} {\bibfnamefont {B.}~\bibnamefont {Murphy}}, \bibinfo {author}
  {\bibfnamefont {C.}~\bibnamefont {Bostedtc}},\ and\ \bibinfo {author}
  {\bibfnamefont {J.}~\bibnamefont {Bozek}},\ }\bibfield  {title} {\bibinfo
  {title} {Multiphoton ionization and fragmentation of molecules with the
  lclsx-ray fel},\ }\href {https://doi.org/10.1016/j.elspec.2013.10.009}
  {\bibfield  {journal} {\bibinfo  {journal} {Journal of Electron Spectroscopy
  and Related Phenomena}\ }\textbf {\bibinfo {volume} {196}},\ \bibinfo {pages}
  {34} (\bibinfo {year} {2014})}\BibitemShut {NoStop}%
\bibitem [{\citenamefont {Fukuzawa}\ \emph {et~al.}(2013)\citenamefont
  {Fukuzawa}, \citenamefont {Son}, \citenamefont {Motomura}, \citenamefont
  {Mondal}, \citenamefont {Nagaya}, \citenamefont {Wada}, \citenamefont {Liu},
  \citenamefont {Feifel}, \citenamefont {Tachibana}, \citenamefont {Ito},
  \citenamefont {Kimura}, \citenamefont {Sakai}, \citenamefont {Matsunami},
  \citenamefont {Hayashita}, \citenamefont {Kajikawa}, \citenamefont
  {Johnsson}, \citenamefont {Siano}, \citenamefont {Kukk}, \citenamefont
  {Rudek}, \citenamefont {Erk}, \citenamefont {Foucar}, \citenamefont {Robert},
  \citenamefont {Miron}, \citenamefont {Tono}, \citenamefont {Inubushi},
  \citenamefont {Hatsui}, \citenamefont {Yabashi}, \citenamefont {Yao},
  \citenamefont {Santra},\ and\ \citenamefont {Ueda}}]{Fukuzawa2013}%
  \BibitemOpen
  \bibfield  {author} {\bibinfo {author} {\bibfnamefont {H.}~\bibnamefont
  {Fukuzawa}}, \bibinfo {author} {\bibfnamefont {S.-K.}\ \bibnamefont {Son}},
  \bibinfo {author} {\bibfnamefont {K.}~\bibnamefont {Motomura}}, \bibinfo
  {author} {\bibfnamefont {S.}~\bibnamefont {Mondal}}, \bibinfo {author}
  {\bibfnamefont {K.}~\bibnamefont {Nagaya}}, \bibinfo {author} {\bibfnamefont
  {S.}~\bibnamefont {Wada}}, \bibinfo {author} {\bibfnamefont {X.-J.}\
  \bibnamefont {Liu}}, \bibinfo {author} {\bibfnamefont {R.}~\bibnamefont
  {Feifel}}, \bibinfo {author} {\bibfnamefont {T.}~\bibnamefont {Tachibana}},
  \bibinfo {author} {\bibfnamefont {Y.}~\bibnamefont {Ito}}, \bibinfo {author}
  {\bibfnamefont {M.}~\bibnamefont {Kimura}}, \bibinfo {author} {\bibfnamefont
  {T.}~\bibnamefont {Sakai}}, \bibinfo {author} {\bibfnamefont
  {K.}~\bibnamefont {Matsunami}}, \bibinfo {author} {\bibfnamefont
  {H.}~\bibnamefont {Hayashita}}, \bibinfo {author} {\bibfnamefont
  {J.}~\bibnamefont {Kajikawa}}, \bibinfo {author} {\bibfnamefont
  {P.}~\bibnamefont {Johnsson}}, \bibinfo {author} {\bibfnamefont
  {M.}~\bibnamefont {Siano}}, \bibinfo {author} {\bibfnamefont
  {E.}~\bibnamefont {Kukk}}, \bibinfo {author} {\bibfnamefont {B.}~\bibnamefont
  {Rudek}}, \bibinfo {author} {\bibfnamefont {B.}~\bibnamefont {Erk}}, \bibinfo
  {author} {\bibfnamefont {L.}~\bibnamefont {Foucar}}, \bibinfo {author}
  {\bibfnamefont {E.}~\bibnamefont {Robert}}, \bibinfo {author} {\bibfnamefont
  {C.}~\bibnamefont {Miron}}, \bibinfo {author} {\bibfnamefont
  {K.}~\bibnamefont {Tono}}, \bibinfo {author} {\bibfnamefont {Y.}~\bibnamefont
  {Inubushi}}, \bibinfo {author} {\bibfnamefont {T.}~\bibnamefont {Hatsui}},
  \bibinfo {author} {\bibfnamefont {M.}~\bibnamefont {Yabashi}}, \bibinfo
  {author} {\bibfnamefont {M.}~\bibnamefont {Yao}}, \bibinfo {author}
  {\bibfnamefont {R.}~\bibnamefont {Santra}},\ and\ \bibinfo {author}
  {\bibfnamefont {K.}~\bibnamefont {Ueda}},\ }\bibfield  {title} {\bibinfo
  {title} {Deep inner-shell multiphoton ionization by intense x-ray
  free-electron laser pulses},\ }\href
  {https://doi.org/10.1103/PhysRevLett.110.173005} {\bibfield  {journal}
  {\bibinfo  {journal} {Phys. Rev. Lett.}\ }\textbf {\bibinfo {volume} {110}},\
  \bibinfo {pages} {173005} (\bibinfo {year} {2013})}\BibitemShut {NoStop}%
\bibitem [{\citenamefont {Southworth}\ \emph {et~al.}(2019)\citenamefont
  {Southworth}, \citenamefont {Dunford}, \citenamefont {Ray}, \citenamefont
  {Kanter}, \citenamefont {Doumy}, \citenamefont {March}, \citenamefont {Ho},
  \citenamefont {Krässig}, \citenamefont {Gao}, \citenamefont {Lehmann},
  \citenamefont {Picón}, \citenamefont {Young}, \citenamefont {Walko},\ and\
  \citenamefont {Cheng}}]{Southworth2019}%
  \BibitemOpen
  \bibfield  {author} {\bibinfo {author} {\bibfnamefont {S.~H.}\ \bibnamefont
  {Southworth}}, \bibinfo {author} {\bibfnamefont {R.~W.}\ \bibnamefont
  {Dunford}}, \bibinfo {author} {\bibfnamefont {D.}~\bibnamefont {Ray}},
  \bibinfo {author} {\bibfnamefont {E.~P.}\ \bibnamefont {Kanter}}, \bibinfo
  {author} {\bibfnamefont {G.}~\bibnamefont {Doumy}}, \bibinfo {author}
  {\bibfnamefont {A.~M.}\ \bibnamefont {March}}, \bibinfo {author}
  {\bibfnamefont {P.~J.}\ \bibnamefont {Ho}}, \bibinfo {author} {\bibfnamefont
  {B.}~\bibnamefont {Krässig}}, \bibinfo {author} {\bibfnamefont
  {Y.}~\bibnamefont {Gao}}, \bibinfo {author} {\bibfnamefont {C.~S.}\
  \bibnamefont {Lehmann}}, \bibinfo {author} {\bibfnamefont {A.}~\bibnamefont
  {Picón}}, \bibinfo {author} {\bibfnamefont {L.}~\bibnamefont {Young}},
  \bibinfo {author} {\bibfnamefont {D.~A.}\ \bibnamefont {Walko}},\ and\
  \bibinfo {author} {\bibfnamefont {L.}~\bibnamefont {Cheng}},\ }\bibfield
  {title} {\bibinfo {title} {Observing pre-edge k-shell resonances in kr, xe,
  and xef$_2$},\ }\href {https://doi.org/10.1103/PhysRevA.100.022507}
  {\bibfield  {journal} {\bibinfo  {journal} {Phys. Rev. A}\ }\textbf {\bibinfo
  {volume} {100}},\ \bibinfo {pages} {022507} (\bibinfo {year}
  {2019})}\BibitemShut {NoStop}%
\bibitem [{\citenamefont {Kurka}\ \emph {et~al.}(2009)\citenamefont {Kurka},
  \citenamefont {Rudenko}, \citenamefont {Foucar}, \citenamefont {K{\"u}hnel},
  \citenamefont {Jiang}, \citenamefont {Ergler}, \citenamefont {Havermeier},
  \citenamefont {Smolarski}, \citenamefont {Sch{\"o}ssler}, \citenamefont
  {Cole}, \citenamefont {Sch{\"o}ffler}, \citenamefont {D{\"o}rner},
  \citenamefont {Gensch}, \citenamefont {D{\"u}sterer}, \citenamefont
  {Treusch}, \citenamefont {Fritzsche}, \citenamefont {Grum-Grzhimailo},
  \citenamefont {Gryzlova}, \citenamefont {Kabachnik}, \citenamefont
  {Schr{\"o}ter}, \citenamefont {Moshammer},\ and\ \citenamefont
  {Ullrich}}]{Kurka2009}%
  \BibitemOpen
  \bibfield  {author} {\bibinfo {author} {\bibfnamefont {M.}~\bibnamefont
  {Kurka}}, \bibinfo {author} {\bibfnamefont {A.}~\bibnamefont {Rudenko}},
  \bibinfo {author} {\bibfnamefont {L.}~\bibnamefont {Foucar}}, \bibinfo
  {author} {\bibfnamefont {K.~U.}\ \bibnamefont {K{\"u}hnel}}, \bibinfo
  {author} {\bibfnamefont {Y.~H.}\ \bibnamefont {Jiang}}, \bibinfo {author}
  {\bibfnamefont {T.}~\bibnamefont {Ergler}}, \bibinfo {author} {\bibfnamefont
  {T.}~\bibnamefont {Havermeier}}, \bibinfo {author} {\bibfnamefont
  {M.}~\bibnamefont {Smolarski}}, \bibinfo {author} {\bibfnamefont
  {S.}~\bibnamefont {Sch{\"o}ssler}}, \bibinfo {author} {\bibfnamefont
  {K.}~\bibnamefont {Cole}}, \bibinfo {author} {\bibfnamefont {M.}~\bibnamefont
  {Sch{\"o}ffler}}, \bibinfo {author} {\bibfnamefont {R.}~\bibnamefont
  {D{\"o}rner}}, \bibinfo {author} {\bibfnamefont {M.}~\bibnamefont {Gensch}},
  \bibinfo {author} {\bibfnamefont {S.}~\bibnamefont {D{\"u}sterer}}, \bibinfo
  {author} {\bibfnamefont {R.}~\bibnamefont {Treusch}}, \bibinfo {author}
  {\bibfnamefont {S.}~\bibnamefont {Fritzsche}}, \bibinfo {author}
  {\bibfnamefont {A.~N.}\ \bibnamefont {Grum-Grzhimailo}}, \bibinfo {author}
  {\bibfnamefont {E.~V.}\ \bibnamefont {Gryzlova}}, \bibinfo {author}
  {\bibfnamefont {N.~M.}\ \bibnamefont {Kabachnik}}, \bibinfo {author}
  {\bibfnamefont {C.~D.}\ \bibnamefont {Schr{\"o}ter}}, \bibinfo {author}
  {\bibfnamefont {R.}~\bibnamefont {Moshammer}},\ and\ \bibinfo {author}
  {\bibfnamefont {J.}~\bibnamefont {Ullrich}},\ }\bibfield  {title} {\bibinfo
  {title} {Two-photon double ionization of {Ne} by free-electron laser
  radiation: a kinematically complete experiment},\ }\href
  {https://doi.org/10.1088/0953-4075/42/14/141002} {\bibfield  {journal}
  {\bibinfo  {journal} {J. Phys. B: At. Mol. Opt. Ph.}\ }\textbf {\bibinfo
  {volume} {42}},\ \bibinfo {pages} {141002} (\bibinfo {year}
  {2009})}\BibitemShut {NoStop}%
\bibitem [{\citenamefont {Braune}\ \emph {et~al.}(2016)\citenamefont {Braune},
  \citenamefont {Hartmann}, \citenamefont {Ilchen}, \citenamefont {Knie},
  \citenamefont {Lischke}, \citenamefont {Reink\"oster}, \citenamefont
  {Meissner}, \citenamefont {Deinert}, \citenamefont {Glaser}, \citenamefont
  {Al-Dossary}, \citenamefont {Ehresmann}, \citenamefont {Kheifets},\ and\
  \citenamefont {Viefhaus}}]{Braune2016}%
  \BibitemOpen
  \bibfield  {author} {\bibinfo {author} {\bibfnamefont {M.}~\bibnamefont
  {Braune}}, \bibinfo {author} {\bibfnamefont {G.}~\bibnamefont {Hartmann}},
  \bibinfo {author} {\bibfnamefont {M.}~\bibnamefont {Ilchen}}, \bibinfo
  {author} {\bibfnamefont {A.}~\bibnamefont {Knie}}, \bibinfo {author}
  {\bibfnamefont {T.}~\bibnamefont {Lischke}}, \bibinfo {author} {\bibfnamefont
  {A.}~\bibnamefont {Reink\"oster}}, \bibinfo {author} {\bibfnamefont
  {A.}~\bibnamefont {Meissner}}, \bibinfo {author} {\bibfnamefont
  {S.}~\bibnamefont {Deinert}}, \bibinfo {author} {\bibfnamefont
  {L.}~\bibnamefont {Glaser}}, \bibinfo {author} {\bibfnamefont
  {O.}~\bibnamefont {Al-Dossary}}, \bibinfo {author} {\bibfnamefont
  {A.}~\bibnamefont {Ehresmann}}, \bibinfo {author} {\bibfnamefont {A.~S.}\
  \bibnamefont {Kheifets}},\ and\ \bibinfo {author} {\bibfnamefont
  {J.}~\bibnamefont {Viefhaus}},\ }\bibfield  {title} {\bibinfo {title}
  {Electron angular distributions of noble gases in sequential two-photon
  double ionization},\ }\href {https://doi.org/10.1080/09500340.2015.1047422}
  {\bibfield  {journal} {\bibinfo  {journal} {J. Mod. Opt.}\ }\textbf {\bibinfo
  {volume} {63}},\ \bibinfo {pages} {324} (\bibinfo {year} {2016})}\BibitemShut
  {NoStop}%
\bibitem [{\citenamefont {Carpeggiani}\ \emph {et~al.}(2019)\citenamefont
  {Carpeggiani}, \citenamefont {Gryzlova}, \citenamefont {Reduzzi},
  \citenamefont {Dubrouil}, \citenamefont {Faccial{\'a}}, \citenamefont
  {Negro}, \citenamefont {Ueda}, \citenamefont {Burkov}, \citenamefont
  {Frassetto}, \citenamefont {Stienkemeier}, \citenamefont {Ovcharenko},
  \citenamefont {Meyer}, \citenamefont {Plekan}, \citenamefont {Finetti},
  \citenamefont {Prince}, \citenamefont {Callegari}, \citenamefont
  {Grum-Grzhimailo},\ and\ \citenamefont {Sansone}}]{Carpeggiani2019}%
  \BibitemOpen
  \bibfield  {author} {\bibinfo {author} {\bibfnamefont {P.~A.}\ \bibnamefont
  {Carpeggiani}}, \bibinfo {author} {\bibfnamefont {E.~V.}\ \bibnamefont
  {Gryzlova}}, \bibinfo {author} {\bibfnamefont {M.}~\bibnamefont {Reduzzi}},
  \bibinfo {author} {\bibfnamefont {A.}~\bibnamefont {Dubrouil}}, \bibinfo
  {author} {\bibfnamefont {D.}~\bibnamefont {Faccial{\'a}}}, \bibinfo {author}
  {\bibfnamefont {M.}~\bibnamefont {Negro}}, \bibinfo {author} {\bibfnamefont
  {K.}~\bibnamefont {Ueda}}, \bibinfo {author} {\bibfnamefont {S.~M.}\
  \bibnamefont {Burkov}}, \bibinfo {author} {\bibfnamefont {F.}~\bibnamefont
  {Frassetto}}, \bibinfo {author} {\bibfnamefont {F.}~\bibnamefont
  {Stienkemeier}}, \bibinfo {author} {\bibfnamefont {Y.}~\bibnamefont
  {Ovcharenko}}, \bibinfo {author} {\bibfnamefont {M.}~\bibnamefont {Meyer}},
  \bibinfo {author} {\bibfnamefont {O.}~\bibnamefont {Plekan}}, \bibinfo
  {author} {\bibfnamefont {P.}~\bibnamefont {Finetti}}, \bibinfo {author}
  {\bibfnamefont {K.~C.}\ \bibnamefont {Prince}}, \bibinfo {author}
  {\bibfnamefont {C.}~\bibnamefont {Callegari}}, \bibinfo {author}
  {\bibfnamefont {A.~N.}\ \bibnamefont {Grum-Grzhimailo}},\ and\ \bibinfo
  {author} {\bibfnamefont {G.}~\bibnamefont {Sansone}},\ }\bibfield  {title}
  {\bibinfo {title} {Complete reconstruction of bound and unbound electronic
  wavefunctions in two-photon double ionization},\ }\href
  {https://doi.org/10.1038/s41567-018-0340-4} {\bibfield  {journal} {\bibinfo
  {journal} {Nat. Phys.}\ }\textbf {\bibinfo {volume} {15}},\ \bibinfo {pages}
  {170} (\bibinfo {year} {2019})}\BibitemShut {NoStop}%
\bibitem [{\citenamefont {Mazza}\ \emph {et~al.}(2020)\citenamefont {Mazza},
  \citenamefont {Ilchen}, \citenamefont {Kiselev}, \citenamefont {Gryzlova},
  \citenamefont {Baumann}, \citenamefont {Boll}, \citenamefont {De~Fanis},
  \citenamefont {Grychtol}, \citenamefont {Monta\~no}, \citenamefont {Music},
  \citenamefont {Ovcharenko}, \citenamefont {Rennhack}, \citenamefont {Rivas},
  \citenamefont {Schmidt}, \citenamefont {Wagner}, \citenamefont {Ziolkowski},
  \citenamefont {Berrah}, \citenamefont {Erk}, \citenamefont {Johnsson},
  \citenamefont {K\"ustner-Wetekam}, \citenamefont {Marder}, \citenamefont
  {Martins}, \citenamefont {Ott}, \citenamefont {Pathak}, \citenamefont
  {Pfeifer}, \citenamefont {Rolles}, \citenamefont {Zatsarinny}, \citenamefont
  {Grum-Grzhimailo},\ and\ \citenamefont {Meyer}}]{Mazza2020}%
  \BibitemOpen
  \bibfield  {author} {\bibinfo {author} {\bibfnamefont {T.}~\bibnamefont
  {Mazza}}, \bibinfo {author} {\bibfnamefont {M.}~\bibnamefont {Ilchen}},
  \bibinfo {author} {\bibfnamefont {M.~D.}\ \bibnamefont {Kiselev}}, \bibinfo
  {author} {\bibfnamefont {E.~V.}\ \bibnamefont {Gryzlova}}, \bibinfo {author}
  {\bibfnamefont {T.~M.}\ \bibnamefont {Baumann}}, \bibinfo {author}
  {\bibfnamefont {R.}~\bibnamefont {Boll}}, \bibinfo {author} {\bibfnamefont
  {A.}~\bibnamefont {De~Fanis}}, \bibinfo {author} {\bibfnamefont
  {P.}~\bibnamefont {Grychtol}}, \bibinfo {author} {\bibfnamefont
  {J.}~\bibnamefont {Monta\~no}}, \bibinfo {author} {\bibfnamefont
  {V.}~\bibnamefont {Music}}, \bibinfo {author} {\bibfnamefont
  {Y.}~\bibnamefont {Ovcharenko}}, \bibinfo {author} {\bibfnamefont
  {N.}~\bibnamefont {Rennhack}}, \bibinfo {author} {\bibfnamefont {D.~E.}\
  \bibnamefont {Rivas}}, \bibinfo {author} {\bibfnamefont {P.}~\bibnamefont
  {Schmidt}}, \bibinfo {author} {\bibfnamefont {R.}~\bibnamefont {Wagner}},
  \bibinfo {author} {\bibfnamefont {P.}~\bibnamefont {Ziolkowski}}, \bibinfo
  {author} {\bibfnamefont {N.}~\bibnamefont {Berrah}}, \bibinfo {author}
  {\bibfnamefont {B.}~\bibnamefont {Erk}}, \bibinfo {author} {\bibfnamefont
  {P.}~\bibnamefont {Johnsson}}, \bibinfo {author} {\bibfnamefont
  {C.}~\bibnamefont {K\"ustner-Wetekam}}, \bibinfo {author} {\bibfnamefont
  {L.}~\bibnamefont {Marder}}, \bibinfo {author} {\bibfnamefont
  {M.}~\bibnamefont {Martins}}, \bibinfo {author} {\bibfnamefont
  {C.}~\bibnamefont {Ott}}, \bibinfo {author} {\bibfnamefont {S.}~\bibnamefont
  {Pathak}}, \bibinfo {author} {\bibfnamefont {T.}~\bibnamefont {Pfeifer}},
  \bibinfo {author} {\bibfnamefont {D.}~\bibnamefont {Rolles}}, \bibinfo
  {author} {\bibfnamefont {O.}~\bibnamefont {Zatsarinny}}, \bibinfo {author}
  {\bibfnamefont {A.~N.}\ \bibnamefont {Grum-Grzhimailo}},\ and\ \bibinfo
  {author} {\bibfnamefont {M.}~\bibnamefont {Meyer}},\ }\bibfield  {title}
  {\bibinfo {title} {Mapping resonance structures in transient core-ionized
  atoms},\ }\href {https://doi.org/10.1103/PhysRevX.10.041056} {\bibfield
  {journal} {\bibinfo  {journal} {Phys. Rev. X}\ }\textbf {\bibinfo {volume}
  {10}},\ \bibinfo {pages} {041056} (\bibinfo {year} {2020})}\BibitemShut
  {NoStop}%
\bibitem [{\citenamefont {Kheifets}(2007)}]{Kheifets2007}%
  \BibitemOpen
  \bibfield  {author} {\bibinfo {author} {\bibfnamefont {A.~S.}\ \bibnamefont
  {Kheifets}},\ }\bibfield  {title} {\bibinfo {title} {Sequential two-photon
  double ionization of noble gas atoms},\ }\href
  {https://doi.org/10.1088/0953-4075/40/22/f02} {\bibfield  {journal} {\bibinfo
   {journal} {Journal of Physics B: Atomic, Molecular and Optical Physics}\
  }\textbf {\bibinfo {volume} {40}},\ \bibinfo {pages} {F313} (\bibinfo {year}
  {2007})}\BibitemShut {NoStop}%
\bibitem [{\citenamefont {Grum-Grzhimailo}\ \emph {et~al.}(2016)\citenamefont
  {Grum-Grzhimailo}, \citenamefont {Gryzlova}, \citenamefont {Fritzsche},\ and\
  \citenamefont {Kabachnik}}]{Grum2016}%
  \BibitemOpen
  \bibfield  {author} {\bibinfo {author} {\bibfnamefont {A.~N.}\ \bibnamefont
  {Grum-Grzhimailo}}, \bibinfo {author} {\bibfnamefont {E.~V.}\ \bibnamefont
  {Gryzlova}}, \bibinfo {author} {\bibfnamefont {S.}~\bibnamefont
  {Fritzsche}},\ and\ \bibinfo {author} {\bibfnamefont {N.~M.}\ \bibnamefont
  {Kabachnik}},\ }\bibfield  {title} {\bibinfo {title} {Photoelectron angular
  distributions and correlations in sequential double and triple atomic
  ionization by free electron lasers},\ }\href
  {https://doi.org/10.1080/09500340.2015.1047805} {\bibfield  {journal}
  {\bibinfo  {journal} {Journal of Modern Optics}\ }\textbf {\bibinfo {volume}
  {63}},\ \bibinfo {pages} {334} (\bibinfo {year} {2016})},\ \Eprint
  {https://arxiv.org/abs/https://doi.org/10.1080/09500340.2015.1047805}
  {https://doi.org/10.1080/09500340.2015.1047805} \BibitemShut {NoStop}%
\bibitem [{\citenamefont {Mondal}\ \emph
  {et~al.}(2013{\natexlab{a}})\citenamefont {Mondal}, \citenamefont {Ma},
  \citenamefont {Motomura}, \citenamefont {Fukuzawa}, \citenamefont {Yamada},
  \citenamefont {Nagaya}, \citenamefont {Yase}, \citenamefont {Mizoguchi},
  \citenamefont {Yao}, \citenamefont {Rouz{\'{e}}e}, \citenamefont
  {Hundertmark}, \citenamefont {Vrakking}, \citenamefont {Johnsson},
  \citenamefont {Nagasono}, \citenamefont {Tono}, \citenamefont {Togashi},
  \citenamefont {Senba}, \citenamefont {Ohashi}, \citenamefont {Yabashi},
  \citenamefont {Ishikawa}, \citenamefont {Sazhina}, \citenamefont {Fritzsche},
  \citenamefont {Kabachnik},\ and\ \citenamefont {Ueda}}]{Mondal_2013}%
  \BibitemOpen
  \bibfield  {author} {\bibinfo {author} {\bibfnamefont {S.}~\bibnamefont
  {Mondal}}, \bibinfo {author} {\bibfnamefont {R.}~\bibnamefont {Ma}}, \bibinfo
  {author} {\bibfnamefont {K.}~\bibnamefont {Motomura}}, \bibinfo {author}
  {\bibfnamefont {H.}~\bibnamefont {Fukuzawa}}, \bibinfo {author}
  {\bibfnamefont {A.}~\bibnamefont {Yamada}}, \bibinfo {author} {\bibfnamefont
  {K.}~\bibnamefont {Nagaya}}, \bibinfo {author} {\bibfnamefont
  {S.}~\bibnamefont {Yase}}, \bibinfo {author} {\bibfnamefont {Y.}~\bibnamefont
  {Mizoguchi}}, \bibinfo {author} {\bibfnamefont {M.}~\bibnamefont {Yao}},
  \bibinfo {author} {\bibfnamefont {A.}~\bibnamefont {Rouz{\'{e}}e}}, \bibinfo
  {author} {\bibfnamefont {A.}~\bibnamefont {Hundertmark}}, \bibinfo {author}
  {\bibfnamefont {M.~J.~J.}\ \bibnamefont {Vrakking}}, \bibinfo {author}
  {\bibfnamefont {P.}~\bibnamefont {Johnsson}}, \bibinfo {author}
  {\bibfnamefont {M.}~\bibnamefont {Nagasono}}, \bibinfo {author}
  {\bibfnamefont {K.}~\bibnamefont {Tono}}, \bibinfo {author} {\bibfnamefont
  {T.}~\bibnamefont {Togashi}}, \bibinfo {author} {\bibfnamefont
  {Y.}~\bibnamefont {Senba}}, \bibinfo {author} {\bibfnamefont
  {H.}~\bibnamefont {Ohashi}}, \bibinfo {author} {\bibfnamefont
  {M.}~\bibnamefont {Yabashi}}, \bibinfo {author} {\bibfnamefont
  {T.}~\bibnamefont {Ishikawa}}, \bibinfo {author} {\bibfnamefont {I.~P.}\
  \bibnamefont {Sazhina}}, \bibinfo {author} {\bibfnamefont {S.}~\bibnamefont
  {Fritzsche}}, \bibinfo {author} {\bibfnamefont {N.~M.}\ \bibnamefont
  {Kabachnik}},\ and\ \bibinfo {author} {\bibfnamefont {K.}~\bibnamefont
  {Ueda}},\ }\bibfield  {title} {\bibinfo {title} {Photoelectron angular
  distributions for the two-photon sequential double ionization of xenon by
  ultrashort extreme ultraviolet free electron laser pulses},\ }\href
  {https://doi.org/10.1088/0953-4075/46/16/164022} {\bibfield  {journal}
  {\bibinfo  {journal} {Journal of Physics B: Atomic, Molecular and Optical
  Physics}\ }\textbf {\bibinfo {volume} {46}},\ \bibinfo {pages} {164022}
  (\bibinfo {year} {2013}{\natexlab{a}})}\BibitemShut {NoStop}%
\bibitem [{\citenamefont {Ilchen}\ \emph {et~al.}(2018)\citenamefont {Ilchen},
  \citenamefont {Hartmann}, \citenamefont {Gryzlova}, \citenamefont {Achner},
  \citenamefont {Allaria}, \citenamefont {Beckmann}, \citenamefont {Braune},
  \citenamefont {Buck}, \citenamefont {Callegari}, \citenamefont {Coffee},
  \citenamefont {Cucini}, \citenamefont {Danailov}, \citenamefont {De~Fanis},
  \citenamefont {Demidovich}, \citenamefont {Ferrari}, \citenamefont {Finetti},
  \citenamefont {Glaser}, \citenamefont {Knie}, \citenamefont {Lindahl},
  \citenamefont {Plekan}, \citenamefont {Mahne}, \citenamefont {Mazza},
  \citenamefont {Raimondi}, \citenamefont {Roussel}, \citenamefont {Seltmann},
  \citenamefont {Shevchuk}, \citenamefont {Svetina}, \citenamefont {Walter},
  \citenamefont {Zangrando}, \citenamefont {Viefhaus}, \citenamefont
  {Grum-Grzhimailo},\ and\ \citenamefont {Meyer}}]{Ilchen2018}%
  \BibitemOpen
  \bibfield  {author} {\bibinfo {author} {\bibfnamefont {M.}~\bibnamefont
  {Ilchen}}, \bibinfo {author} {\bibfnamefont {G.}~\bibnamefont {Hartmann}},
  \bibinfo {author} {\bibfnamefont {E.~V.}\ \bibnamefont {Gryzlova}}, \bibinfo
  {author} {\bibfnamefont {A.}~\bibnamefont {Achner}}, \bibinfo {author}
  {\bibfnamefont {E.}~\bibnamefont {Allaria}}, \bibinfo {author} {\bibfnamefont
  {A.}~\bibnamefont {Beckmann}}, \bibinfo {author} {\bibfnamefont
  {M.}~\bibnamefont {Braune}}, \bibinfo {author} {\bibfnamefont
  {J.}~\bibnamefont {Buck}}, \bibinfo {author} {\bibfnamefont {C.}~\bibnamefont
  {Callegari}}, \bibinfo {author} {\bibfnamefont {R.~N.}\ \bibnamefont
  {Coffee}}, \bibinfo {author} {\bibfnamefont {R.}~\bibnamefont {Cucini}},
  \bibinfo {author} {\bibfnamefont {M.}~\bibnamefont {Danailov}}, \bibinfo
  {author} {\bibfnamefont {A.}~\bibnamefont {De~Fanis}}, \bibinfo {author}
  {\bibfnamefont {A.}~\bibnamefont {Demidovich}}, \bibinfo {author}
  {\bibfnamefont {E.}~\bibnamefont {Ferrari}}, \bibinfo {author} {\bibfnamefont
  {P.}~\bibnamefont {Finetti}}, \bibinfo {author} {\bibfnamefont
  {L.}~\bibnamefont {Glaser}}, \bibinfo {author} {\bibfnamefont
  {A.}~\bibnamefont {Knie}}, \bibinfo {author} {\bibfnamefont {A.~O.}\
  \bibnamefont {Lindahl}}, \bibinfo {author} {\bibfnamefont {O.}~\bibnamefont
  {Plekan}}, \bibinfo {author} {\bibfnamefont {N.}~\bibnamefont {Mahne}},
  \bibinfo {author} {\bibfnamefont {T.}~\bibnamefont {Mazza}}, \bibinfo
  {author} {\bibfnamefont {L.}~\bibnamefont {Raimondi}}, \bibinfo {author}
  {\bibfnamefont {E.}~\bibnamefont {Roussel}}, \bibinfo {author} {\bibfnamefont
  {F.~S.~J.}\ \bibnamefont {Seltmann}}, \bibinfo {author} {\bibfnamefont
  {I.}~\bibnamefont {Shevchuk}}, \bibinfo {author} {\bibfnamefont
  {C.}~\bibnamefont {Svetina}}, \bibinfo {author} {\bibfnamefont
  {P.}~\bibnamefont {Walter}}, \bibinfo {author} {\bibfnamefont
  {M.}~\bibnamefont {Zangrando}}, \bibinfo {author} {\bibfnamefont
  {J.}~\bibnamefont {Viefhaus}}, \bibinfo {author} {\bibfnamefont {A.~N.}\
  \bibnamefont {Grum-Grzhimailo}},\ and\ \bibinfo {author} {\bibfnamefont
  {M.}~\bibnamefont {Meyer}},\ }\bibfield  {title} {\bibinfo {title} {Symmetry
  breakdown of electron emission in extreme ultraviolet photoionization of
  argon},\ }\href {https://doi.org/10.1038/s41467-018-07152-7} {\bibfield
  {journal} {\bibinfo  {journal} {Nat. Commun.}\ }\textbf {\bibinfo {volume}
  {8}},\ \bibinfo {pages} {4659} (\bibinfo {year} {2018})}\BibitemShut
  {NoStop}%
\bibitem [{\citenamefont {Augustin}\ \emph {et~al.}(2018)\citenamefont
  {Augustin}, \citenamefont {Schulz}, \citenamefont {Schmid}, \citenamefont
  {Schnorr}, \citenamefont {Gryzlova}, \citenamefont {Lindenblatt},
  \citenamefont {Meister}, \citenamefont {Liu}, \citenamefont {Trost},
  \citenamefont {Fechner}, \citenamefont {Grum-Grzhimailo}, \citenamefont
  {Burkov}, \citenamefont {Braune}, \citenamefont {Treusch}, \citenamefont
  {Gisselbrecht}, \citenamefont {Schr{\"o}ter}, \citenamefont {Pfeifer},\ and\
  \citenamefont {Moshammer}}]{Augustin2018}%
  \BibitemOpen
  \bibfield  {author} {\bibinfo {author} {\bibfnamefont {S.}~\bibnamefont
  {Augustin}}, \bibinfo {author} {\bibfnamefont {M.}~\bibnamefont {Schulz}},
  \bibinfo {author} {\bibfnamefont {G.}~\bibnamefont {Schmid}}, \bibinfo
  {author} {\bibfnamefont {K.}~\bibnamefont {Schnorr}}, \bibinfo {author}
  {\bibfnamefont {E.~V.}\ \bibnamefont {Gryzlova}}, \bibinfo {author}
  {\bibfnamefont {H.}~\bibnamefont {Lindenblatt}}, \bibinfo {author}
  {\bibfnamefont {S.}~\bibnamefont {Meister}}, \bibinfo {author} {\bibfnamefont
  {Y.~F.}\ \bibnamefont {Liu}}, \bibinfo {author} {\bibfnamefont
  {F.}~\bibnamefont {Trost}}, \bibinfo {author} {\bibfnamefont
  {L.}~\bibnamefont {Fechner}}, \bibinfo {author} {\bibfnamefont {A.~N.}\
  \bibnamefont {Grum-Grzhimailo}}, \bibinfo {author} {\bibfnamefont {S.~M.}\
  \bibnamefont {Burkov}}, \bibinfo {author} {\bibfnamefont {M.}~\bibnamefont
  {Braune}}, \bibinfo {author} {\bibfnamefont {R.}~\bibnamefont {Treusch}},
  \bibinfo {author} {\bibfnamefont {M.}~\bibnamefont {Gisselbrecht}}, \bibinfo
  {author} {\bibfnamefont {C.~D.}\ \bibnamefont {Schr{\"o}ter}}, \bibinfo
  {author} {\bibfnamefont {T.}~\bibnamefont {Pfeifer}},\ and\ \bibinfo {author}
  {\bibfnamefont {R.}~\bibnamefont {Moshammer}},\ }\bibfield  {title} {\bibinfo
  {title} {Signatures of autoionization in the angular electron distribution in
  two-photon double ionization of ar},\ }\href
  {https://doi.org/10.1103/PhysRevA.98.033408} {\bibfield  {journal} {\bibinfo
  {journal} {Phys. Rev. A}\ }\textbf {\bibinfo {volume} {98}},\ \bibinfo
  {pages} {033408} (\bibinfo {year} {2018})}\BibitemShut {NoStop}%
\bibitem [{\citenamefont {Rouzee}\ \emph {et~al.}(2011)\citenamefont {Rouzee},
  \citenamefont {Johnsson}, \citenamefont {Gryzlova}, \citenamefont {Fukuzawa},
  \citenamefont {Yamada}, \citenamefont {Siu}, \citenamefont {Huismans},
  \citenamefont {Louis}, \citenamefont {Bijkerk}, \citenamefont {Holland},
  \citenamefont {Grum-Grzhimailo}, \citenamefont {Kabachnik}, \citenamefont
  {Vrakking},\ and\ \citenamefont {Ueda}}]{Rouzee2011}%
  \BibitemOpen
  \bibfield  {author} {\bibinfo {author} {\bibfnamefont {A.}~\bibnamefont
  {Rouzee}}, \bibinfo {author} {\bibfnamefont {P.}~\bibnamefont {Johnsson}},
  \bibinfo {author} {\bibfnamefont {E.}~\bibnamefont {Gryzlova}}, \bibinfo
  {author} {\bibfnamefont {H.}~\bibnamefont {Fukuzawa}}, \bibinfo {author}
  {\bibfnamefont {A.}~\bibnamefont {Yamada}}, \bibinfo {author} {\bibfnamefont
  {W.}~\bibnamefont {Siu}}, \bibinfo {author} {\bibfnamefont {Y.}~\bibnamefont
  {Huismans}}, \bibinfo {author} {\bibfnamefont {E.}~\bibnamefont {Louis}},
  \bibinfo {author} {\bibfnamefont {F.}~\bibnamefont {Bijkerk}}, \bibinfo
  {author} {\bibfnamefont {D.}~\bibnamefont {Holland}}, \bibinfo {author}
  {\bibfnamefont {A.}~\bibnamefont {Grum-Grzhimailo}}, \bibinfo {author}
  {\bibfnamefont {N.}~\bibnamefont {Kabachnik}}, \bibinfo {author}
  {\bibfnamefont {M.}~\bibnamefont {Vrakking}},\ and\ \bibinfo {author}
  {\bibfnamefont {K.}~\bibnamefont {Ueda}},\ }\bibfield  {title} {\bibinfo
  {title} {Angle-resolved photoelectron spectroscopy of sequential three-photon
  triple ionization of neon at 90.5 ev photon energy},\ }\href
  {https://doi.org/10.1103/PhysRevA.83.031401} {\bibfield  {journal} {\bibinfo
  {journal} {Phys. Rev. A}\ }\textbf {\bibinfo {volume} {83}},\ \bibinfo
  {pages} {031401(R)} (\bibinfo {year} {2011})}\BibitemShut {NoStop}%
\bibitem [{\citenamefont {Kabachnik}\ \emph {et~al.}(2007)\citenamefont
  {Kabachnik}, \citenamefont {Fritzsche}, \citenamefont {Grum-Grzhimailo},
  \citenamefont {Meyer},\ and\ \citenamefont {Ueda}}]{Kabachnik2007}%
  \BibitemOpen
  \bibfield  {author} {\bibinfo {author} {\bibfnamefont {N.~M.}\ \bibnamefont
  {Kabachnik}}, \bibinfo {author} {\bibfnamefont {S.}~\bibnamefont
  {Fritzsche}}, \bibinfo {author} {\bibfnamefont {A.~N.}\ \bibnamefont
  {Grum-Grzhimailo}}, \bibinfo {author} {\bibfnamefont {M.}~\bibnamefont
  {Meyer}},\ and\ \bibinfo {author} {\bibfnamefont {K.}~\bibnamefont {Ueda}},\
  }\bibfield  {title} {\bibinfo {title} {Coherence and correlations in
  photoinduced auger and fluorescence cascades in atoms},\ }\href@noop {}
  {\bibfield  {journal} {\bibinfo  {journal} {Phys. Rep.}\ }\textbf {\bibinfo
  {volume} {451}},\ \bibinfo {pages} {155} (\bibinfo {year}
  {2007})}\BibitemShut {NoStop}%
\bibitem [{\citenamefont {O'Keeffe}\ \emph {et~al.}(2013)\citenamefont
  {O'Keeffe}, \citenamefont {Gryzlova}, \citenamefont {Cubaynes}, \citenamefont
  {Garcia}, \citenamefont {Nahon}, \citenamefont {Grum-Grzhimailo},\ and\
  \citenamefont {Meyer}}]{OKeeffe2013}%
  \BibitemOpen
  \bibfield  {author} {\bibinfo {author} {\bibfnamefont {P.}~\bibnamefont
  {O'Keeffe}}, \bibinfo {author} {\bibfnamefont {E.~V.}\ \bibnamefont
  {Gryzlova}}, \bibinfo {author} {\bibfnamefont {D.}~\bibnamefont {Cubaynes}},
  \bibinfo {author} {\bibfnamefont {G.~A.}\ \bibnamefont {Garcia}}, \bibinfo
  {author} {\bibfnamefont {L.}~\bibnamefont {Nahon}}, \bibinfo {author}
  {\bibfnamefont {A.~N.}\ \bibnamefont {Grum-Grzhimailo}},\ and\ \bibinfo
  {author} {\bibfnamefont {M.}~\bibnamefont {Meyer}},\ }\bibfield  {title}
  {\bibinfo {title} {Isotopically resolved photoelectron imaging unravels
  complex atomic autoionization dynamics by two-color resonant ionization},\
  }\href {https://doi.org/10.1103/PhysRevLett.111.243002} {\bibfield  {journal}
  {\bibinfo  {journal} {Phys. Rev. Lett.}\ }\textbf {\bibinfo {volume} {111}},\
  \bibinfo {pages} {243002} (\bibinfo {year} {2013})}\BibitemShut {NoStop}%
\bibitem [{\citenamefont {Wernet}\ \emph {et~al.}(2001)\citenamefont {Wernet},
  \citenamefont {Schulz}, \citenamefont {Sonntag}, \citenamefont {Godehusen},
  \citenamefont {Zimmermann}, \citenamefont {Grum-Grzhimailo}, \citenamefont
  {Kabachnik},\ and\ \citenamefont {Martins}}]{Wernet2001}%
  \BibitemOpen
  \bibfield  {author} {\bibinfo {author} {\bibfnamefont {P.}~\bibnamefont
  {Wernet}}, \bibinfo {author} {\bibfnamefont {J.}~\bibnamefont {Schulz}},
  \bibinfo {author} {\bibfnamefont {B.}~\bibnamefont {Sonntag}}, \bibinfo
  {author} {\bibfnamefont {K.}~\bibnamefont {Godehusen}}, \bibinfo {author}
  {\bibfnamefont {P.}~\bibnamefont {Zimmermann}}, \bibinfo {author}
  {\bibfnamefont {A.~N.}\ \bibnamefont {Grum-Grzhimailo}}, \bibinfo {author}
  {\bibfnamefont {N.~M.}\ \bibnamefont {Kabachnik}},\ and\ \bibinfo {author}
  {\bibfnamefont {M.}~\bibnamefont {Martins}},\ }\bibfield  {title} {\bibinfo
  {title} {$2p$ photoelectron spectra and linear alignment dichroism of atomic
  cr},\ }\href {https://doi.org/10.1103/PhysRevA.64.042707} {\bibfield
  {journal} {\bibinfo  {journal} {Phys. Rev. A}\ }\textbf {\bibinfo {volume}
  {64}},\ \bibinfo {pages} {042707} (\bibinfo {year} {2001})}\BibitemShut
  {NoStop}%
\bibitem [{\citenamefont {Meyer}\ \emph {et~al.}(2011)\citenamefont {Meyer},
  \citenamefont {Grum-Grzhimailo}, \citenamefont {Cubaynes}, \citenamefont
  {Felfli}, \citenamefont {Heinecke}, \citenamefont {Manson},\ and\
  \citenamefont {Zimmermann}}]{Meyer2011}%
  \BibitemOpen
  \bibfield  {author} {\bibinfo {author} {\bibfnamefont {M.}~\bibnamefont
  {Meyer}}, \bibinfo {author} {\bibfnamefont {A.~N.}\ \bibnamefont
  {Grum-Grzhimailo}}, \bibinfo {author} {\bibfnamefont {D.}~\bibnamefont
  {Cubaynes}}, \bibinfo {author} {\bibfnamefont {Z.}~\bibnamefont {Felfli}},
  \bibinfo {author} {\bibfnamefont {E.}~\bibnamefont {Heinecke}}, \bibinfo
  {author} {\bibfnamefont {S.~T.}\ \bibnamefont {Manson}},\ and\ \bibinfo
  {author} {\bibfnamefont {P.}~\bibnamefont {Zimmermann}},\ }\bibfield  {title}
  {\bibinfo {title} {Magnetic dichroism in $k$-shell photoemission from laser
  excited li atoms},\ }\href {https://doi.org/10.1103/PhysRevLett.107.213001}
  {\bibfield  {journal} {\bibinfo  {journal} {Phys. Rev. Lett.}\ }\textbf
  {\bibinfo {volume} {107}},\ \bibinfo {pages} {213001} (\bibinfo {year}
  {2011})}\BibitemShut {NoStop}%
\bibitem [{\citenamefont {Wedowski}\ \emph {et~al.}(1997)\citenamefont
  {Wedowski}, \citenamefont {Godehusen}, \citenamefont {Weisbarth},
  \citenamefont {Zimmermann}, \citenamefont {Martins}, \citenamefont
  {Dohrmann}, \citenamefont {von~dem Borne}, \citenamefont {Sonntag},\ and\
  \citenamefont {Grum-Grzhimailo}}]{Wedowski1997}%
  \BibitemOpen
  \bibfield  {author} {\bibinfo {author} {\bibfnamefont {M.}~\bibnamefont
  {Wedowski}}, \bibinfo {author} {\bibfnamefont {K.}~\bibnamefont {Godehusen}},
  \bibinfo {author} {\bibfnamefont {F.}~\bibnamefont {Weisbarth}}, \bibinfo
  {author} {\bibfnamefont {P.}~\bibnamefont {Zimmermann}}, \bibinfo {author}
  {\bibfnamefont {M.}~\bibnamefont {Martins}}, \bibinfo {author} {\bibfnamefont
  {T.}~\bibnamefont {Dohrmann}}, \bibinfo {author} {\bibfnamefont
  {A.}~\bibnamefont {von~dem Borne}}, \bibinfo {author} {\bibfnamefont
  {B.}~\bibnamefont {Sonntag}},\ and\ \bibinfo {author} {\bibfnamefont {A.~N.}\
  \bibnamefont {Grum-Grzhimailo}},\ }\bibfield  {title} {\bibinfo {title}
  {Vacuum-ultraviolet photoelectron spectroscopy of laser-excited aligned ca
  atomsin the 3p-3d resonance region},\ }\href
  {https://doi.org/10.1103/PhysRevA.55.1922} {\bibfield  {journal} {\bibinfo
  {journal} {Phys. Rev. A}\ }\textbf {\bibinfo {volume} {55}},\ \bibinfo
  {pages} {1922} (\bibinfo {year} {1997})}\BibitemShut {NoStop}%
\bibitem [{\citenamefont {Ilchen}\ \emph {et~al.}(2017)\citenamefont {Ilchen},
  \citenamefont {Douguet}, \citenamefont {Mazza}, \citenamefont {Rafipoor},
  \citenamefont {Callegari}, \citenamefont {Finetti}, \citenamefont {Plekan},
  \citenamefont {Prince}, \citenamefont {Demidovich}, \citenamefont {Grazioli},
  \citenamefont {Avaldi}, \citenamefont {Bolognesi}, \citenamefont {Coreno},
  \citenamefont {Di~Fraia}, \citenamefont {Devetta}, \citenamefont
  {Ovcharenko}, \citenamefont {D\"usterer}, \citenamefont {Ueda}, \citenamefont
  {Bartschat}, \citenamefont {Grum-Grzhimailo}, \citenamefont {Bozhevolnov},
  \citenamefont {Kazansky}, \citenamefont {Kabachnik},\ and\ \citenamefont
  {Meyer}}]{Ilchen2017}%
  \BibitemOpen
  \bibfield  {author} {\bibinfo {author} {\bibfnamefont {M.}~\bibnamefont
  {Ilchen}}, \bibinfo {author} {\bibfnamefont {N.}~\bibnamefont {Douguet}},
  \bibinfo {author} {\bibfnamefont {T.}~\bibnamefont {Mazza}}, \bibinfo
  {author} {\bibfnamefont {A.~J.}\ \bibnamefont {Rafipoor}}, \bibinfo {author}
  {\bibfnamefont {C.}~\bibnamefont {Callegari}}, \bibinfo {author}
  {\bibfnamefont {P.}~\bibnamefont {Finetti}}, \bibinfo {author} {\bibfnamefont
  {O.}~\bibnamefont {Plekan}}, \bibinfo {author} {\bibfnamefont {K.~C.}\
  \bibnamefont {Prince}}, \bibinfo {author} {\bibfnamefont {A.}~\bibnamefont
  {Demidovich}}, \bibinfo {author} {\bibfnamefont {C.}~\bibnamefont
  {Grazioli}}, \bibinfo {author} {\bibfnamefont {L.}~\bibnamefont {Avaldi}},
  \bibinfo {author} {\bibfnamefont {P.}~\bibnamefont {Bolognesi}}, \bibinfo
  {author} {\bibfnamefont {M.}~\bibnamefont {Coreno}}, \bibinfo {author}
  {\bibfnamefont {M.}~\bibnamefont {Di~Fraia}}, \bibinfo {author}
  {\bibfnamefont {M.}~\bibnamefont {Devetta}}, \bibinfo {author} {\bibfnamefont
  {Y.}~\bibnamefont {Ovcharenko}}, \bibinfo {author} {\bibfnamefont
  {S.}~\bibnamefont {D\"usterer}}, \bibinfo {author} {\bibfnamefont
  {K.}~\bibnamefont {Ueda}}, \bibinfo {author} {\bibfnamefont {K.}~\bibnamefont
  {Bartschat}}, \bibinfo {author} {\bibfnamefont {A.~N.}\ \bibnamefont
  {Grum-Grzhimailo}}, \bibinfo {author} {\bibfnamefont {A.~V.}\ \bibnamefont
  {Bozhevolnov}}, \bibinfo {author} {\bibfnamefont {A.~K.}\ \bibnamefont
  {Kazansky}}, \bibinfo {author} {\bibfnamefont {N.~M.}\ \bibnamefont
  {Kabachnik}},\ and\ \bibinfo {author} {\bibfnamefont {M.}~\bibnamefont
  {Meyer}},\ }\bibfield  {title} {\bibinfo {title} {Circular dichroism in
  multiphoton ionization of resonantly excited ${\mathrm{he}}^{+}$ ions},\
  }\href {https://doi.org/10.1103/PhysRevLett.118.013002} {\bibfield  {journal}
  {\bibinfo  {journal} {Phys. Rev. Lett.}\ }\textbf {\bibinfo {volume} {118}},\
  \bibinfo {pages} {013002} (\bibinfo {year} {2017})}\BibitemShut {NoStop}%
\bibitem [{\citenamefont {Cubaynes}\ \emph {et~al.}(2004)\citenamefont
  {Cubaynes}, \citenamefont {Meyer}, \citenamefont {Grum-Grzhimailo},
  \citenamefont {Bizau}, \citenamefont {Kennedy}, \citenamefont {Bozek},
  \citenamefont {Martins}, \citenamefont {Canton}, \citenamefont {Rude},
  \citenamefont {Berrah},\ and\ \citenamefont {Wuilleumier}}]{Cubaynes2004}%
  \BibitemOpen
  \bibfield  {author} {\bibinfo {author} {\bibfnamefont {D.}~\bibnamefont
  {Cubaynes}}, \bibinfo {author} {\bibfnamefont {M.}~\bibnamefont {Meyer}},
  \bibinfo {author} {\bibfnamefont {A.~N.}\ \bibnamefont {Grum-Grzhimailo}},
  \bibinfo {author} {\bibfnamefont {J.-M.}\ \bibnamefont {Bizau}}, \bibinfo
  {author} {\bibfnamefont {E.~T.}\ \bibnamefont {Kennedy}}, \bibinfo {author}
  {\bibfnamefont {J.}~\bibnamefont {Bozek}}, \bibinfo {author} {\bibfnamefont
  {M.}~\bibnamefont {Martins}}, \bibinfo {author} {\bibfnamefont
  {S.}~\bibnamefont {Canton}}, \bibinfo {author} {\bibfnamefont
  {B.}~\bibnamefont {Rude}}, \bibinfo {author} {\bibfnamefont {N.}~\bibnamefont
  {Berrah}},\ and\ \bibinfo {author} {\bibfnamefont {F.~J.}\ \bibnamefont
  {Wuilleumier}},\ }\bibfield  {title} {\bibinfo {title} {Dynamically and
  quasiforbidden transitions in photoionization of open-shell atoms: A combined
  experimental and theoretical study},\ }\href
  {https://doi.org/10.1103/PhysRevLett.92.233002} {\bibfield  {journal}
  {\bibinfo  {journal} {Phys. Rev. Lett.}\ }\textbf {\bibinfo {volume} {92}},\
  \bibinfo {pages} {233002} (\bibinfo {year} {2004})}\BibitemShut {NoStop}%
\bibitem [{\citenamefont {N\"ortersh\"auser}\ \emph {et~al.}(2021)\citenamefont
  {N\"ortersh\"auser}, \citenamefont {Surzhykov}, \citenamefont {S\'anchez},
  \citenamefont {Botermann}, \citenamefont {Gwinner}, \citenamefont {Huber},
  \citenamefont {Karpuk}, \citenamefont {K\"uhl}, \citenamefont {Novotny},
  \citenamefont {Reinhardt}, \citenamefont {Saathoff}, \citenamefont
  {St\"ohlker},\ and\ \citenamefont {Wolf}}]{Nortershauser2021}%
  \BibitemOpen
  \bibfield  {author} {\bibinfo {author} {\bibfnamefont {W.}~\bibnamefont
  {N\"ortersh\"auser}}, \bibinfo {author} {\bibfnamefont {A.}~\bibnamefont
  {Surzhykov}}, \bibinfo {author} {\bibfnamefont {R.}~\bibnamefont
  {S\'anchez}}, \bibinfo {author} {\bibfnamefont {B.}~\bibnamefont
  {Botermann}}, \bibinfo {author} {\bibfnamefont {G.}~\bibnamefont {Gwinner}},
  \bibinfo {author} {\bibfnamefont {G.}~\bibnamefont {Huber}}, \bibinfo
  {author} {\bibfnamefont {S.}~\bibnamefont {Karpuk}}, \bibinfo {author}
  {\bibfnamefont {T.}~\bibnamefont {K\"uhl}}, \bibinfo {author} {\bibfnamefont
  {C.}~\bibnamefont {Novotny}}, \bibinfo {author} {\bibfnamefont
  {S.}~\bibnamefont {Reinhardt}}, \bibinfo {author} {\bibfnamefont
  {G.}~\bibnamefont {Saathoff}}, \bibinfo {author} {\bibfnamefont
  {T.}~\bibnamefont {St\"ohlker}},\ and\ \bibinfo {author} {\bibfnamefont
  {A.}~\bibnamefont {Wolf}},\ }\bibfield  {title} {\bibinfo {title}
  {Polarization-dependent disappearance of a resonance signal: Indication for
  optical pumping in a storage ring?},\ }\href
  {https://doi.org/10.1103/PhysRevAccelBeams.24.024701} {\bibfield  {journal}
  {\bibinfo  {journal} {Phys. Rev. Accel. Beams}\ }\textbf {\bibinfo {volume}
  {24}},\ \bibinfo {pages} {024701} (\bibinfo {year} {2021})}\BibitemShut
  {NoStop}%
\bibitem [{\citenamefont {Karamatskos}\ \emph {et~al.}(2013)\citenamefont
  {Karamatskos}, \citenamefont {Markellos},\ and\ \citenamefont
  {Lambropoulos}}]{Karamatskos2013}%
  \BibitemOpen
  \bibfield  {author} {\bibinfo {author} {\bibfnamefont {E.~T.}\ \bibnamefont
  {Karamatskos}}, \bibinfo {author} {\bibfnamefont {D.}~\bibnamefont
  {Markellos}},\ and\ \bibinfo {author} {\bibfnamefont {P.}~\bibnamefont
  {Lambropoulos}},\ }\bibfield  {title} {\bibinfo {title} {Multiple ionization
  of argon under 123 {eV} {FEL} radiation and the creation of 3s-hollow ions},\
  }\href {https://doi.org/10.1088/0953-4075/46/16/164011} {\bibfield  {journal}
  {\bibinfo  {journal} {Journal of Physics B: Atomic, Molecular and Optical
  Physics}\ }\textbf {\bibinfo {volume} {46}},\ \bibinfo {pages} {164011}
  (\bibinfo {year} {2013})}\BibitemShut {NoStop}%
\bibitem [{\citenamefont {Nakajima}\ and\ \citenamefont
  {Nikolopoulos}(2002)}]{Nakajima2002}%
  \BibitemOpen
  \bibfield  {author} {\bibinfo {author} {\bibfnamefont {T.}~\bibnamefont
  {Nakajima}}\ and\ \bibinfo {author} {\bibfnamefont {L.~A.~A.}\ \bibnamefont
  {Nikolopoulos}},\ }\bibfield  {title} {\bibinfo {title} {Use of helium double
  ionization for autocorrelation of an xuv pulse},\ }\href
  {https://doi.org/10.1103/PhysRevA.66.041402} {\bibfield  {journal} {\bibinfo
  {journal} {Phys. Rev. A}\ }\textbf {\bibinfo {volume} {66}},\ \bibinfo
  {pages} {041402R} (\bibinfo {year} {2002})}\BibitemShut {NoStop}%
\bibitem [{\citenamefont {Makris}\ \emph {et~al.}(2009)\citenamefont {Makris},
  \citenamefont {Lambropoulos},\ and\ \citenamefont
  {Miheli\v{c}}}]{Makris2009}%
  \BibitemOpen
  \bibfield  {author} {\bibinfo {author} {\bibfnamefont {M.~G.}\ \bibnamefont
  {Makris}}, \bibinfo {author} {\bibfnamefont {P.}~\bibnamefont
  {Lambropoulos}},\ and\ \bibinfo {author} {\bibfnamefont {A.}~\bibnamefont
  {Miheli\v{c}}},\ }\bibfield  {title} {\bibinfo {title} {Theory of multiphoton
  multielectron ionization of xenon under strong 93-{eV} radiation},\ }\href
  {https://doi.org/10.1103/PhysRevLett.102.033002} {\bibfield  {journal}
  {\bibinfo  {journal} {Phys. Rev. Lett.}\ }\textbf {\bibinfo {volume} {102}},\
  \bibinfo {pages} {033002} (\bibinfo {year} {2009})}\BibitemShut {NoStop}%
\bibitem [{\citenamefont {Son}\ and\ \citenamefont {Santra}(2011)}]{Son2011}%
  \BibitemOpen
  \bibfield  {author} {\bibinfo {author} {\bibfnamefont {S.-K.}\ \bibnamefont
  {Son}}\ and\ \bibinfo {author} {\bibfnamefont {R.}~\bibnamefont {Santra}},\
  }\bibfield  {title} {\bibinfo {title} {Impact of hollow-atom formation on
  coherent {x}-ray scattering at high intensity},\ }\href
  {https://doi.org/10.1103/PhysRevA.83.033402} {\bibfield  {journal} {\bibinfo
  {journal} {Phys. Rev. A}\ }\textbf {\bibinfo {volume} {83}},\ \bibinfo
  {pages} {033402} (\bibinfo {year} {2011})}\BibitemShut {NoStop}%
\bibitem [{\citenamefont {Son}\ and\ \citenamefont {Santra}(2012)}]{Son2012}%
  \BibitemOpen
  \bibfield  {author} {\bibinfo {author} {\bibfnamefont {S.-K.}\ \bibnamefont
  {Son}}\ and\ \bibinfo {author} {\bibfnamefont {R.}~\bibnamefont {Santra}},\
  }\bibfield  {title} {\bibinfo {title} {{M}onte {C}arlo calculation of ion,
  electron, and photon spectra of xenon atoms in x-ray free-electron laser
  pulses},\ }\href {https://doi.org/10.1103/PhysRevA.85.063415} {\bibfield
  {journal} {\bibinfo  {journal} {Phys. Rev. A}\ }\textbf {\bibinfo {volume}
  {85}},\ \bibinfo {pages} {063415} (\bibinfo {year} {2012})}\BibitemShut
  {NoStop}%
\bibitem [{\citenamefont {Lorenz}\ \emph {et~al.}(2012)\citenamefont {Lorenz},
  \citenamefont {Kabachnik}, \citenamefont {Weckert},\ and\ \citenamefont
  {Vartanyants}}]{Lorenz2012}%
  \BibitemOpen
  \bibfield  {author} {\bibinfo {author} {\bibfnamefont {U.}~\bibnamefont
  {Lorenz}}, \bibinfo {author} {\bibfnamefont {N.~M.}\ \bibnamefont
  {Kabachnik}}, \bibinfo {author} {\bibfnamefont {E.}~\bibnamefont {Weckert}},\
  and\ \bibinfo {author} {\bibfnamefont {I.~A.}\ \bibnamefont {Vartanyants}},\
  }\bibfield  {title} {\bibinfo {title} {Impact of ultrafast electronic damage
  in single-particle x-ray imaging experiments},\ }\href
  {https://doi.org/10.1103/PhysRevE.86.051911} {\bibfield  {journal} {\bibinfo
  {journal} {Phys. Rev. E}\ }\textbf {\bibinfo {volume} {86}},\ \bibinfo
  {pages} {051911} (\bibinfo {year} {2012})}\BibitemShut {NoStop}%
\bibitem [{\citenamefont {Lunin}\ \emph {et~al.}(2015)\citenamefont {Lunin},
  \citenamefont {Grum-Grzhimailo}, \citenamefont {Gryzlova}, \citenamefont
  {Sinitsyn}, \citenamefont {Petrova}, \citenamefont {Lunina}, \citenamefont
  {Balabaev}, \citenamefont {Tereshkina}, \citenamefont {Stepanov},\ and\
  \citenamefont {Krupyanskii}}]{Lunin2015}%
  \BibitemOpen
  \bibfield  {author} {\bibinfo {author} {\bibfnamefont {V.~Y.}\ \bibnamefont
  {Lunin}}, \bibinfo {author} {\bibfnamefont {A.~N.}\ \bibnamefont
  {Grum-Grzhimailo}}, \bibinfo {author} {\bibfnamefont {E.~V.}\ \bibnamefont
  {Gryzlova}}, \bibinfo {author} {\bibfnamefont {D.~O.}\ \bibnamefont
  {Sinitsyn}}, \bibinfo {author} {\bibfnamefont {T.~E.}\ \bibnamefont
  {Petrova}}, \bibinfo {author} {\bibfnamefont {N.~L.}\ \bibnamefont {Lunina}},
  \bibinfo {author} {\bibfnamefont {N.~K.}\ \bibnamefont {Balabaev}}, \bibinfo
  {author} {\bibfnamefont {K.~B.}\ \bibnamefont {Tereshkina}}, \bibinfo
  {author} {\bibfnamefont {A.~S.}\ \bibnamefont {Stepanov}},\ and\ \bibinfo
  {author} {\bibfnamefont {Y.~F.}\ \bibnamefont {Krupyanskii}},\ }\bibfield
  {title} {\bibinfo {title} {Efficient calculation of diffracted intensities in
  the case of nonstationary scattering by biological macromolecules under
  {XFEL} pulses},\ }\href {https://doi.org/10.1107/S1399004714025450}
  {\bibfield  {journal} {\bibinfo  {journal} {Acta Cryst. D}\ }\textbf
  {\bibinfo {volume} {71}},\ \bibinfo {pages} {293} (\bibinfo {year}
  {2015})}\BibitemShut {NoStop}%
\bibitem [{\citenamefont {Serkez}\ \emph {et~al.}(2018)\citenamefont {Serkez},
  \citenamefont {Geloni}, \citenamefont {Tomin}, \citenamefont {Feng},
  \citenamefont {Gryzlova}, \citenamefont {Grum-Grzhimailo},\ and\
  \citenamefont {Meyer}}]{Serkez2018}%
  \BibitemOpen
  \bibfield  {author} {\bibinfo {author} {\bibfnamefont {S.}~\bibnamefont
  {Serkez}}, \bibinfo {author} {\bibfnamefont {G.}~\bibnamefont {Geloni}},
  \bibinfo {author} {\bibfnamefont {S.}~\bibnamefont {Tomin}}, \bibinfo
  {author} {\bibfnamefont {G.}~\bibnamefont {Feng}}, \bibinfo {author}
  {\bibfnamefont {E.~V.}\ \bibnamefont {Gryzlova}}, \bibinfo {author}
  {\bibfnamefont {A.~N.}\ \bibnamefont {Grum-Grzhimailo}},\ and\ \bibinfo
  {author} {\bibfnamefont {M.}~\bibnamefont {Meyer}},\ }\bibfield  {title}
  {\bibinfo {title} {Overview of options for generating high-brightness
  attosecond x-ray pulses at free-electron lasers and applications at the
  {E}uropean {XFEL}},\ }\href {https://doi.org/10.1088/2040-8986/aa9f4f}
  {\bibfield  {journal} {\bibinfo  {journal} {J. Opt.}\ }\textbf {\bibinfo
  {volume} {20}},\ \bibinfo {pages} {024005} (\bibinfo {year}
  {2018})}\BibitemShut {NoStop}%
\bibitem [{\citenamefont {Buth}\ \emph {et~al.}(2018)\citenamefont {Buth},
  \citenamefont {Beerwerth}, \citenamefont {Obaid}, \citenamefont {Berrah},
  \citenamefont {Cederbaum},\ and\ \citenamefont {Fritzsche}}]{Buth2018}%
  \BibitemOpen
  \bibfield  {author} {\bibinfo {author} {\bibfnamefont {C.}~\bibnamefont
  {Buth}}, \bibinfo {author} {\bibfnamefont {R.}~\bibnamefont {Beerwerth}},
  \bibinfo {author} {\bibfnamefont {R.}~\bibnamefont {Obaid}}, \bibinfo
  {author} {\bibfnamefont {N.}~\bibnamefont {Berrah}}, \bibinfo {author}
  {\bibfnamefont {L.~S.}\ \bibnamefont {Cederbaum}},\ and\ \bibinfo {author}
  {\bibfnamefont {S.}~\bibnamefont {Fritzsche}},\ }\bibfield  {title} {\bibinfo
  {title} {Neon in ultrashort and intense x-rays from free electron lasers},\
  }\href {https://doi.org/10.1088/1361-6455/aaa39a} {\bibfield  {journal}
  {\bibinfo  {journal} {J. Phys. B: At. Mol. Opt. Ph.}\ }\textbf {\bibinfo
  {volume} {51}},\ \bibinfo {pages} {055602} (\bibinfo {year}
  {2018})}\BibitemShut {NoStop}%
\bibitem [{\citenamefont {King}\ \emph {et~al.}(1977)\citenamefont {King},
  \citenamefont {Tronc}, \citenamefont {Read},\ and\ \citenamefont
  {Bradford}}]{King1977}%
  \BibitemOpen
  \bibfield  {author} {\bibinfo {author} {\bibfnamefont {G.~C.}\ \bibnamefont
  {King}}, \bibinfo {author} {\bibfnamefont {M.}~\bibnamefont {Tronc}},
  \bibinfo {author} {\bibfnamefont {F.~H.}\ \bibnamefont {Read}},\ and\
  \bibinfo {author} {\bibfnamefont {R.~C.}\ \bibnamefont {Bradford}},\
  }\bibfield  {title} {\bibinfo {title} {An investigation of the structure near
  the l$_{2,3}$ edges of argon, the m$_{4,5}$ edges of krypton and the
  n$_{4,5}$ edges of xenon, using electron impact with high resolution},\
  }\href {https://doi.org/10.1088/0022-3700/10/12/026} {\bibfield  {journal}
  {\bibinfo  {journal} {J. Phys. B: At. Mol. Ph.}\ }\textbf {\bibinfo {volume}
  {10}},\ \bibinfo {pages} {2479} (\bibinfo {year} {1977})}\BibitemShut
  {NoStop}%
\bibitem [{\citenamefont {Ilchen}\ \emph {et~al.}(2016)\citenamefont {Ilchen},
  \citenamefont {Mazza}, \citenamefont {Karamatskos}, \citenamefont
  {Markellos}, \citenamefont {Bakhtiarzadeh}, \citenamefont {Rafipoor},
  \citenamefont {Kelly}, \citenamefont {Walsh}, \citenamefont {Costello},
  \citenamefont {O'Keeffe}, \citenamefont {Gerken}, \citenamefont {Martins},
  \citenamefont {Lambropoulos},\ and\ \citenamefont {Meyer}}]{Ilchen2016}%
  \BibitemOpen
  \bibfield  {author} {\bibinfo {author} {\bibfnamefont {M.}~\bibnamefont
  {Ilchen}}, \bibinfo {author} {\bibfnamefont {T.}~\bibnamefont {Mazza}},
  \bibinfo {author} {\bibfnamefont {E.~T.}\ \bibnamefont {Karamatskos}},
  \bibinfo {author} {\bibfnamefont {D.}~\bibnamefont {Markellos}}, \bibinfo
  {author} {\bibfnamefont {S.}~\bibnamefont {Bakhtiarzadeh}}, \bibinfo {author}
  {\bibfnamefont {A.~J.}\ \bibnamefont {Rafipoor}}, \bibinfo {author}
  {\bibfnamefont {T.~J.}\ \bibnamefont {Kelly}}, \bibinfo {author}
  {\bibfnamefont {N.}~\bibnamefont {Walsh}}, \bibinfo {author} {\bibfnamefont
  {J.~T.}\ \bibnamefont {Costello}}, \bibinfo {author} {\bibfnamefont
  {P.}~\bibnamefont {O'Keeffe}}, \bibinfo {author} {\bibfnamefont
  {N.}~\bibnamefont {Gerken}}, \bibinfo {author} {\bibfnamefont
  {M.}~\bibnamefont {Martins}}, \bibinfo {author} {\bibfnamefont
  {P.}~\bibnamefont {Lambropoulos}},\ and\ \bibinfo {author} {\bibfnamefont
  {M.}~\bibnamefont {Meyer}},\ }\bibfield  {title} {\bibinfo {title}
  {Two-electron processes in multiple ionization under strong soft-x-ray
  radiation},\ }\href {https://doi.org/10.1103/PhysRevA.94.013413} {\bibfield
  {journal} {\bibinfo  {journal} {Phys. Rev. A}\ }\textbf {\bibinfo {volume}
  {94}},\ \bibinfo {pages} {013413} (\bibinfo {year} {2016})}\BibitemShut
  {NoStop}%
\bibitem [{\citenamefont {Gryzlova}\ \emph {et~al.}(2020)\citenamefont
  {Gryzlova}, \citenamefont {Kiselev}, \citenamefont {Popova}, \citenamefont
  {Zubekhin}, \citenamefont {Sansone},\ and\ \citenamefont
  {Grum-Grzhimailo}}]{Gryzlova2020}%
  \BibitemOpen
  \bibfield  {author} {\bibinfo {author} {\bibfnamefont {E.~V.}\ \bibnamefont
  {Gryzlova}}, \bibinfo {author} {\bibfnamefont {M.~D.}\ \bibnamefont
  {Kiselev}}, \bibinfo {author} {\bibfnamefont {M.~M.}\ \bibnamefont {Popova}},
  \bibinfo {author} {\bibfnamefont {A.~A.}\ \bibnamefont {Zubekhin}}, \bibinfo
  {author} {\bibfnamefont {G.}~\bibnamefont {Sansone}},\ and\ \bibinfo {author}
  {\bibfnamefont {A.~N.}\ \bibnamefont {Grum-Grzhimailo}},\ }\bibfield  {title}
  {\bibinfo {title} {Multiple sequential ionization of valence n = 4 shell of
  krypton by intense femtosecond xuv pulses},\ }\bibfield  {journal} {\bibinfo
  {journal} {Atoms}\ }\textbf {\bibinfo {volume} {8}},\ \href
  {https://doi.org/10.3390/atoms8040080} {10.3390/atoms8040080} (\bibinfo
  {year} {2020})\BibitemShut {NoStop}%
\bibitem [{NIS(2020)}]{NIST}%
  \BibitemOpen
  \href@noop {} {\bibinfo {title} {{NIST} {A}tomic {S}pectra {D}atabase
  (version 5.8). {A}vailable: https://physics.nist.gov/asd [{M}ay 18 2020].
  {N}ational {I}nstitute of {S}tandards and {T}echnology, {G}aithersburg,
  {MD}}} (\bibinfo {year} {2020})\BibitemShut {NoStop}%
\bibitem [{\citenamefont {Balashov}\ \emph {et~al.}(2000)\citenamefont
  {Balashov}, \citenamefont {Grum-Grzhimailo},\ and\ \citenamefont
  {Kabachnik}}]{Balashov2000}%
  \BibitemOpen
  \bibfield  {author} {\bibinfo {author} {\bibfnamefont {V.~V.}\ \bibnamefont
  {Balashov}}, \bibinfo {author} {\bibfnamefont {A.~N.}\ \bibnamefont
  {Grum-Grzhimailo}},\ and\ \bibinfo {author} {\bibfnamefont {.~M.}\
  \bibnamefont {Kabachnik}},\ }\href@noop {} {\emph {\bibinfo {title}
  {Polarization and Correlation Phenomena in Atomic Collisions. A Practical
  Theory Course}}}\ (\bibinfo  {publisher} {Kluwer Plenum, New York},\ \bibinfo
  {year} {2000})\BibitemShut {NoStop}%
\bibitem [{\citenamefont {Blum}(1996)}]{Blum96}%
  \BibitemOpen
  \bibfield  {author} {\bibinfo {author} {\bibfnamefont {K.}~\bibnamefont
  {Blum}},\ }\href@noop {} {\emph {\bibinfo {title} {Density Matrix Theory and
  Applications}}}\ (\bibinfo  {publisher} {Plenum, New York},\ \bibinfo {year}
  {1996})\BibitemShut {NoStop}%
\bibitem [{\citenamefont {Allaria}\ \emph {et~al.}(2012)\citenamefont
  {Allaria}, \citenamefont {Appio}, \citenamefont {Badano}, \citenamefont
  {Barletta}, \citenamefont {Bassanese}, \citenamefont {Biedron}, \citenamefont
  {Borga}, \citenamefont {Busetto}, \citenamefont {Castronovo}, \citenamefont
  {Cinquegrana}, \citenamefont {Cleva}, \citenamefont {Cocco}, \citenamefont
  {Cornacchia}, \citenamefont {Craievich}, \citenamefont {Cudin}, \citenamefont
  {D'Auria}, \citenamefont {Dal~Forno}, \citenamefont {Danailov}, \citenamefont
  {De~Monte}, \citenamefont {De~Ninno}, \citenamefont {Delgiusto},
  \citenamefont {Demidovich}, \citenamefont {Di~Mitri}, \citenamefont
  {Diviacco}, \citenamefont {Fabris}, \citenamefont {Fabris}, \citenamefont
  {Fawley}, \citenamefont {Ferianis}, \citenamefont {Ferrari}, \citenamefont
  {Ferry}, \citenamefont {Froehlich}, \citenamefont {Furlan}, \citenamefont
  {Gaio}, \citenamefont {Gelmetti}, \citenamefont {Giannessi}, \citenamefont
  {Giannini}, \citenamefont {Gobessi}, \citenamefont {Ivanov}, \citenamefont
  {Karantzoulis}, \citenamefont {Lonza}, \citenamefont {Lutman}, \citenamefont
  {Mahieu}, \citenamefont {Milloch}, \citenamefont {Milton}, \citenamefont
  {Musardo}, \citenamefont {Nikolov}, \citenamefont {Noe}, \citenamefont
  {Parmigiani}, \citenamefont {Penco}, \citenamefont {Petronio}, \citenamefont
  {Pivetta}, \citenamefont {Predonzani}, \citenamefont {Rossi}, \citenamefont
  {Rumiz}, \citenamefont {Salom}, \citenamefont {Scafuri}, \citenamefont
  {Serpico}, \citenamefont {Sigalotti}, \citenamefont {Spampinati},
  \citenamefont {Spezzani}, \citenamefont {Svandrlik}, \citenamefont {Svetina},
  \citenamefont {Tazzari}, \citenamefont {Trovo}, \citenamefont {Umer},
  \citenamefont {Vascotto}, \citenamefont {Veronese}, \citenamefont
  {Visintini}, \citenamefont {Zaccaria}, \citenamefont {Zangrando},\ and\
  \citenamefont {Zangrando}}]{Allaria2012}%
  \BibitemOpen
  \bibfield  {author} {\bibinfo {author} {\bibfnamefont {E.}~\bibnamefont
  {Allaria}}, \bibinfo {author} {\bibfnamefont {R.}~\bibnamefont {Appio}},
  \bibinfo {author} {\bibfnamefont {L.}~\bibnamefont {Badano}}, \bibinfo
  {author} {\bibfnamefont {W.~A.}\ \bibnamefont {Barletta}}, \bibinfo {author}
  {\bibfnamefont {S.}~\bibnamefont {Bassanese}}, \bibinfo {author}
  {\bibfnamefont {S.~G.}\ \bibnamefont {Biedron}}, \bibinfo {author}
  {\bibfnamefont {A.}~\bibnamefont {Borga}}, \bibinfo {author} {\bibfnamefont
  {E.}~\bibnamefont {Busetto}}, \bibinfo {author} {\bibfnamefont
  {D.}~\bibnamefont {Castronovo}}, \bibinfo {author} {\bibfnamefont
  {P.}~\bibnamefont {Cinquegrana}}, \bibinfo {author} {\bibfnamefont
  {S.}~\bibnamefont {Cleva}}, \bibinfo {author} {\bibfnamefont
  {D.}~\bibnamefont {Cocco}}, \bibinfo {author} {\bibfnamefont
  {M.}~\bibnamefont {Cornacchia}}, \bibinfo {author} {\bibfnamefont
  {P.}~\bibnamefont {Craievich}}, \bibinfo {author} {\bibfnamefont
  {I.}~\bibnamefont {Cudin}}, \bibinfo {author} {\bibfnamefont
  {G.}~\bibnamefont {D'Auria}}, \bibinfo {author} {\bibfnamefont
  {M.}~\bibnamefont {Dal~Forno}}, \bibinfo {author} {\bibfnamefont {M.~B.}\
  \bibnamefont {Danailov}}, \bibinfo {author} {\bibfnamefont {R.}~\bibnamefont
  {De~Monte}}, \bibinfo {author} {\bibfnamefont {G.}~\bibnamefont {De~Ninno}},
  \bibinfo {author} {\bibfnamefont {P.}~\bibnamefont {Delgiusto}}, \bibinfo
  {author} {\bibfnamefont {A.}~\bibnamefont {Demidovich}}, \bibinfo {author}
  {\bibfnamefont {S.}~\bibnamefont {Di~Mitri}}, \bibinfo {author}
  {\bibfnamefont {B.}~\bibnamefont {Diviacco}}, \bibinfo {author}
  {\bibfnamefont {A.}~\bibnamefont {Fabris}}, \bibinfo {author} {\bibfnamefont
  {R.}~\bibnamefont {Fabris}}, \bibinfo {author} {\bibfnamefont
  {W.}~\bibnamefont {Fawley}}, \bibinfo {author} {\bibfnamefont
  {M.}~\bibnamefont {Ferianis}}, \bibinfo {author} {\bibfnamefont
  {E.}~\bibnamefont {Ferrari}}, \bibinfo {author} {\bibfnamefont
  {S.}~\bibnamefont {Ferry}}, \bibinfo {author} {\bibfnamefont
  {L.}~\bibnamefont {Froehlich}}, \bibinfo {author} {\bibfnamefont
  {P.}~\bibnamefont {Furlan}}, \bibinfo {author} {\bibfnamefont
  {G.}~\bibnamefont {Gaio}}, \bibinfo {author} {\bibfnamefont {F.}~\bibnamefont
  {Gelmetti}}, \bibinfo {author} {\bibfnamefont {L.}~\bibnamefont {Giannessi}},
  \bibinfo {author} {\bibfnamefont {M.}~\bibnamefont {Giannini}}, \bibinfo
  {author} {\bibfnamefont {R.}~\bibnamefont {Gobessi}}, \bibinfo {author}
  {\bibfnamefont {R.}~\bibnamefont {Ivanov}}, \bibinfo {author} {\bibfnamefont
  {E.}~\bibnamefont {Karantzoulis}}, \bibinfo {author} {\bibfnamefont
  {M.}~\bibnamefont {Lonza}}, \bibinfo {author} {\bibfnamefont
  {A.}~\bibnamefont {Lutman}}, \bibinfo {author} {\bibfnamefont
  {B.}~\bibnamefont {Mahieu}}, \bibinfo {author} {\bibfnamefont
  {M.}~\bibnamefont {Milloch}}, \bibinfo {author} {\bibfnamefont {S.~V.}\
  \bibnamefont {Milton}}, \bibinfo {author} {\bibfnamefont {M.}~\bibnamefont
  {Musardo}}, \bibinfo {author} {\bibfnamefont {I.}~\bibnamefont {Nikolov}},
  \bibinfo {author} {\bibfnamefont {S.}~\bibnamefont {Noe}}, \bibinfo {author}
  {\bibfnamefont {F.}~\bibnamefont {Parmigiani}}, \bibinfo {author}
  {\bibfnamefont {G.}~\bibnamefont {Penco}}, \bibinfo {author} {\bibfnamefont
  {M.}~\bibnamefont {Petronio}}, \bibinfo {author} {\bibfnamefont
  {L.}~\bibnamefont {Pivetta}}, \bibinfo {author} {\bibfnamefont
  {M.}~\bibnamefont {Predonzani}}, \bibinfo {author} {\bibfnamefont
  {F.}~\bibnamefont {Rossi}}, \bibinfo {author} {\bibfnamefont
  {L.}~\bibnamefont {Rumiz}}, \bibinfo {author} {\bibfnamefont
  {A.}~\bibnamefont {Salom}}, \bibinfo {author} {\bibfnamefont
  {C.}~\bibnamefont {Scafuri}}, \bibinfo {author} {\bibfnamefont
  {C.}~\bibnamefont {Serpico}}, \bibinfo {author} {\bibfnamefont
  {P.}~\bibnamefont {Sigalotti}}, \bibinfo {author} {\bibfnamefont
  {S.}~\bibnamefont {Spampinati}}, \bibinfo {author} {\bibfnamefont
  {C.}~\bibnamefont {Spezzani}}, \bibinfo {author} {\bibfnamefont
  {M.}~\bibnamefont {Svandrlik}}, \bibinfo {author} {\bibfnamefont
  {C.}~\bibnamefont {Svetina}}, \bibinfo {author} {\bibfnamefont
  {S.}~\bibnamefont {Tazzari}}, \bibinfo {author} {\bibfnamefont
  {M.}~\bibnamefont {Trovo}}, \bibinfo {author} {\bibfnamefont
  {R.}~\bibnamefont {Umer}}, \bibinfo {author} {\bibfnamefont {A.}~\bibnamefont
  {Vascotto}}, \bibinfo {author} {\bibfnamefont {M.}~\bibnamefont {Veronese}},
  \bibinfo {author} {\bibfnamefont {R.}~\bibnamefont {Visintini}}, \bibinfo
  {author} {\bibfnamefont {M.}~\bibnamefont {Zaccaria}}, \bibinfo {author}
  {\bibfnamefont {D.}~\bibnamefont {Zangrando}},\ and\ \bibinfo {author}
  {\bibfnamefont {M.}~\bibnamefont {Zangrando}},\ }\bibfield  {title} {\bibinfo
  {title} {Highly coherent and stable pulses from the {FERMI} seeded
  free-electron laser in the extreme ultraviolet},\ }\href
  {https://doi.org/10.1038/nphoton.2012.233} {\bibfield  {journal} {\bibinfo
  {journal} {Nat. Photonics}\ }\textbf {\bibinfo {volume} {6}},\ \bibinfo
  {pages} {699} (\bibinfo {year} {2012})}\BibitemShut {NoStop}%
\bibitem [{\citenamefont {Finetti}\ \emph {et~al.}(2017)\citenamefont
  {Finetti}, \citenamefont {H\"oppner}, \citenamefont {Allaria}, \citenamefont
  {Callegari}, \citenamefont {Capotondi}, \citenamefont {Cinquegrana},
  \citenamefont {Coreno}, \citenamefont {Cucini}, \citenamefont {Danailov},
  \citenamefont {Demidovich}, \citenamefont {De~Ninno}, \citenamefont
  {Di~Fraia}, \citenamefont {Feifel}, \citenamefont {Ferrari}, \citenamefont
  {Fr\"ohlich}, \citenamefont {Gauthier}, \citenamefont {Golz}, \citenamefont
  {Grazioli}, \citenamefont {Kai}, \citenamefont {Kurdi}, \citenamefont
  {Mahne}, \citenamefont {Manfredda}, \citenamefont {Medvedev}, \citenamefont
  {Nikolov}, \citenamefont {Pedersoli}, \citenamefont {Penco}, \citenamefont
  {Plekan}, \citenamefont {Prandolini}, \citenamefont {Prince}, \citenamefont
  {Raimondi}, \citenamefont {Rebernik}, \citenamefont {Riedel}, \citenamefont
  {Roussel}, \citenamefont {Sigalotti}, \citenamefont {Squibb}, \citenamefont
  {Stojanovic}, \citenamefont {Stranges}, \citenamefont {Svetina},
  \citenamefont {Tanikawa}, \citenamefont {Teubner}, \citenamefont {Tkachenko},
  \citenamefont {Toleikis}, \citenamefont {Zangrando}, \citenamefont {Ziaja},
  \citenamefont {Tavella},\ and\ \citenamefont {Giannessi}}]{Finetti2017}%
  \BibitemOpen
  \bibfield  {author} {\bibinfo {author} {\bibfnamefont {P.}~\bibnamefont
  {Finetti}}, \bibinfo {author} {\bibfnamefont {H.}~\bibnamefont {H\"oppner}},
  \bibinfo {author} {\bibfnamefont {E.}~\bibnamefont {Allaria}}, \bibinfo
  {author} {\bibfnamefont {C.}~\bibnamefont {Callegari}}, \bibinfo {author}
  {\bibfnamefont {F.}~\bibnamefont {Capotondi}}, \bibinfo {author}
  {\bibfnamefont {P.}~\bibnamefont {Cinquegrana}}, \bibinfo {author}
  {\bibfnamefont {M.}~\bibnamefont {Coreno}}, \bibinfo {author} {\bibfnamefont
  {R.}~\bibnamefont {Cucini}}, \bibinfo {author} {\bibfnamefont {M.~B.}\
  \bibnamefont {Danailov}}, \bibinfo {author} {\bibfnamefont {A.}~\bibnamefont
  {Demidovich}}, \bibinfo {author} {\bibfnamefont {G.}~\bibnamefont
  {De~Ninno}}, \bibinfo {author} {\bibfnamefont {M.}~\bibnamefont {Di~Fraia}},
  \bibinfo {author} {\bibfnamefont {R.}~\bibnamefont {Feifel}}, \bibinfo
  {author} {\bibfnamefont {E.}~\bibnamefont {Ferrari}}, \bibinfo {author}
  {\bibfnamefont {L.}~\bibnamefont {Fr\"ohlich}}, \bibinfo {author}
  {\bibfnamefont {D.}~\bibnamefont {Gauthier}}, \bibinfo {author}
  {\bibfnamefont {T.}~\bibnamefont {Golz}}, \bibinfo {author} {\bibfnamefont
  {C.}~\bibnamefont {Grazioli}}, \bibinfo {author} {\bibfnamefont
  {Y.}~\bibnamefont {Kai}}, \bibinfo {author} {\bibfnamefont {G.}~\bibnamefont
  {Kurdi}}, \bibinfo {author} {\bibfnamefont {N.}~\bibnamefont {Mahne}},
  \bibinfo {author} {\bibfnamefont {M.}~\bibnamefont {Manfredda}}, \bibinfo
  {author} {\bibfnamefont {N.}~\bibnamefont {Medvedev}}, \bibinfo {author}
  {\bibfnamefont {I.~P.}\ \bibnamefont {Nikolov}}, \bibinfo {author}
  {\bibfnamefont {E.}~\bibnamefont {Pedersoli}}, \bibinfo {author}
  {\bibfnamefont {G.}~\bibnamefont {Penco}}, \bibinfo {author} {\bibfnamefont
  {O.}~\bibnamefont {Plekan}}, \bibinfo {author} {\bibfnamefont {M.~J.}\
  \bibnamefont {Prandolini}}, \bibinfo {author} {\bibfnamefont {K.~C.}\
  \bibnamefont {Prince}}, \bibinfo {author} {\bibfnamefont {L.}~\bibnamefont
  {Raimondi}}, \bibinfo {author} {\bibfnamefont {P.}~\bibnamefont {Rebernik}},
  \bibinfo {author} {\bibfnamefont {R.}~\bibnamefont {Riedel}}, \bibinfo
  {author} {\bibfnamefont {E.}~\bibnamefont {Roussel}}, \bibinfo {author}
  {\bibfnamefont {P.}~\bibnamefont {Sigalotti}}, \bibinfo {author}
  {\bibfnamefont {R.}~\bibnamefont {Squibb}}, \bibinfo {author} {\bibfnamefont
  {N.}~\bibnamefont {Stojanovic}}, \bibinfo {author} {\bibfnamefont
  {S.}~\bibnamefont {Stranges}}, \bibinfo {author} {\bibfnamefont
  {C.}~\bibnamefont {Svetina}}, \bibinfo {author} {\bibfnamefont
  {T.}~\bibnamefont {Tanikawa}}, \bibinfo {author} {\bibfnamefont
  {U.}~\bibnamefont {Teubner}}, \bibinfo {author} {\bibfnamefont
  {V.}~\bibnamefont {Tkachenko}}, \bibinfo {author} {\bibfnamefont
  {S.}~\bibnamefont {Toleikis}}, \bibinfo {author} {\bibfnamefont
  {M.}~\bibnamefont {Zangrando}}, \bibinfo {author} {\bibfnamefont
  {B.}~\bibnamefont {Ziaja}}, \bibinfo {author} {\bibfnamefont
  {F.}~\bibnamefont {Tavella}},\ and\ \bibinfo {author} {\bibfnamefont
  {L.}~\bibnamefont {Giannessi}},\ }\bibfield  {title} {\bibinfo {title} {Pulse
  duration of seeded free-electron lasers},\ }\href
  {https://doi.org/10.1103/PhysRevX.7.021043} {\bibfield  {journal} {\bibinfo
  {journal} {Phys. Rev. X}\ }\textbf {\bibinfo {volume} {7}},\ \bibinfo {pages}
  {021043} (\bibinfo {year} {2017})}\BibitemShut {NoStop}%
\bibitem [{\citenamefont {Zatsarinny}(2006)}]{Zatsarinny2006}%
  \BibitemOpen
  \bibfield  {author} {\bibinfo {author} {\bibfnamefont {O.}~\bibnamefont
  {Zatsarinny}},\ }\bibfield  {title} {\bibinfo {title} {{BSR}: {B}-spline
  atomic {R}-matrix codes},\ }\href {https://doi.org/10.1016/j.cpc.2005.10.006}
  {\bibfield  {journal} {\bibinfo  {journal} {Comput. Phys. Commun.}\ }\textbf
  {\bibinfo {volume} {174}},\ \bibinfo {pages} {273} (\bibinfo {year}
  {2006})}\BibitemShut {NoStop}%
\bibitem [{\citenamefont {Kleiman}\ and\ \citenamefont
  {Lohmann}(2003)}]{KLEIMAN200329}%
  \BibitemOpen
  \bibfield  {author} {\bibinfo {author} {\bibfnamefont {U.}~\bibnamefont
  {Kleiman}}\ and\ \bibinfo {author} {\bibfnamefont {B.}~\bibnamefont
  {Lohmann}},\ }\bibfield  {title} {\bibinfo {title} {Photoionization of
  closed-shell atoms: Hartree-fock calculations of orientation and alignment},\
  }\href {https://doi.org/https://doi.org/10.1016/S0368-2048(03)00034-3}
  {\bibfield  {journal} {\bibinfo  {journal} {Journal of Electron Spectroscopy
  and Related Phenomena}\ }\textbf {\bibinfo {volume} {131-132}},\ \bibinfo
  {pages} {29} (\bibinfo {year} {2003})}\BibitemShut {NoStop}%
\bibitem [{\citenamefont {Amusia}\ \emph {et~al.}(2012)\citenamefont {Amusia},
  \citenamefont {Chernysheva},\ and\ \citenamefont {Yarzhemsky}}]{Amusia2012}%
  \BibitemOpen
  \bibfield  {author} {\bibinfo {author} {\bibfnamefont {M.}~\bibnamefont
  {Amusia}}, \bibinfo {author} {\bibfnamefont {L.}~\bibnamefont
  {Chernysheva}},\ and\ \bibinfo {author} {\bibfnamefont {V.}~\bibnamefont
  {Yarzhemsky}},\ }\href@noop {} {\emph {\bibinfo {title} {Handbook of
  theoretical atomic physics: data for photon absorption, electron scattering,
  and vacancies decay}}}\ (\bibinfo  {publisher} {Springer Science \& Business
  Media},\ \bibinfo {year} {2012})\BibitemShut {NoStop}%
\bibitem [{\citenamefont {Johnson}\ and\ \citenamefont
  {Cheng}(1979)}]{Johnson1979}%
  \BibitemOpen
  \bibfield  {author} {\bibinfo {author} {\bibfnamefont {W.~R.}\ \bibnamefont
  {Johnson}}\ and\ \bibinfo {author} {\bibfnamefont {K.~T.}\ \bibnamefont
  {Cheng}},\ }\bibfield  {title} {\bibinfo {title} {Photoionization of the
  outer shells of neon, argon, krypton, and xenon using the relativistic
  random-phase approximation},\ }\href
  {https://doi.org/10.1103/PhysRevA.20.978} {\bibfield  {journal} {\bibinfo
  {journal} {Phys. Rev. A}\ }\textbf {\bibinfo {volume} {20}},\ \bibinfo
  {pages} {978} (\bibinfo {year} {1979})}\BibitemShut {NoStop}%
\bibitem [{\citenamefont {Huang}\ \emph {et~al.}(1981)\citenamefont {Huang},
  \citenamefont {Johnson},\ and\ \citenamefont {Cheng}}]{Huang1981}%
  \BibitemOpen
  \bibfield  {author} {\bibinfo {author} {\bibfnamefont {K.-N.}\ \bibnamefont
  {Huang}}, \bibinfo {author} {\bibfnamefont {W.}~\bibnamefont {Johnson}},\
  and\ \bibinfo {author} {\bibfnamefont {K.}~\bibnamefont {Cheng}},\ }\bibfield
   {title} {\bibinfo {title} {Theoretical photoionization parameters for the
  noble gases argon, krypton, and xenon},\ }\href
  {https://doi.org/https://doi.org/10.1016/0092-640X(81)90010-3} {\bibfield
  {journal} {\bibinfo  {journal} {Atomic Data and Nuclear Data Tables}\
  }\textbf {\bibinfo {volume} {26}},\ \bibinfo {pages} {33 } (\bibinfo {year}
  {1981})}\BibitemShut {NoStop}%
\bibitem [{\citenamefont {Aksela}\ \emph {et~al.}(1987)\citenamefont {Aksela},
  \citenamefont {Aksela}, \citenamefont {Levasalmi}, \citenamefont {Tan},\ and\
  \citenamefont {Bancroft}}]{Aksela1987}%
  \BibitemOpen
  \bibfield  {author} {\bibinfo {author} {\bibfnamefont {S.}~\bibnamefont
  {Aksela}}, \bibinfo {author} {\bibfnamefont {H.}~\bibnamefont {Aksela}},
  \bibinfo {author} {\bibfnamefont {M.}~\bibnamefont {Levasalmi}}, \bibinfo
  {author} {\bibfnamefont {K.~H.}\ \bibnamefont {Tan}},\ and\ \bibinfo {author}
  {\bibfnamefont {G.~M.}\ \bibnamefont {Bancroft}},\ }\bibfield  {title}
  {\bibinfo {title} {Partial photoionization cross sections of kr 3d, 4s, and
  4p levels in the photon energy range 37--160 ev},\ }\href
  {https://doi.org/10.1103/PhysRevA.36.3449} {\bibfield  {journal} {\bibinfo
  {journal} {Phys. Rev. A}\ }\textbf {\bibinfo {volume} {36}},\ \bibinfo
  {pages} {3449} (\bibinfo {year} {1987})}\BibitemShut {NoStop}%
\bibitem [{\citenamefont {Amusia}\ \emph {et~al.}(1972)\citenamefont {Amusia},
  \citenamefont {Ivanov}, \citenamefont {Cherepkov},\ and\ \citenamefont
  {Chernysheva}}]{Amusia1972}%
  \BibitemOpen
  \bibfield  {author} {\bibinfo {author} {\bibfnamefont {M.~Y.}\ \bibnamefont
  {Amusia}}, \bibinfo {author} {\bibfnamefont {V.~K.}\ \bibnamefont {Ivanov}},
  \bibinfo {author} {\bibfnamefont {N.~A.}\ \bibnamefont {Cherepkov}},\ and\
  \bibinfo {author} {\bibfnamefont {L.~V.}\ \bibnamefont {Chernysheva}},\
  }\bibfield  {title} {\bibinfo {title} {Interference effects in
  photoionization of noble gas atoms outer s-shells},\ }\href
  {https://doi.org/10.1016/0375-9601(72)90531-2} {\bibfield  {journal}
  {\bibinfo  {journal} {Physics Letters}\ }\textbf {\bibinfo {volume} {40A}},\
  \bibinfo {pages} {361} (\bibinfo {year} {1972})}\BibitemShut {NoStop}%
\bibitem [{\citenamefont {Berrah}\ \emph {et~al.}(1997)\citenamefont {Berrah},
  \citenamefont {Farhat}, \citenamefont {Langer}, \citenamefont {Lagutin},
  \citenamefont {Demekhin}, \citenamefont {Petrov}, \citenamefont {Sukhorukov},
  \citenamefont {Wehlitz}, \citenamefont {Whitfield}, \citenamefont
  {Viefhaus},\ and\ \citenamefont {Becker}}]{Berrah1997}%
  \BibitemOpen
  \bibfield  {author} {\bibinfo {author} {\bibfnamefont {N.}~\bibnamefont
  {Berrah}}, \bibinfo {author} {\bibfnamefont {A.}~\bibnamefont {Farhat}},
  \bibinfo {author} {\bibfnamefont {B.}~\bibnamefont {Langer}}, \bibinfo
  {author} {\bibfnamefont {B.~M.}\ \bibnamefont {Lagutin}}, \bibinfo {author}
  {\bibfnamefont {P.~V.}\ \bibnamefont {Demekhin}}, \bibinfo {author}
  {\bibfnamefont {I.~D.}\ \bibnamefont {Petrov}}, \bibinfo {author}
  {\bibfnamefont {V.~L.}\ \bibnamefont {Sukhorukov}}, \bibinfo {author}
  {\bibfnamefont {R.}~\bibnamefont {Wehlitz}}, \bibinfo {author} {\bibfnamefont
  {S.~B.}\ \bibnamefont {Whitfield}}, \bibinfo {author} {\bibfnamefont
  {J.}~\bibnamefont {Viefhaus}},\ and\ \bibinfo {author} {\bibfnamefont
  {U.}~\bibnamefont {Becker}},\ }\bibfield  {title} {\bibinfo {title}
  {Angle-resolved energy dependence of the ${4p}^{4}{nd(}^{2}{S}_{1/2})$
  $(n=4--7)$ correlation satellites in kr from 38.5 to 250 ev: Experiment and
  theory},\ }\href {https://doi.org/10.1103/PhysRevA.56.4545} {\bibfield
  {journal} {\bibinfo  {journal} {Phys. Rev. A}\ }\textbf {\bibinfo {volume}
  {56}},\ \bibinfo {pages} {4545} (\bibinfo {year} {1997})}\BibitemShut
  {NoStop}%
\bibitem [{\citenamefont {Ehresmann}\ \emph {et~al.}(1994)\citenamefont
  {Ehresmann}, \citenamefont {Vollweiler}, \citenamefont {Schmoranzer},
  \citenamefont {Sukhorukov}, \citenamefont {Lagutin}, \citenamefont {Petrov},
  \citenamefont {Mentzel},\ and\ \citenamefont {Schartner}}]{Ehresmann1994}%
  \BibitemOpen
  \bibfield  {author} {\bibinfo {author} {\bibfnamefont {A.}~\bibnamefont
  {Ehresmann}}, \bibinfo {author} {\bibfnamefont {F.}~\bibnamefont
  {Vollweiler}}, \bibinfo {author} {\bibfnamefont {H.}~\bibnamefont
  {Schmoranzer}}, \bibinfo {author} {\bibfnamefont {V.~L.}\ \bibnamefont
  {Sukhorukov}}, \bibinfo {author} {\bibfnamefont {B.~M.}\ \bibnamefont
  {Lagutin}}, \bibinfo {author} {\bibfnamefont {I.~D.}\ \bibnamefont {Petrov}},
  \bibinfo {author} {\bibfnamefont {G.}~\bibnamefont {Mentzel}},\ and\ \bibinfo
  {author} {\bibfnamefont {K.-H.}\ \bibnamefont {Schartner}},\ }\bibfield
  {title} {\bibinfo {title} {Photoionization of kr 4s: Iii. detailed and
  extended measurements of the kr 4s-electron ionization cross section},\
  }\href {https://doi.org/10.1088/0953-4075/27/8/010} {\bibfield  {journal}
  {\bibinfo  {journal} {J. Phys. B: At. Mol. Opt. Ph.}\ }\textbf {\bibinfo
  {volume} {27}},\ \bibinfo {pages} {1489} (\bibinfo {year}
  {1994})}\BibitemShut {NoStop}%
\bibitem [{\citenamefont {Fritzsche}\ \emph {et~al.}(2008)\citenamefont
  {Fritzsche}, \citenamefont {Grum-Grzhimailo}, \citenamefont {Gryzlova},\ and\
  \citenamefont {Kabachnik}}]{Fritzsche2008}%
  \BibitemOpen
  \bibfield  {author} {\bibinfo {author} {\bibfnamefont {S.}~\bibnamefont
  {Fritzsche}}, \bibinfo {author} {\bibfnamefont {A.~N.}\ \bibnamefont
  {Grum-Grzhimailo}}, \bibinfo {author} {\bibfnamefont {E.~V.}\ \bibnamefont
  {Gryzlova}},\ and\ \bibinfo {author} {\bibfnamefont {N.~M.}\ \bibnamefont
  {Kabachnik}},\ }\bibfield  {title} {\bibinfo {title} {Angular distributions
  and angular correlations in sequential two-photon double ionization of
  atoms},\ }\href {https://doi.org/10.1088/0953-4075/41/16/165601} {\bibfield
  {journal} {\bibinfo  {journal} {J. Phys. B: At. Mol. Opt. Ph.}\ }\textbf
  {\bibinfo {volume} {41}},\ \bibinfo {pages} {165601} (\bibinfo {year}
  {2008})}\BibitemShut {NoStop}%
\bibitem [{\citenamefont {Fritzsche}\ \emph {et~al.}(2009)\citenamefont
  {Fritzsche}, \citenamefont {Grum-Grzhimailo}, \citenamefont {Gryzlova},\ and\
  \citenamefont {Kabachnik}}]{Fritzsche2009}%
  \BibitemOpen
  \bibfield  {author} {\bibinfo {author} {\bibfnamefont {S.}~\bibnamefont
  {Fritzsche}}, \bibinfo {author} {\bibfnamefont {A.~N.}\ \bibnamefont
  {Grum-Grzhimailo}}, \bibinfo {author} {\bibfnamefont {E.~V.}\ \bibnamefont
  {Gryzlova}},\ and\ \bibinfo {author} {\bibfnamefont {N.~M.}\ \bibnamefont
  {Kabachnik}},\ }\bibfield  {title} {\bibinfo {title} {Sequential two-photon
  double ionization of {Kr} atoms},\ }\href
  {https://doi.org/10.1088/0953-4075/42/14/145602} {\bibfield  {journal}
  {\bibinfo  {journal} {J. Phys. B: At. Mol. Opt. Ph.}\ }\textbf {\bibinfo
  {volume} {42}},\ \bibinfo {pages} {145602} (\bibinfo {year}
  {2009})}\BibitemShut {NoStop}%
\bibitem [{\citenamefont {Grum-Grzhimailo}\ \emph {et~al.}(2012)\citenamefont
  {Grum-Grzhimailo}, \citenamefont {Gryzlova},\ and\ \citenamefont
  {Meyer}}]{Grum2012}%
  \BibitemOpen
  \bibfield  {author} {\bibinfo {author} {\bibfnamefont {A.~N.}\ \bibnamefont
  {Grum-Grzhimailo}}, \bibinfo {author} {\bibfnamefont {E.~V.}\ \bibnamefont
  {Gryzlova}},\ and\ \bibinfo {author} {\bibfnamefont {M.}~\bibnamefont
  {Meyer}},\ }\bibfield  {title} {\bibinfo {title} {Non-dipole effects in the
  angular distribution of photoelectrons in sequential two-photon atomic double
  ionization},\ }\href {https://doi.org/10.1088/0953-4075/45/21/215602}
  {\bibfield  {journal} {\bibinfo  {journal} {J. Phys. B: At. Mol. Opt. Ph.}\
  }\textbf {\bibinfo {volume} {45}},\ \bibinfo {pages} {215602} (\bibinfo
  {year} {2012})}\BibitemShut {NoStop}%
\bibitem [{\citenamefont {Gryzlova}\ \emph {et~al.}(2010)\citenamefont
  {Gryzlova}, \citenamefont {Grum-Grzhimailo}, \citenamefont {Fritzsche},\ and\
  \citenamefont {Kabachnik}}]{Gryzlova2010}%
  \BibitemOpen
  \bibfield  {author} {\bibinfo {author} {\bibfnamefont {E.~V.}\ \bibnamefont
  {Gryzlova}}, \bibinfo {author} {\bibfnamefont {A.~N.}\ \bibnamefont
  {Grum-Grzhimailo}}, \bibinfo {author} {\bibfnamefont {S.}~\bibnamefont
  {Fritzsche}},\ and\ \bibinfo {author} {\bibfnamefont {N.~M.}\ \bibnamefont
  {Kabachnik}},\ }\bibfield  {title} {\bibinfo {title} {Angular correlations
  between two electrons emitted in the sequential two-photon double ionization
  of atoms},\ }\href {https://doi.org/10.1088/0953-4075/43/22/225602}
  {\bibfield  {journal} {\bibinfo  {journal} {J. Phys. B: At. Mol. Opt. Ph.}\
  }\textbf {\bibinfo {volume} {43}},\ \bibinfo {pages} {225602} (\bibinfo
  {year} {2010})}\BibitemShut {NoStop}%
\bibitem [{\citenamefont {Kiselev}\ \emph {et~al.}(2020)\citenamefont
  {Kiselev}, \citenamefont {Carpeggiani}, \citenamefont {Gryzlova},
  \citenamefont {Burkov}, \citenamefont {Reduzzi}, \citenamefont {Dubrouil},
  \citenamefont {Facciala}, \citenamefont {Negro}, \citenamefont {Ueda},
  \citenamefont {Frassetto}, \citenamefont {Stienkemeier}, \citenamefont
  {Ovcharenko}, \citenamefont {Meyer}, \citenamefont {Fraia}, \citenamefont
  {Plekan}, \citenamefont {Prince},\ and\ \citenamefont
  {Callegari1}}]{Kiselev2020}%
  \BibitemOpen
  \bibfield  {author} {\bibinfo {author} {\bibfnamefont {M.~D.}\ \bibnamefont
  {Kiselev}}, \bibinfo {author} {\bibfnamefont {P.~A.}\ \bibnamefont
  {Carpeggiani}}, \bibinfo {author} {\bibfnamefont {E.~V.}\ \bibnamefont
  {Gryzlova}}, \bibinfo {author} {\bibfnamefont {S.~M.}\ \bibnamefont
  {Burkov}}, \bibinfo {author} {\bibfnamefont {M.}~\bibnamefont {Reduzzi}},
  \bibinfo {author} {\bibfnamefont {A.}~\bibnamefont {Dubrouil}}, \bibinfo
  {author} {\bibfnamefont {D.}~\bibnamefont {Facciala}}, \bibinfo {author}
  {\bibfnamefont {M.}~\bibnamefont {Negro}}, \bibinfo {author} {\bibfnamefont
  {K.}~\bibnamefont {Ueda}}, \bibinfo {author} {\bibfnamefont {F.}~\bibnamefont
  {Frassetto}}, \bibinfo {author} {\bibfnamefont {F.}~\bibnamefont
  {Stienkemeier}}, \bibinfo {author} {\bibfnamefont {Y.}~\bibnamefont
  {Ovcharenko}}, \bibinfo {author} {\bibfnamefont {M.}~\bibnamefont {Meyer}},
  \bibinfo {author} {\bibfnamefont {M.~D.}\ \bibnamefont {Fraia}}, \bibinfo
  {author} {\bibfnamefont {O.}~\bibnamefont {Plekan}}, \bibinfo {author}
  {\bibfnamefont {K.~C.}\ \bibnamefont {Prince}},\ and\ \bibinfo {author}
  {\bibfnamefont {C.}~\bibnamefont {Callegari1}},\ }\bibfield  {title}
  {\bibinfo {title} {Photoelectron spectra and angular distribution in
  sequential two-photon double ionization in the region of autoionizing
  resonances of arii and krii},\ }\href@noop {} {\bibfield  {journal} {\bibinfo
   {journal} {J. Phys. B: At. Mol. Ph.}\ }\textbf {\bibinfo {volume}
  {accepted}} (\bibinfo {year} {2020})}\BibitemShut {NoStop}%
\bibitem [{\citenamefont {Lynch}\ \emph {et~al.}(1973)\citenamefont {Lynch},
  \citenamefont {Gardner}, \citenamefont {Codling},\ and\ \citenamefont
  {Marr}}]{Lynch1973}%
  \BibitemOpen
  \bibfield  {author} {\bibinfo {author} {\bibfnamefont {M.~J.}\ \bibnamefont
  {Lynch}}, \bibinfo {author} {\bibfnamefont {A.~B.}\ \bibnamefont {Gardner}},
  \bibinfo {author} {\bibfnamefont {K.}~\bibnamefont {Codling}},\ and\ \bibinfo
  {author} {\bibfnamefont {G.~V.}\ \bibnamefont {Marr}},\ }\bibfield  {title}
  {\bibinfo {title} {The photoionization of the 3s subshell of argon in the
  threshold region by photoelectron spectroscopy},\ }\href
  {https://doi.org/10.1016/0375-9601(73)90286-7} {\bibfield  {journal}
  {\bibinfo  {journal} {Physics Letters}\ }\textbf {\bibinfo {volume} {43A}},\
  \bibinfo {pages} {237} (\bibinfo {year} {1973})}\BibitemShut {NoStop}%
\bibitem [{\citenamefont {Mondal}\ \emph
  {et~al.}(2013{\natexlab{b}})\citenamefont {Mondal}, \citenamefont {Ma},
  \citenamefont {Motomura}, \citenamefont {Fukuzawa}, \citenamefont {Yamada},
  \citenamefont {Nagaya}, \citenamefont {Yase}, \citenamefont {Mizoguchi},
  \citenamefont {Yao}, \citenamefont {Rouzée}, \citenamefont {Hundertmark},
  \citenamefont {Vrakking}, \citenamefont {Johnsson}, \citenamefont {Nagasono},
  \citenamefont {Tono}, \citenamefont {Togashi}, \citenamefont {Senba},
  \citenamefont {Ohashi}, \citenamefont {Yabashi}, \citenamefont {Ishikawa},
  \citenamefont {Sazhina}, \citenamefont {Fritzsche}, \citenamefont
  {Kabachnik},\ and\ \citenamefont {Ueda}}]{Mondal2013}%
  \BibitemOpen
  \bibfield  {author} {\bibinfo {author} {\bibfnamefont {S.}~\bibnamefont
  {Mondal}}, \bibinfo {author} {\bibfnamefont {R.}~\bibnamefont {Ma}}, \bibinfo
  {author} {\bibfnamefont {K.}~\bibnamefont {Motomura}}, \bibinfo {author}
  {\bibfnamefont {H.}~\bibnamefont {Fukuzawa}}, \bibinfo {author}
  {\bibfnamefont {A.}~\bibnamefont {Yamada}}, \bibinfo {author} {\bibfnamefont
  {K.}~\bibnamefont {Nagaya}}, \bibinfo {author} {\bibfnamefont
  {S.}~\bibnamefont {Yase}}, \bibinfo {author} {\bibfnamefont {Y.}~\bibnamefont
  {Mizoguchi}}, \bibinfo {author} {\bibfnamefont {M.}~\bibnamefont {Yao}},
  \bibinfo {author} {\bibfnamefont {A.}~\bibnamefont {Rouzée}}, \bibinfo
  {author} {\bibfnamefont {A.}~\bibnamefont {Hundertmark}}, \bibinfo {author}
  {\bibfnamefont {M.~J.~J.}\ \bibnamefont {Vrakking}}, \bibinfo {author}
  {\bibfnamefont {P.}~\bibnamefont {Johnsson}}, \bibinfo {author}
  {\bibfnamefont {M.}~\bibnamefont {Nagasono}}, \bibinfo {author}
  {\bibfnamefont {K.}~\bibnamefont {Tono}}, \bibinfo {author} {\bibfnamefont
  {T.}~\bibnamefont {Togashi}}, \bibinfo {author} {\bibfnamefont
  {Y.}~\bibnamefont {Senba}}, \bibinfo {author} {\bibfnamefont
  {H.}~\bibnamefont {Ohashi}}, \bibinfo {author} {\bibfnamefont
  {M.}~\bibnamefont {Yabashi}}, \bibinfo {author} {\bibfnamefont
  {T.}~\bibnamefont {Ishikawa}}, \bibinfo {author} {\bibfnamefont {I.~P.}\
  \bibnamefont {Sazhina}}, \bibinfo {author} {\bibfnamefont {S.}~\bibnamefont
  {Fritzsche}}, \bibinfo {author} {\bibfnamefont {N.~M.}\ \bibnamefont
  {Kabachnik}},\ and\ \bibinfo {author} {\bibfnamefont {K.}~\bibnamefont
  {Ueda}},\ }\bibfield  {title} {\bibinfo {title} {Photoelectron angular
  distributions for the two-photon sequential double ionization of xenon by
  ultrashort extreme ultraviolet free electron laser pulses},\ }\href
  {https://doi.org/10.1088/0953-4075/46/16/164022} {\bibfield  {journal}
  {\bibinfo  {journal} {J. Phys. B: At. Mol. Opt. Phys}\ }\textbf {\bibinfo
  {volume} {46}},\ \bibinfo {pages} {164022} (\bibinfo {year}
  {2013}{\natexlab{b}})}\BibitemShut {NoStop}%
\bibitem [{\citenamefont {Rudek}\ \emph {et~al.}(2012)\citenamefont {Rudek},
  \citenamefont {Son}, \citenamefont {Foucar}, \citenamefont {Epp},
  \citenamefont {Erk}, \citenamefont {Hartmann}, \citenamefont {Adolph},
  \citenamefont {Andritschke}, \citenamefont {Aquila}, \citenamefont {Berrah},
  \citenamefont {Bostedt}, \citenamefont {Bozek}, \citenamefont {Coppola},
  \citenamefont {Filsinger}, \citenamefont {Gorke}, \citenamefont {Gorkhover},
  \citenamefont {Graafsma}, \citenamefont {Gumprecht}, \citenamefont
  {Hartmann}, \citenamefont {Hauser}, \citenamefont {Herrmann}, \citenamefont
  {Hirsemann}, \citenamefont {Holl}, \citenamefont {H{\"o}mke}, \citenamefont
  {Journel}, \citenamefont {Kaiser}, \citenamefont {Kimmel}, \citenamefont
  {Krasniqi}, \citenamefont {K{\"u}hnel}, \citenamefont {Matysek},
  \citenamefont {Messerschmidt}, \citenamefont {Miesner}, \citenamefont
  {M{\"o}ller}, \citenamefont {Moshammer}, \citenamefont {Nagaya},
  \citenamefont {Nilsson}, \citenamefont {Potdevin}, \citenamefont
  {Pietschner}, \citenamefont {Reich}, \citenamefont {Rupp}, \citenamefont
  {Schaller}, \citenamefont {Schlichting}, \citenamefont {Schmidt},
  \citenamefont {Schopper}, \citenamefont {Schorb}, \citenamefont
  {Schr{\"o}ter}, \citenamefont {Schulz}, \citenamefont {Simon}, \citenamefont
  {Soltau}, \citenamefont {Str{\"u}der}, \citenamefont {Ueda}, \citenamefont
  {Weidenspointner}, \citenamefont {Santra}, \citenamefont {Ullrich},
  \citenamefont {Rudenko},\ and\ \citenamefont {Rolles}}]{Rudek2012}%
  \BibitemOpen
  \bibfield  {author} {\bibinfo {author} {\bibfnamefont {B.}~\bibnamefont
  {Rudek}}, \bibinfo {author} {\bibfnamefont {S.-K.}\ \bibnamefont {Son}},
  \bibinfo {author} {\bibfnamefont {L.}~\bibnamefont {Foucar}}, \bibinfo
  {author} {\bibfnamefont {S.~W.}\ \bibnamefont {Epp}}, \bibinfo {author}
  {\bibfnamefont {B.}~\bibnamefont {Erk}}, \bibinfo {author} {\bibfnamefont
  {R.}~\bibnamefont {Hartmann}}, \bibinfo {author} {\bibfnamefont
  {M.}~\bibnamefont {Adolph}}, \bibinfo {author} {\bibfnamefont
  {R.}~\bibnamefont {Andritschke}}, \bibinfo {author} {\bibfnamefont
  {A.}~\bibnamefont {Aquila}}, \bibinfo {author} {\bibfnamefont
  {N.}~\bibnamefont {Berrah}}, \bibinfo {author} {\bibfnamefont
  {C.}~\bibnamefont {Bostedt}}, \bibinfo {author} {\bibfnamefont
  {J.}~\bibnamefont {Bozek}}, \bibinfo {author} {\bibfnamefont
  {N.}~\bibnamefont {Coppola}}, \bibinfo {author} {\bibfnamefont
  {F.}~\bibnamefont {Filsinger}}, \bibinfo {author} {\bibfnamefont
  {H.}~\bibnamefont {Gorke}}, \bibinfo {author} {\bibfnamefont
  {T.}~\bibnamefont {Gorkhover}}, \bibinfo {author} {\bibfnamefont
  {H.}~\bibnamefont {Graafsma}}, \bibinfo {author} {\bibfnamefont
  {L.}~\bibnamefont {Gumprecht}}, \bibinfo {author} {\bibfnamefont
  {A.}~\bibnamefont {Hartmann}}, \bibinfo {author} {\bibfnamefont
  {G.}~\bibnamefont {Hauser}}, \bibinfo {author} {\bibfnamefont
  {S.}~\bibnamefont {Herrmann}}, \bibinfo {author} {\bibfnamefont
  {H.}~\bibnamefont {Hirsemann}}, \bibinfo {author} {\bibfnamefont
  {P.}~\bibnamefont {Holl}}, \bibinfo {author} {\bibfnamefont {A.}~\bibnamefont
  {H{\"o}mke}}, \bibinfo {author} {\bibfnamefont {L.}~\bibnamefont {Journel}},
  \bibinfo {author} {\bibfnamefont {C.}~\bibnamefont {Kaiser}}, \bibinfo
  {author} {\bibfnamefont {N.}~\bibnamefont {Kimmel}}, \bibinfo {author}
  {\bibfnamefont {F.}~\bibnamefont {Krasniqi}}, \bibinfo {author}
  {\bibfnamefont {K.-U.}\ \bibnamefont {K{\"u}hnel}}, \bibinfo {author}
  {\bibfnamefont {M.}~\bibnamefont {Matysek}}, \bibinfo {author} {\bibfnamefont
  {M.}~\bibnamefont {Messerschmidt}}, \bibinfo {author} {\bibfnamefont
  {D.}~\bibnamefont {Miesner}}, \bibinfo {author} {\bibfnamefont
  {T.}~\bibnamefont {M{\"o}ller}}, \bibinfo {author} {\bibfnamefont
  {R.}~\bibnamefont {Moshammer}}, \bibinfo {author} {\bibfnamefont
  {K.}~\bibnamefont {Nagaya}}, \bibinfo {author} {\bibfnamefont
  {B.}~\bibnamefont {Nilsson}}, \bibinfo {author} {\bibfnamefont
  {G.}~\bibnamefont {Potdevin}}, \bibinfo {author} {\bibfnamefont
  {D.}~\bibnamefont {Pietschner}}, \bibinfo {author} {\bibfnamefont
  {C.}~\bibnamefont {Reich}}, \bibinfo {author} {\bibfnamefont
  {D.}~\bibnamefont {Rupp}}, \bibinfo {author} {\bibfnamefont {G.}~\bibnamefont
  {Schaller}}, \bibinfo {author} {\bibfnamefont {I.}~\bibnamefont
  {Schlichting}}, \bibinfo {author} {\bibfnamefont {C.}~\bibnamefont
  {Schmidt}}, \bibinfo {author} {\bibfnamefont {F.}~\bibnamefont {Schopper}},
  \bibinfo {author} {\bibfnamefont {S.}~\bibnamefont {Schorb}}, \bibinfo
  {author} {\bibfnamefont {C.-D.}\ \bibnamefont {Schr{\"o}ter}}, \bibinfo
  {author} {\bibfnamefont {J.}~\bibnamefont {Schulz}}, \bibinfo {author}
  {\bibfnamefont {M.}~\bibnamefont {Simon}}, \bibinfo {author} {\bibfnamefont
  {H.}~\bibnamefont {Soltau}}, \bibinfo {author} {\bibfnamefont
  {L.}~\bibnamefont {Str{\"u}der}}, \bibinfo {author} {\bibfnamefont
  {K.}~\bibnamefont {Ueda}}, \bibinfo {author} {\bibfnamefont {G.}~\bibnamefont
  {Weidenspointner}}, \bibinfo {author} {\bibfnamefont {R.}~\bibnamefont
  {Santra}}, \bibinfo {author} {\bibfnamefont {J.}~\bibnamefont {Ullrich}},
  \bibinfo {author} {\bibfnamefont {A.}~\bibnamefont {Rudenko}},\ and\ \bibinfo
  {author} {\bibfnamefont {D.}~\bibnamefont {Rolles}},\ }\bibfield  {title}
  {\bibinfo {title} {Ultra-efficient ionization of heavy atoms by intense x-ray
  free-electron laser pulses},\ }\href
  {https://doi.org/10.1038/nphoton.2012.261} {\bibfield  {journal} {\bibinfo
  {journal} {Nature Photonics}\ }\textbf {\bibinfo {volume} {6}},\ \bibinfo
  {pages} {858} (\bibinfo {year} {2012})}\BibitemShut {NoStop}%
\bibitem [{\citenamefont {Rudek}\ \emph {et~al.}(2013)\citenamefont {Rudek},
  \citenamefont {Rolles}, \citenamefont {Son}, \citenamefont {Foucar},
  \citenamefont {Erk}, \citenamefont {Epp}, \citenamefont {Boll}, \citenamefont
  {Anielski}, \citenamefont {Bostedt}, \citenamefont {Schorb}, \citenamefont
  {Coffee}, \citenamefont {Bozek}, \citenamefont {Trippel}, \citenamefont
  {Marchenko}, \citenamefont {Simon}, \citenamefont {Christensen},
  \citenamefont {De}, \citenamefont {Wada}, \citenamefont {Ueda}, \citenamefont
  {Schlichting}, \citenamefont {Santra}, \citenamefont {Ullrich},\ and\
  \citenamefont {Rudenko}}]{Rudek2013}%
  \BibitemOpen
  \bibfield  {author} {\bibinfo {author} {\bibfnamefont {B.}~\bibnamefont
  {Rudek}}, \bibinfo {author} {\bibfnamefont {D.}~\bibnamefont {Rolles}},
  \bibinfo {author} {\bibfnamefont {S.-K.}\ \bibnamefont {Son}}, \bibinfo
  {author} {\bibfnamefont {L.}~\bibnamefont {Foucar}}, \bibinfo {author}
  {\bibfnamefont {B.}~\bibnamefont {Erk}}, \bibinfo {author} {\bibfnamefont
  {S.}~\bibnamefont {Epp}}, \bibinfo {author} {\bibfnamefont {R.}~\bibnamefont
  {Boll}}, \bibinfo {author} {\bibfnamefont {D.}~\bibnamefont {Anielski}},
  \bibinfo {author} {\bibfnamefont {C.}~\bibnamefont {Bostedt}}, \bibinfo
  {author} {\bibfnamefont {S.}~\bibnamefont {Schorb}}, \bibinfo {author}
  {\bibfnamefont {R.}~\bibnamefont {Coffee}}, \bibinfo {author} {\bibfnamefont
  {J.}~\bibnamefont {Bozek}}, \bibinfo {author} {\bibfnamefont
  {S.}~\bibnamefont {Trippel}}, \bibinfo {author} {\bibfnamefont
  {T.}~\bibnamefont {Marchenko}}, \bibinfo {author} {\bibfnamefont
  {M.}~\bibnamefont {Simon}}, \bibinfo {author} {\bibfnamefont
  {L.}~\bibnamefont {Christensen}}, \bibinfo {author} {\bibfnamefont
  {S.}~\bibnamefont {De}}, \bibinfo {author} {\bibfnamefont {S.-i.}\
  \bibnamefont {Wada}}, \bibinfo {author} {\bibfnamefont {K.}~\bibnamefont
  {Ueda}}, \bibinfo {author} {\bibfnamefont {I.}~\bibnamefont {Schlichting}},
  \bibinfo {author} {\bibfnamefont {R.}~\bibnamefont {Santra}}, \bibinfo
  {author} {\bibfnamefont {J.}~\bibnamefont {Ullrich}},\ and\ \bibinfo {author}
  {\bibfnamefont {A.}~\bibnamefont {Rudenko}},\ }\bibfield  {title} {\bibinfo
  {title} {Resonance-enhanced multiple ionization of krypton at an x-ray
  free-electron laser},\ }\href {https://doi.org/10.1103/PhysRevA.87.023413}
  {\bibfield  {journal} {\bibinfo  {journal} {Phys. Rev. A}\ }\textbf {\bibinfo
  {volume} {87}},\ \bibinfo {pages} {023413} (\bibinfo {year}
  {2013})}\BibitemShut {NoStop}%
\bibitem [{\citenamefont {Samson}\ and\ \citenamefont
  {Gardner}(1974)}]{Samson1974}%
  \BibitemOpen
  \bibfield  {author} {\bibinfo {author} {\bibfnamefont {J.~A.~R.}\
  \bibnamefont {Samson}}\ and\ \bibinfo {author} {\bibfnamefont {J.~L.}\
  \bibnamefont {Gardner}},\ }\bibfield  {title} {\bibinfo {title}
  {Photoionization cross sections of the outer $s$-subshell electrons in the
  rare gases},\ }\href {https://doi.org/10.1103/PhysRevLett.33.671} {\bibfield
  {journal} {\bibinfo  {journal} {Phys. Rev. Lett.}\ }\textbf {\bibinfo
  {volume} {33}},\ \bibinfo {pages} {671} (\bibinfo {year} {1974})}\BibitemShut
  {NoStop}%
\bibitem [{\citenamefont {Sukhorukov}\ \emph {et~al.}(1994)\citenamefont
  {Sukhorukov}, \citenamefont {Lagutin}, \citenamefont {Petrov}, \citenamefont
  {Schmoranzer}, \citenamefont {Ehresmann},\ and\ \citenamefont
  {Schartner}}]{Sukhorukov1994}%
  \BibitemOpen
  \bibfield  {author} {\bibinfo {author} {\bibfnamefont {V.~L.}\ \bibnamefont
  {Sukhorukov}}, \bibinfo {author} {\bibfnamefont {B.~M.}\ \bibnamefont
  {Lagutin}}, \bibinfo {author} {\bibfnamefont {I.~D.}\ \bibnamefont {Petrov}},
  \bibinfo {author} {\bibfnamefont {H.}~\bibnamefont {Schmoranzer}}, \bibinfo
  {author} {\bibfnamefont {A.}~\bibnamefont {Ehresmann}},\ and\ \bibinfo
  {author} {\bibfnamefont {K.~H.}\ \bibnamefont {Schartner}},\ }\bibfield
  {title} {\bibinfo {title} {Photoionization of kr near 4s threshold. {II}.
  intermediate-coupling theory},\ }\href
  {https://doi.org/10.1088/0953-4075/27/2/003} {\bibfield  {journal} {\bibinfo
  {journal} {J. Phys. B: At. Mol. Opt. Ph.}\ }\textbf {\bibinfo {volume}
  {27}},\ \bibinfo {pages} {241} (\bibinfo {year} {1994})}\BibitemShut
  {NoStop}%
\bibitem [{\citenamefont {Tulkki}\ \emph {et~al.}(1992)\citenamefont {Tulkki},
  \citenamefont {Aksela}, \citenamefont {Aksela}, \citenamefont {Shigemasa},
  \citenamefont {Yagishita},\ and\ \citenamefont {Furusawa}}]{Tulkki1992}%
  \BibitemOpen
  \bibfield  {author} {\bibinfo {author} {\bibfnamefont {J.}~\bibnamefont
  {Tulkki}}, \bibinfo {author} {\bibfnamefont {S.}~\bibnamefont {Aksela}},
  \bibinfo {author} {\bibfnamefont {H.}~\bibnamefont {Aksela}}, \bibinfo
  {author} {\bibfnamefont {E.}~\bibnamefont {Shigemasa}}, \bibinfo {author}
  {\bibfnamefont {A.}~\bibnamefont {Yagishita}},\ and\ \bibinfo {author}
  {\bibfnamefont {Y.}~\bibnamefont {Furusawa}},\ }\bibfield  {title} {\bibinfo
  {title} {Krypton 4p, 4s, and 3d partial photoionization cross sections below
  a photon energy of 260 ev},\ }\href
  {https://doi.org/10.1103/PhysRevA.45.4640} {\bibfield  {journal} {\bibinfo
  {journal} {Phys. Rev. A}\ }\textbf {\bibinfo {volume} {45}},\ \bibinfo
  {pages} {4640} (\bibinfo {year} {1992})}\BibitemShut {NoStop}%
\bibitem [{\citenamefont {Nikolopoulos}(2013)}]{Nikolopoulos2013}%
  \BibitemOpen
  \bibfield  {author} {\bibinfo {author} {\bibfnamefont {L.~A.~A.}\
  \bibnamefont {Nikolopoulos}},\ }\bibfield  {title} {\bibinfo {title}
  {Time-dependent theory of angular correlations in sequential double
  ionization},\ }\href {https://doi.org/10.1103/PhysRevLett.111.093001}
  {\bibfield  {journal} {\bibinfo  {journal} {Phys. Rev. Lett.}\ }\textbf
  {\bibinfo {volume} {111}},\ \bibinfo {pages} {093001} (\bibinfo {year}
  {2013})}\BibitemShut {NoStop}%
\bibitem [{\citenamefont {Gryzlova}\ \emph {et~al.}(2011)\citenamefont
  {Gryzlova}, \citenamefont {Ma}, \citenamefont {Fukuzawa}, \citenamefont
  {Motomura}, \citenamefont {Yamada}, \citenamefont {Ueda}, \citenamefont
  {Grum-Grzhimailo}, \citenamefont {Kabachnik}, \citenamefont {Strakhova},
  \citenamefont {Rouz\'ee}, \citenamefont {Hundermark}, \citenamefont
  {Vrakking}, \citenamefont {Johnsson}, \citenamefont {Nagaya}, \citenamefont
  {Yase}, \citenamefont {Mizoguchi}, \citenamefont {Yao}, \citenamefont
  {Nagasono}, \citenamefont {Tono}, \citenamefont {Togashi}, \citenamefont
  {Senba}, \citenamefont {Ohashi}, \citenamefont {Yabashi},\ and\ \citenamefont
  {Ishikawa}}]{Gryzlova2011}%
  \BibitemOpen
  \bibfield  {author} {\bibinfo {author} {\bibfnamefont {E.~V.}\ \bibnamefont
  {Gryzlova}}, \bibinfo {author} {\bibfnamefont {R.}~\bibnamefont {Ma}},
  \bibinfo {author} {\bibfnamefont {H.}~\bibnamefont {Fukuzawa}}, \bibinfo
  {author} {\bibfnamefont {K.}~\bibnamefont {Motomura}}, \bibinfo {author}
  {\bibfnamefont {A.}~\bibnamefont {Yamada}}, \bibinfo {author} {\bibfnamefont
  {K.}~\bibnamefont {Ueda}}, \bibinfo {author} {\bibfnamefont {A.~N.}\
  \bibnamefont {Grum-Grzhimailo}}, \bibinfo {author} {\bibfnamefont {N.~M.}\
  \bibnamefont {Kabachnik}}, \bibinfo {author} {\bibfnamefont {S.~I.}\
  \bibnamefont {Strakhova}}, \bibinfo {author} {\bibfnamefont {A.}~\bibnamefont
  {Rouz\'ee}}, \bibinfo {author} {\bibfnamefont {A.}~\bibnamefont
  {Hundermark}}, \bibinfo {author} {\bibfnamefont {M.~J.~J.}\ \bibnamefont
  {Vrakking}}, \bibinfo {author} {\bibfnamefont {P.}~\bibnamefont {Johnsson}},
  \bibinfo {author} {\bibfnamefont {K.}~\bibnamefont {Nagaya}}, \bibinfo
  {author} {\bibfnamefont {S.}~\bibnamefont {Yase}}, \bibinfo {author}
  {\bibfnamefont {Y.}~\bibnamefont {Mizoguchi}}, \bibinfo {author}
  {\bibfnamefont {M.}~\bibnamefont {Yao}}, \bibinfo {author} {\bibfnamefont
  {M.}~\bibnamefont {Nagasono}}, \bibinfo {author} {\bibfnamefont
  {K.}~\bibnamefont {Tono}}, \bibinfo {author} {\bibfnamefont {T.}~\bibnamefont
  {Togashi}}, \bibinfo {author} {\bibfnamefont {Y.}~\bibnamefont {Senba}},
  \bibinfo {author} {\bibfnamefont {H.}~\bibnamefont {Ohashi}}, \bibinfo
  {author} {\bibfnamefont {M.}~\bibnamefont {Yabashi}},\ and\ \bibinfo {author}
  {\bibfnamefont {T.}~\bibnamefont {Ishikawa}},\ }\bibfield  {title} {\bibinfo
  {title} {Doubly resonant three-photon double ionization of ar atoms induced
  by an euv free-electron laser},\ }\href
  {https://doi.org/10.1103/PhysRevA.84.063405} {\bibfield  {journal} {\bibinfo
  {journal} {Phys. Rev. A}\ }\textbf {\bibinfo {volume} {84}},\ \bibinfo
  {pages} {063405} (\bibinfo {year} {2011})}\BibitemShut {NoStop}%
\bibitem [{\citenamefont {Shwartz}\ \emph {et~al.}(2014)\citenamefont
  {Shwartz}, \citenamefont {Fuchs}, \citenamefont {Hastings}, \citenamefont
  {Inubushi}, \citenamefont {Ishikawa}, \citenamefont {Katayama}, \citenamefont
  {Reis}, \citenamefont {Sato}, \citenamefont {Tono}, \citenamefont {Yabashi},
  \citenamefont {Yudovich},\ and\ \citenamefont {Harris}}]{Shwartz2014}%
  \BibitemOpen
  \bibfield  {author} {\bibinfo {author} {\bibfnamefont {S.}~\bibnamefont
  {Shwartz}}, \bibinfo {author} {\bibfnamefont {M.}~\bibnamefont {Fuchs}},
  \bibinfo {author} {\bibfnamefont {J.~B.}\ \bibnamefont {Hastings}}, \bibinfo
  {author} {\bibfnamefont {Y.}~\bibnamefont {Inubushi}}, \bibinfo {author}
  {\bibfnamefont {T.}~\bibnamefont {Ishikawa}}, \bibinfo {author}
  {\bibfnamefont {T.}~\bibnamefont {Katayama}}, \bibinfo {author}
  {\bibfnamefont {D.~A.}\ \bibnamefont {Reis}}, \bibinfo {author}
  {\bibfnamefont {T.}~\bibnamefont {Sato}}, \bibinfo {author} {\bibfnamefont
  {K.}~\bibnamefont {Tono}}, \bibinfo {author} {\bibfnamefont {M.}~\bibnamefont
  {Yabashi}}, \bibinfo {author} {\bibfnamefont {S.}~\bibnamefont {Yudovich}},\
  and\ \bibinfo {author} {\bibfnamefont {S.~E.}\ \bibnamefont {Harris}},\
  }\bibfield  {title} {\bibinfo {title} {X-ray second harmonic generation},\
  }\href {https://doi.org/10.1103/PhysRevLett.112.163901} {\bibfield  {journal}
  {\bibinfo  {journal} {Phys. Rev. Lett.}\ }\textbf {\bibinfo {volume} {112}},\
  \bibinfo {pages} {163901} (\bibinfo {year} {2014})}\BibitemShut {NoStop}%
\bibitem [{\citenamefont {Kettle}\ \emph {et~al.}(2021)\citenamefont {Kettle},
  \citenamefont {Aquila}, \citenamefont {Boutet}, \citenamefont {Bucksbaum},
  \citenamefont {Carini}, \citenamefont {Feng}, \citenamefont {Gamboa},
  \citenamefont {Ghimire}, \citenamefont {Glenzer}, \citenamefont {Hart},
  \citenamefont {Hastings}, \citenamefont {Henighan}, \citenamefont {Hunter},
  \citenamefont {Koglin}, \citenamefont {Kozina}, \citenamefont {Liu},
  \citenamefont {MacDonald}, \citenamefont {Trigo}, \citenamefont {Reis},\ and\
  \citenamefont {Fuchs}}]{Kettle2021}%
  \BibitemOpen
  \bibfield  {author} {\bibinfo {author} {\bibfnamefont {B.}~\bibnamefont
  {Kettle}}, \bibinfo {author} {\bibfnamefont {A.}~\bibnamefont {Aquila}},
  \bibinfo {author} {\bibfnamefont {S.}~\bibnamefont {Boutet}}, \bibinfo
  {author} {\bibfnamefont {P.~H.}\ \bibnamefont {Bucksbaum}}, \bibinfo {author}
  {\bibfnamefont {G.}~\bibnamefont {Carini}}, \bibinfo {author} {\bibfnamefont
  {Y.}~\bibnamefont {Feng}}, \bibinfo {author} {\bibfnamefont {E.}~\bibnamefont
  {Gamboa}}, \bibinfo {author} {\bibfnamefont {S.}~\bibnamefont {Ghimire}},
  \bibinfo {author} {\bibfnamefont {S.}~\bibnamefont {Glenzer}}, \bibinfo
  {author} {\bibfnamefont {P.}~\bibnamefont {Hart}}, \bibinfo {author}
  {\bibfnamefont {J.~B.}\ \bibnamefont {Hastings}}, \bibinfo {author}
  {\bibfnamefont {T.}~\bibnamefont {Henighan}}, \bibinfo {author}
  {\bibfnamefont {M.}~\bibnamefont {Hunter}}, \bibinfo {author} {\bibfnamefont
  {J.}~\bibnamefont {Koglin}}, \bibinfo {author} {\bibfnamefont
  {M.}~\bibnamefont {Kozina}}, \bibinfo {author} {\bibfnamefont
  {H.}~\bibnamefont {Liu}}, \bibinfo {author} {\bibfnamefont {M.~J.}\
  \bibnamefont {MacDonald}}, \bibinfo {author} {\bibfnamefont {M.}~\bibnamefont
  {Trigo}}, \bibinfo {author} {\bibfnamefont {D.~A.}\ \bibnamefont {Reis}},\
  and\ \bibinfo {author} {\bibfnamefont {M.}~\bibnamefont {Fuchs}},\ }\bibfield
   {title} {\bibinfo {title} {Anomalous two-photon compton scattering},\ }\href
  {https://doi.org/10.1088/1367-2630/ac3553} {\bibfield  {journal} {\bibinfo
  {journal} {New Journal of Physics}\ }\textbf {\bibinfo {volume} {23}},\
  \bibinfo {pages} {115008} (\bibinfo {year} {2021})}\BibitemShut {NoStop}%
\end{thebibliography}%

\end{document}